\documentclass[twocolumn,english,preprintnumbers,nofootinbib,showpacs,showkeys,pra]{revtex4}

\usepackage{cancel}
\usepackage{epsfig}
\usepackage{mathrsfs}
\usepackage{graphicx}
\usepackage{dcolumn}
\usepackage{bm}
\usepackage{fancyhdr} 
\usepackage{verbatim}
\usepackage[all, knot]{xy}
\xyoption{arc}
\usepackage{amsmath}
\usepackage{amssymb}
\usepackage{revsymb}
\usepackage{slashed}

\usepackage[usenames]{xcolor}

\definecolor{red_ud}{rgb}{1,0,0} 
\definecolor{red_cs}{rgb}{.75,0,0} 
\definecolor{red_tb}{rgb}{.5,0,0} 

\definecolor{blue_ud}{rgb}{0,0,1} 
\definecolor{blue_cs}{rgb}{0,0,.75} 
\definecolor{blue_tb}{rgb}{0,0,.5} 

\definecolor{green_ud}{rgb}{0,1,0} 
\definecolor{green_cs}{rgb}{0,.75,0} 
\definecolor{green_tb}{rgb}{0,.5,0} 

\definecolor{lilac}{rgb}{.78, .05, .78}

\usepackage{subfigure} 

\begin{document}

\title{Quantum lattice gas model of Dirac particles in 1+1 dimensions}

\author{Jeffrey Yepez}
\affiliation{Air Force Research Laboratory/Directed Energy, Air Force Maui Optical \& Supercomputing Observatory, Kihei, Hawai`i  96753\\
and the Department of Physics and Astronomy, University of  Hawai`i at Manoa\\
Watanabe Hall, 2505 Correa Road, Honolulu, Hawai`i 96822
 }

\date{June 5, 2013}

\begin{abstract}
Presented is a  quantum computing representation of Dirac particle dynamics.   The  approach employs an operator splitting method that is an analytically closed-form product decomposition of the unitary evolution operator.  This allows the Dirac equation to be cast as a unitary finite-difference equation in a high-energy limit. The split evolution operator (with separate kinetic  and interaction terms) is useful for  efficient quantum simulation.   For pedagogical purposes, here  we restrict the treatment to Dirac particle dynamics in 1+1 spacetime dimensions.  Independent derivations of the quantum algorithm are presented and the model's validity is tested in several quantum simulations by comparing the numerical results against analytical predictions.  Using the relativistic quantum algorithm in the case when $mc^2 \gg pc$, quantum simulations of a nonrelativistic particle in an external scalar square well and parabolic potential is presented. 
\end{abstract}



\maketitle

\section{Introduction}

Here we consider a discrete unitary model of a quantum gas confined to a lattice. The model is called a quantum lattice gas.  The model is useful for simulating many-body systems of strongly-correlated Dirac particles \cite{PhysRevLett_2013_1}---here we treat the simplest version  of the model and thus restrict our study to quantum particle dynamics in 1+1 dimensions.  Quantum lattice gases were one of the earliest quantum algorithms devised \cite{riazanov-spj58,feynman-65-1st-qlga,JSP.53.323,FPL.10.105,yepez-afosr96,yepez_96_tech_report,PhysRevD.49.6920,PhysRevA.54.1106,PhysRevE.55.5261,PhysRevE.57.54}.  Feynman's original representation of a path integral was the first  quantum lattice gas algorithm \cite{feynman-cit46,feynman-rmp48,feynman-65}, 
 commonly known as the Feynman chessboard model  \cite{jacobson-jpamg84}.  This kind of algorithmic representation of quantum mechanics was the foundational idea that led Feynman to conjecture that there should exist a universal quantum computational model for efficient quantum simulation \cite{feynman-ces60,feynman-82,feynman-85}.   The quantum lattice gas model presented here is an improved version of an earlier  model \cite{yepez-qip-05}.  The improved version has  its unitary evolution operator generated by the Dirac Hamiltonian with no spatial error terms.  The evolution operator is represented as a product of a unitary operator for the kinetic part of the evolution and a unitary operator for the particle-particle interaction part of the evolution, but not as a Trotter decomposition \cite{Trotter_JSTOR_1959}.   Thus, in the improved model, we avoid the need for the grid's  space and time scales to be infinitesimal for the model to be a faithful representation of the quantum particle dynamics.

In a quantum lattice gas system, a particle's occurrence at a point is encoded using a complex-valued probability amplitude.  In a quantum lattice gas, at the grid-level, a particle's occurrence is encoded with a qubit.   The grid-level kinetic transport equation is a discrete (or finite-difference) version of a quantum wave equation.  That is, particle motion is  restricted along a finite set of displacement vectors, yet 
 each displacement vector in general  has complex-valued components.  Hence, the lattice defined by this class of displacement vectors constitutes a quantum network.  In such a quantum network, even without a chiral breaking interaction, in two or more spatial dimensions a free massless particle initially localized in space will become delocalized over time, whereas in one spatial dimension, spontaneous delocalization of a wave packet occurs only for massive particles.  The dispersion of a particle's  wave packet behaves quantum mechanically according to the Heisenberg uncertainty principle.  Particle-particle interactions are represented by entangling quantum gates implemented at the grid points, so all the dynamics is strictly unitary and thus manifestly reversible.

\subsection{Overview of the modeling approach}

\begin{figure}[!h!t!b!p]
\begin{center}
\subfigure[  \;$N=60$ and $M=40$]{\includegraphics[width=1.65in]{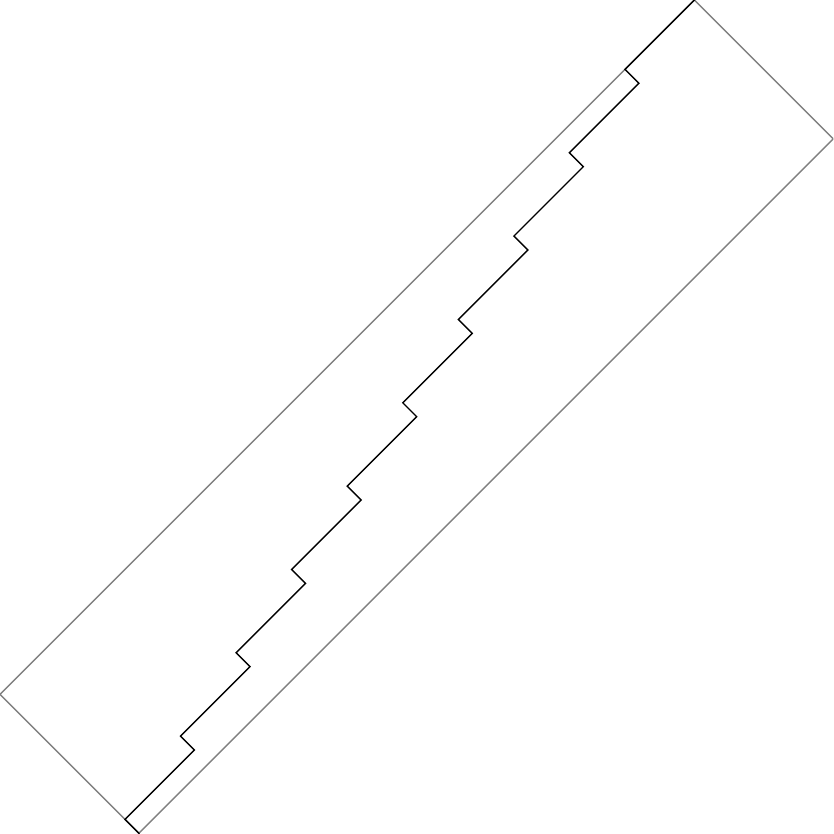}}
\subfigure[ \;$N=600$ and $M=400$]{\includegraphics[width=1.65in]{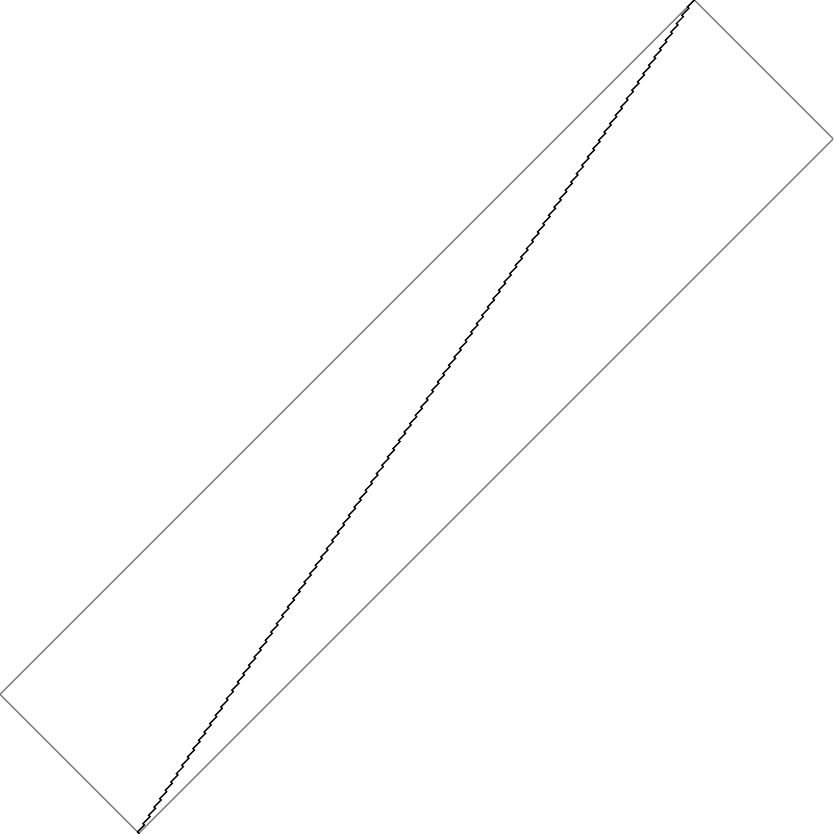}}%
\caption{\label{ZigZagPathRightTriangleN6M2} \footnotesize Example paths at two different energy resolutions $E\sim {1}/{(N\tau)}$ for $N$ time steps where the net displacement in space is two-thirds the displacement in time, $M = 2 N/3$. The light cones are rendered in gray and the mean path (or ``hypothenuse") in black.}
\end{center}
\end{figure}

Let us consider two paths in a 1+1 dimensional spacetime, with both paths starting at point $a=(0, 0)$ and ending at point $b=(N \tau, 0)$, for example, rendered at energy scale $E\sim {1}/{N\tau}$ for $N=60$ time steps
\begin{subequations}
\begin{equation}
\label{ZigZagPathIsoscelesTriangle60}
\xy
(0,-24)*{a};
(0,24)*{b};
(-5,0)*{h};
(13,13)*{y};
(13,-13)*{x};
(0,0)*{\includegraphics[width=1.65in]{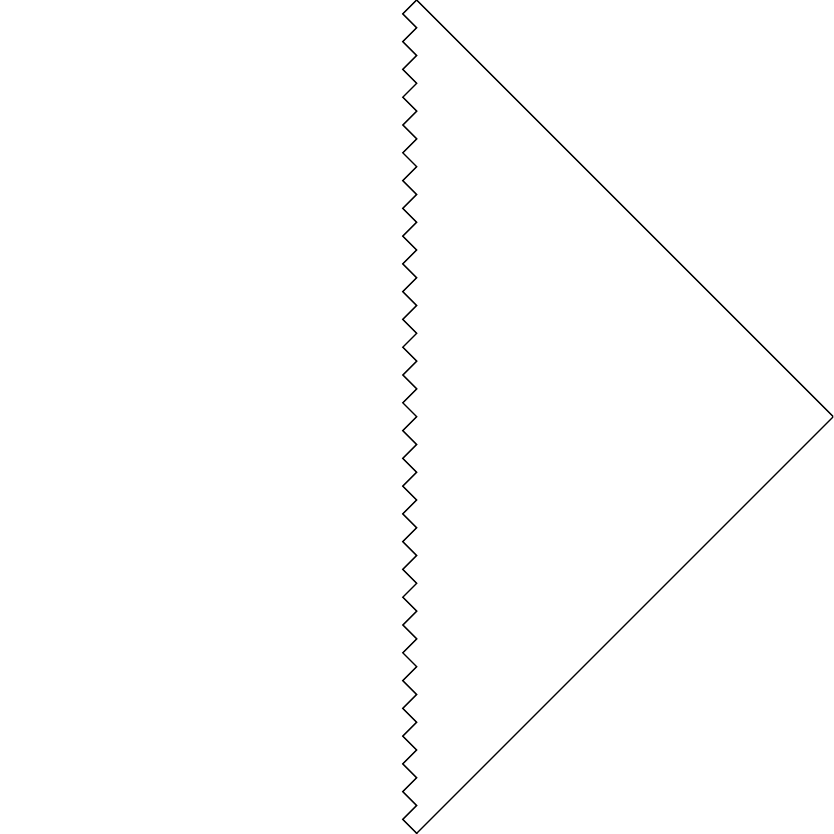}};
\endxy
\end{equation}
with time increasing going upward.  
A particle's trajectory is encoded by the motion of a bit, which moves at the speed of light $c=\ell/\tau$, so each path segment of a trajectory is rendered as a line with slope of $\pm 45$ degrees.  The first path has $N-1=59$ bends and represents the trajectory of a particle a rest;  this path length is $h$.  The second path has $1$ bend and represents the trajectory of a particle moving on the light cone;  this path length  is the sum of the two legs $x + y$.   The two paths shown in (\ref{ZigZagPathIsoscelesTriangle60}) are of equal lengths, so we have the linear identity
\begin{equation}
\label{high_energy_Pythagorean_theorem}
h = x + y.
\end{equation}
\end{subequations}
Now, we may consider the geometry of these two paths at an energy scale at an order of magnitude lower say, $E\sim {1}/{600\tau}$
\begin{subequations}
\begin{equation}
\label{ZigZagPathIsoscelesTriangle600}
\xy
(0,-24)*{a};
(0,24)*{b};
(-5,0)*{h_\text{\tiny L.E.}};
(13,13)*{y_\text{\tiny L.E.}};
(13,-13)*{x_\text{\tiny L.E.}};
(0,0)*{\includegraphics[width=1.65in]{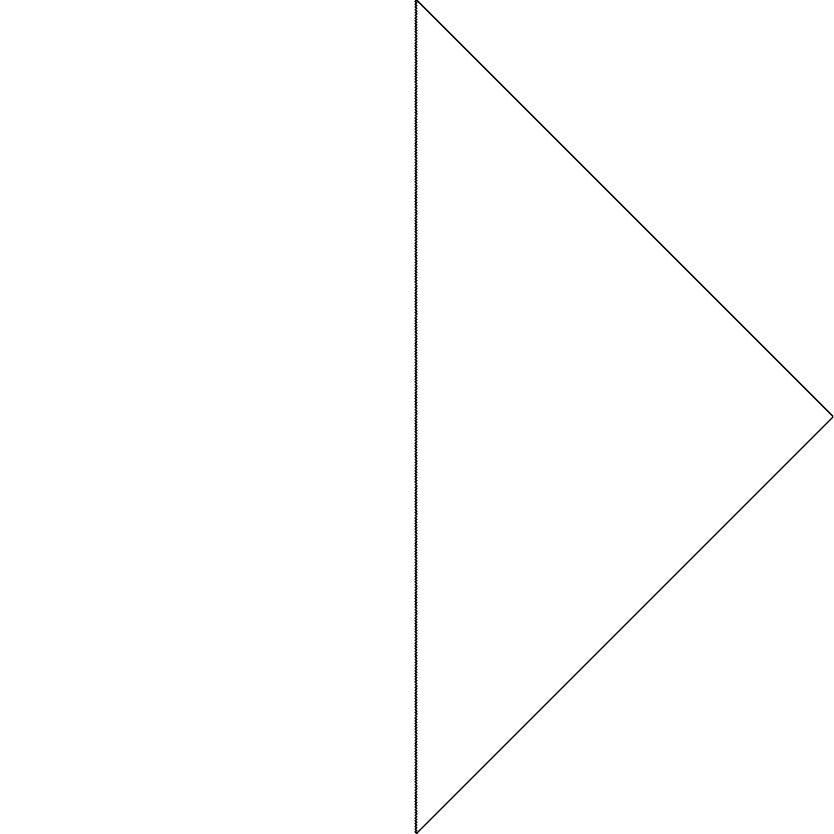}};
\endxy
\end{equation}
One can zoom in (on-line version) and see that both paths here actually have equal lengths according to (\ref{high_energy_Pythagorean_theorem}). However, in the low-energy ({\scriptsize L.E.}) limit, the path lengths satisfy the Pythagorean theorem
\begin{equation}
\label{Pythagorean_theorem}
h_\text{\tiny L.E.} = \sqrt{x_\text{\tiny L.E.}^2  + y_\text{\tiny L.E.}^2 },
\end{equation}
\end{subequations}
so the hypothenuse is  effectively shorter, by a factor of $1/\sqrt{2}$, than the sum of the legs of the isosceles triangle.  Exploiting the anticommutativity of the Pauli matrices $\{\sigma_x, \sigma_y\}=\sigma_x \sigma_y+\sigma_y \sigma_x=0$, we may write  (\ref{Pythagorean_theorem})  in an operator form akin to (\ref{high_energy_Pythagorean_theorem})
\begin{equation}
\label{linear_Pythagorean_theorem_operator_form}
h_\text{\tiny L.E.} =  x_\text{\tiny L.E.} \sigma_x+y_\text{\tiny L.E.}\sigma_y .
\end{equation}
The device of the Dirac matrices allows us to linearize the Pythagorean theorem by taking its square root. 
In the quantum lattice gas model, the trajectory of a Dirac particle is represented as a  superposition of all the paths\footnote{All paths going from initial point to final point have equal length in the high-energy limit.} bounded by the light cone originating at an initial point and the inverse light cone terminating at some final point, for example as shown in Fig.~\ref{ZigZagPathRightTriangleN6M2}.
Furthermore, in the quantum lattice gas model, along the ``hypothenuse" (which in the low-energy limit represents the classical straight-line path connecting the initial and final points), the number of bends, $R$,  is parametrized by the particle's mass, while the number of unbent adjacent pairs of path segments, $\overline{R}=N-R$, is parametrized by the particle's momentum.  The greater the ratio of the momentum eigenvalue to the mass, the greater the spatial displacement.  The maximal displacement for a given number of time steps $N$ occurs for a massless particle, so the classical path falls along the light cone.  At any point, the probability that a particle's path makes a bend is equal to $(m c^2 \tau/\hbar)^2$.  So the probability that the particle's path is unbent at that point is equal to  $1-(m c^2 \tau/\hbar)^2$.

To every path is assigned  a probability amplitude (c-number) according to the rule:  every  pair of path segments contributes a multiplicative factor of $- i m c^2\tau\, e^{-i\xi}/\hbar$ if bent and  $\sqrt{1-(m c^2 \tau/\hbar)^2}$ if unbent, where the phase angle  $\xi \equiv p \ell/\hbar$ is determined by the momentum eigenvalue $p$. Hence, the phase factor $e^{-i\xi}\approx 1$ in the nonrelativistic limit. 
  In 1+1 dimensions, the quantum state at a point is represented by a 2-spinor
\begin{equation}
\psi(t,z) 
=
\begin{pmatrix}
      \psi_{\uparrow}(t,z)     \\
         \psi_{\downarrow}(t,z) 
\end{pmatrix},
\end{equation}
and thus the local unitary update rule applied at each point is 
\begin{subequations}
\label{qlg_steps}
\begin{equation}
\label{collide_step}
\psi'(t,z) = 
\begin{pmatrix}
     \sqrt{1-(m c^2 \tau/\hbar)^2} &   - i m c^2\tau \,e^{-i\xi}/\hbar \\
- i m c^2\tau\,e^{i\xi}/\hbar      &  \sqrt{1-(m c^2 \tau/\hbar)^2}
\end{pmatrix}
\psi(t,z).
\end{equation}
So the unitary matrix in (\ref{collide_step}) represents a chiral symmetry breaking interaction.  To complete the specification of the quantum lattice gas algorithm, one additional unitary operation is applied to the 2-spinor representing kinetic transport of the particles
\begin{equation}
\label{stream_step}
\psi(t+\tau,z) = 
e^{i \sigma_z \hat{p}_z \ell/\hbar}
\psi'(t,z).
\end{equation}
\end{subequations}
This is $\psi(t+\tau,z -\sigma_z \ell) = 
\psi'(t,z)$ since the momentum operator $\hat{p}_z = -i \hbar \partial_z$ in the exponent causes a fixed displacement.  The algorithmic steps (\ref{collide_step}) and (\ref{stream_step}) can be combined into a single equation of motion
\begin{equation}
\label{quantum_lattice_transport_equation}
\psi(t+\tau, z-\sigma_z \ell) = 
{\scriptsize
\begin{pmatrix}
     \sqrt{1-(m c^2 \tau/\hbar)^2} &   - i m c^2\tau\,e^{-i\xi}/\hbar \\
- i m c^2\tau\,e^{i\xi}/\hbar      &  \sqrt{1-(m c^2 \tau/\hbar)^2}
\end{pmatrix}
}
\psi(t,z).
\end{equation}
This type of high-energy representation of a quantum lattice gas \cite{PhysRevE.63.046702}. 
In component form, this is a set of coupled finite-difference equations
\begin{subequations}
\begin{eqnarray}
\nonumber
\psi_{\uparrow}(t+\tau, z-\ell) 
\!\!\!&=& \!\!\!
     \sqrt{1-(m  \tau)^2} \, \psi_{\uparrow}(t,\!z)    - i m \tau e^{-i\xi}\psi_{\downarrow}(t,\! z)   
     \\
     \\
\nonumber
\psi_{\downarrow}(t+\tau, z+\ell) 
\!\!\!&=& \!\!\!
- i m \tau e^{i\xi} \psi_{\uparrow}(t,\!z)   +  \sqrt{1-(m \tau)^2}\, \psi_{\downarrow}(t,\! z) ,
\\
\end{eqnarray}
\end{subequations}
using natural units ($c=1$ and $\hbar=1$).  With the single-particle probability of occurrence defined as $ f_{\uparrow,\downarrow}(t, z) \equiv | \psi_{\uparrow,\downarrow}(t, z) |^2$, multiplying (\ref{quantum_lattice_transport_equation}) by its complex conjugate gives the relativistic lattice Boltzmann  equation 
\begin{equation}
\label{relativistic_lattice_Boltzmann_equation}
f_{\uparrow,\downarrow}(t+\tau, z\mp \ell) =  f_{\uparrow,\downarrow}(t, z) + m^2 \tau^2 \left[f_{\downarrow,\uparrow}(t, z)-f_{\uparrow,\downarrow}(t, z) \right].
\end{equation}
So, as expected, the  quantity $(m c^2 \tau/\hbar)^2$ parametrizes the probability that a particle's path makes a bend at a point, as described above.  The quantum transport equation (\ref{quantum_lattice_transport_equation}) is indeed a square-root equation of (\ref{relativistic_lattice_Boltzmann_equation}). In the low-energy limit, written in differential point form, it is equivalent to the Dirac equation
\begin{equation}
\label{Dirac_equation_1_1_dimensions_no_errors_in_space}
i\hbar \partial_t \psi  + \cdots = (\alpha p_z c + \beta m c^2)\psi
\end{equation}
in the chiral representation with Dirac matrices $\alpha = -\sigma_z$ and $\beta = \sigma_x$.   The righthand side of (\ref{Dirac_equation_1_1_dimensions_no_errors_in_space}) is exact as there are no spatial error terms in the discrete representation.

\subsection{Organization}

Quantum lattice gas models are efficient quantum algorithms for simulating a Fermi system of particles on a quantum computer.  To see why this class of quantum algorithms is also naturally suited for quantum computers,  Sec. II explains many-body quantum lattice gas simulations.  Multiple quantum particles are  handled by the same evolution operator already found in the quantum algorithm for the single particle case. The protocol for quantum gate operations remains the same, independent of the number of particles to be simulated.  So, when implemented on a quantum computer, the quantum lattice gas model is an efficient quantum algorithm -- the number of states needed to encode the particle dynamics grows exponentially in the number of particles, but  the number of quantum gates operations remains fixed and simply proportional to the number of points in the system. 

Equation (\ref{quantum_lattice_transport_equation}) represents the quantum lattice gas algorithm for the Dirac equation.   Below, we derive this quantum lattice gas algorithm using two different analytical approaches, from the viewpoint of a composite rotation of a qubit on the Bloch sphere and  from the viewpoint of a Feynman path summation.  Sec.~\ref{Quantum_lattice_gas_as_composite_rotations} and \ref{Quantum_lattice_gas_as_a_path_summation} are devoted to these two derivations, respectively. 

Next, Sec.~V presents 1+1 dimensional quantum simulation examples of how to use the quantum lattice gas method as a computational physics tool.  The  examples include  simulations of a Dirac 2-spinor field.  These 1+1 dimensional simulations are rich in pedagogy.  One learns how the quantum lattice gas models may be implemented in a traditional digital electronic computing scheme.  Since they are locally computed models (i.e. with unitary transformations occurring independently at each point in the system), it is clear why this class of quantum algorithms is naturally suited for classically parallel computers when the number of Dirac particles in the modeled system is sufficiently small so that the Hilbert space can fit into the available memory of the supercomputer.  

For completeness of the presentation, Appendix A explains what a qubit is and presents a quantum algorithm (constructed from the point of view of composite rotations) for modeling the dynamics of a Dirac particle. 
 Entangling quantum gates are reviewed in Appendix B, and a particular focus on an analytical representation of quantum gates employing qubit creation and annihilation operators is presented. 
Appendix~\ref{Collide_operator_for_the_Dirac_equation} gives a derivation of the chiral symmetry breaking operator for the Dirac equation represented as an entangling quantum gate.  This interaction operator, presented in both its analytical and matrix forms,  is needed to implement many-body quantum simulations.

\section{Quantum lattice gas as a quantum algorithm}

To program a quantum computer to simulate the dynamical behavior of system of Dirac particles,
it is necessary to specify an algorithmic protocol that represents a desired Dirac Hamiltonian as an engineered hermitian generator.  We present an  algorithm  that is second order accurate in space. 
After an algorithmic scheme is formulated, we may then verify a posteriori  by direct numerical simulation that it is indeed at least a second-order convergent numerical scheme.

\subsection{Encoding the wave function}

Present here is a way to  encode the value of each Dirac particle's spinor field. 
For simplicity, we  consider a 1+1 dimensional quantum system, but its generalization to higher dimensions is straightforward.  
In one spatial dimension, there are two physical degrees of freedom:  the Dirac particle can either move to the left or to the right.
Thus, for a one-dimensional lattice with $L$ number of points, the numbered state by the ket
\begin{equation}
\label{numbered_position_spin_basis_states}
 |n^0_0 n^0_1 n^1_0 n^1_1 n^2_0 n^2_1\cdots n^{L-1}_0 n^{L-1}_1\rangle, 
\end{equation}
 where $n^l_a = 0$ or $1$ for all $l$ and $a$, is sufficient to represent the particle motion.  We will use the convention that $n^l_0$ represents a right goer and $n^l_1$ represents a left goer.  The  variables $n^l_a$ are Boolean quantities and are  number variables of a system of spin-$1/2$ particles that obey Fermi-Dirac statistics.  The superscript $l$ indexes to a point on the grid and the subscript $a$ that specifies the direction of  motion can also be interpreted as a spin index for the spin-$1/2$ particle.   Also, the numbered variables correspond to the bit values of a binary encoded integer that labels each numbered state in the range from $0$ to $2^{2L}-1$.

According to the encoding scheme (\ref{numbered_position_spin_basis_states}), to each point in space we may associate a position-spin basis ket denoted by $|x_l, \sigma\rangle$, where $0\ge l\ge L-1$ and $\sigma = 0$ (spin up $\uparrow$) or $\sigma=1$ (spin down $\downarrow$).  The position-spin ket $|x_l, \sigma\rangle$ is the numbered state with $n_\sigma^l\in [0,1]$ with all the other numbered variables being set to zero
\begin{subequations}
\label{position_spin_kets_for_2_qubits_per_point}
\begin{eqnarray}
\label{position_spin_kets_for_2_qubits_per_point_a}
|x_l, 0\rangle & = & | 00\quad  \cdots \underbrace{00}_{l^\text{th} \text{ point}} \cdots \quad 00 \rangle \\
|x_l, \uparrow\rangle & = & | 00\quad  \cdots \underbrace{10}_{l^\text{th} \text{ point}} \cdots \quad 00 \rangle \\
|x_l, \downarrow\rangle & = & | 00\quad  \cdots \underbrace{01}_{l^\text{th} \text{ point}} \cdots \quad 00 \rangle \\
\label{position_spin_kets_for_2_qubits_per_point_d}
|x_l, \uparrow\downarrow\rangle & = & | 00\quad  \cdots \underbrace{11}_{l^\text{th} \text{ point}} \cdots \quad 00 \rangle .
\end{eqnarray}
\end{subequations}
  In the  ket (\ref{position_spin_kets_for_2_qubits_per_point_d}), the doubly occupied point has a particular ordering of the spin degrees of freedom, spin-up followed by spin-down, and only that ordering is encoded. Kets of  type (\ref{position_spin_kets_for_2_qubits_per_point_d}) reside in the two-body subsector of the Hilbert space.  

The notion of position and spin are intrinsically linked in this construction.
The position-spin kets are orthonormal
\begin{equation}
\langle x_l, \sigma | x_m, \sigma'\rangle = \delta_{lm} \delta_{\sigma\sigma'}.
\end{equation}
For the convenience afforded by specifying the quantum algorithm in the one-particle sector, we  use the notation
\begin{equation}
\label{2_spinor_notation_convention}
 |x_l,\uparrow\rangle \equiv |0\rangle_{x_l}
 =
 \begin{pmatrix}
      1   \\
      0  
\end{pmatrix}_{\!\!x_l},
\qquad
 |x_l,\downarrow\rangle \equiv |1\rangle_{x_l}
 =
 \begin{pmatrix}
      0   \\
      1  
\end{pmatrix}_{\!\!x_l}.
\end{equation}

The quantum state representing a Dirac particle is in general a superposition state in both position-space and spin-space. 
Thus, at time $t$, the full quantum state of a Dirac particle, say $|\psi(t)\rangle$ on the grid 
in the position-spin representation, is a quantum superposition state over all the points of the system
\begin{subequations}
\begin{equation}
\label{system_ket_position_rep}
|\psi^{\text{1-body}}(t)\rangle = 
 \sum_{l=0}^{L-1}
  \sum_{\sigma=0}^{1}  \psi_\sigma(t,x_l) |x_l,\sigma\rangle,
\end{equation}
where the probability amplitude $ \psi_\sigma(x_l) =\langle x_l, \sigma|\psi\rangle$ is a complex number.  The general one-body quantum state (\ref{system_ket_position_rep}) may be rewritten as
\begin{equation}
\label{system_ket_position_rep_2}
|\psi^{\text{1-body}}(t)\rangle
\stackrel{(\ref{2_spinor_notation_convention})}{=}
 \sum_{l=0}^{L-1}
\begin{pmatrix}
      \psi_\uparrow(t,x_l)    \\
      \psi_\downarrow(t,x_l)
\end{pmatrix}
,
\end{equation}
%
where we drop  $x_l$ as a subscript on the 2-spinor to avoid reluctancy as the particular position is already specified by the functional dependency of the component probability amplitudes.
 The  set of probability amplitudes $ \psi_\sigma(x_l)$ for integer $l\in[0,L-1]$ is a discrete representation of the continuous quantum field $\psi_\sigma(x)$ in space with $0\le x < L$ for particle with spin $\sigma=\uparrow,\downarrow$.
In other words, the basic approach we use  represents a state of the system  $|\psi\rangle$ with a quantum field  $\psi(x)$ expressed as a sum of all the possible ways a particle can be situated in position-space and spin-space superposition with a probability amplitude $\psi_\sigma(x_l)$ associated with each position-spin ket $|x_l, \sigma\rangle$.    Since the quantum algorithm is a locally computed one, we employ the convention of writing the quantum state in local 2-spinor notation
\begin{equation}
\label{system_ket_position_2_spinor_rep}
\psi^{\text{1-body}}(t, x_l) 
=
\begin{pmatrix}
      \psi_\uparrow(t,x_l)    \\
      \psi_\downarrow(t,x_l)
\end{pmatrix}.
\end{equation}
The form of (\ref{system_ket_position_2_spinor_rep}) is sufficient to specify the quantum lattice gas algorithm.

 So, if we are concerned with modeling a one-particle system, then all we need consider is a finite set of numbered basis states, where each element of the  set has the form (\ref{numbered_position_spin_basis_states}) with only one of the number variables equaling 1 and all the others 0.  This subset of all the possible numbered basis states is called the one-body sector and the quantum states in the one-body sector may be denoted as (\ref{system_ket_position_rep}) or (\ref{system_ket_position_rep_2}).  Yet, there is another and very useful way to denote such quantum states.  Since there are $2L$ probability amplitudes, we may label the numbered state associated with each these probability amplitudes using the binary encoded integer $|2^{2l+a}\rangle$, for $a\in[0,1]$ and $ l\in [0,L)$. Therefore, the system ket in the number representation can be alternatively written as
\begin{equation}
\label{waveeq_system_ket_number_rep}
|\psi^{\text{1-body}}(t)\rangle = \sum_{l=0}^{L-1}\sum_{a=0}^1 \psi_{2l+a}(t) |2^{2l+a}\rangle,
\end{equation}
where each $\psi_{2l+a}(t)$ is a c-number \cite{yepez-cpc01}. That is, the  position-spin kets $|x_l,\sigma\rangle \equiv |2^{2l+\sigma}\rangle$ comprise a set of $2L$ vectors of the 
full Hilbert space of  the quantum computer.  We could equivalently  write (\ref{waveeq_system_ket_number_rep}) in  the compact way
\begin{equation}
\label{waveeq_system_ket_number_rep_alpha}
|\psi^{\text{1-body}}\rangle = \sum_{\alpha=1}^{2L} \psi_{\alpha} |2^{\alpha-1}\rangle,
\end{equation}
\end{subequations}
where we label the qubits starting with 1 say.

The binary integer encoding (\ref{numbered_position_spin_basis_states}) readily accommodates the specification of many-body quantum states.  For example, a general system ket in the two-body sector is simply written as
\begin{equation}
\label{2_body_system_ket_number_rep_alpha_beta}
|\psi^{\text{2-body}}(t)\rangle = \sum_{\alpha=1}^{2L}\sum_{\beta=\alpha+1}^{2L} \psi_{\alpha,\beta}(t) |2^{\alpha-1}+2^{\beta-1}\rangle,
\end{equation}
and in the three-body sector a general system ket is written as
\begin{equation}
\label{3_body_system_ket_number_rep_alpha_beta}
|\psi^{\text{3-body}}\rangle = \sum_{\alpha=1}^{2L}\sum_{\beta=\alpha+1}^{2L}\sum_{\gamma=\beta+1}^{2L} \psi_{\alpha,\beta,\gamma} |2^{\alpha-1}+2^{\beta-1}+2^{\gamma-1}\rangle,
\end{equation}
and so forth.

\subsection{Encoding scheme using qubits}

Since the number of qubits in any quantum computer is necessarily a finite number, each particle's quantum state will have to be approximated  
by representing a physically continuous amplitude field by an ordered and finite set of complex numbers.  

To encode  kets such as (\ref{numbered_position_spin_basis_states}) in a quantum computer, we would need to assign at least two qubits to each point\footnote{To model a Dirac 4-spinor field in 3+1 dimensions, at least four qubits per point are needed so the quantum algorithm is useful for many-body quantum simulation \cite{yepez-qip-05}.
}
  So, at a minimum, one would need $Q=2L$ qubits in the host quantum computer.  The qubits that encode  the $l\hbox{th}$ point are denoted by $|q^l_a\rangle$ for $a\in [0,1]$.  
We consider each qubit to be a container that may or may not be occupied by the quantum particle. 
Since each qubit is a two-level quantum system  
 $|q^l_a\rangle = \alpha^l_a|0\rangle + \beta^l_a |1\rangle$ 
with $|\alpha^l_a|^2 + |\beta^l_a|^2 = 1$. A  review of qubit representations is given in Appendix~\ref{Qubit_review}. A quantum particle is said to occupy the $a\hbox{th}$ spin state at point $x_l$ when $\beta^l_a =1$, and the $a\hbox{th}$ spin state at point $x_l$ is empty when $\beta^1_a=0$.

Consider the simplest many-body example, a two-body system. In addition to redefining the local position-spin kets
\begin{subequations}
\begin{equation}
\label{4_spinor_notation_convention_perpendicular}
 |x_l,\uparrow\rangle \equiv |2\rangle_{x_l}
 =
 \begin{pmatrix}
 0\\
      0   \\
      1\\
      0  
\end{pmatrix}_{\!\!x_l},
\qquad
 |x_l,\downarrow\rangle \equiv |1\rangle_{x_l}
 =
 \begin{pmatrix}
 0\\
      1   \\
      0\\
      0  
\end{pmatrix}_{\!\!x_l},
\end{equation}
we also need to define the empty and doubly occupied local position-spin kets
\begin{equation}
\label{4_spinor_notation_convention_parallel}
 |x_l,0\rangle \equiv |0\rangle_{x_l}
 =
 \begin{pmatrix}
 1\\
      0   \\
      0\\
      0  
\end{pmatrix}_{\!\!x_l},
\qquad
 |x_l,\uparrow\downarrow\rangle \equiv |3\rangle_{x_l}
 =
 \begin{pmatrix}
 0\\
      0   \\
      0\\
      1  
\end{pmatrix}_{\!\!x_l}.
\end{equation}
\end{subequations}
Then, the two-body system ket (\ref{2_body_system_ket_number_rep_alpha_beta}) may be rewritten as
\begin{subequations}
\label{2_body_system_ket_number_4_spionr_rep}
\begin{equation}
\label{2_body_system_ket_number_4_spionr_rep_a}
|\psi^{\text{2-body}}(t)\rangle = \sum_{l=0}^{L-1}  
\begin{pmatrix}
      \psi_0(t, x_l)    \\
            \psi_\uparrow(t, x_l)    \\
      \psi_\downarrow(t, x_l)\\
      \psi_{\uparrow\downarrow}(t, x_l)
\end{pmatrix},
\end{equation}
and the local quantum state may be specified simply by a 4-spinor
\begin{equation}
\label{2_body_system_ket_number_4_spionr_rep_b}
\psi^{\text{2-body}}(t, x_l)
=
\begin{pmatrix}
      \psi_0(t, x_l)    \\
            \psi_\uparrow(t, x_l)    \\
      \psi_\downarrow(t, x_l)\\
      \psi_{\uparrow\downarrow}(t, x_l)
\end{pmatrix}
.
\end{equation}
\end{subequations}
Since only a 2-spinor is needed to represent a Dirac particle in 1+1 dimensions, the  spin-up and and spin-down probability amplitudes 
 in (\ref{2_body_system_ket_number_4_spionr_rep}) are each  encoded in a qubit
\begin{equation}
|q_\sigma(t, x_l)\rangle = \sqrt{1-|\psi_\sigma(t,x_l)|^2}\, |0\rangle + \psi_\sigma (t, x_l)|1\rangle,
\end{equation}
for $\sigma= \uparrow, \downarrow$. 
That is, the expectation value of the number operator gives the  probability of type $\sigma$ occupancy at point $x_l$ according to the formula
\begin{equation}
\langle q_\sigma (t, x_l)|n|q_a(t, x_l)\rangle = |\psi_\sigma(t, x_l)|^2,
\end{equation}
where $n$ is the singleton number operator.  Alternatively, using the quantum state of the entire system with $Q$ qubits, we may write the expectation value by using the multiple qubit number operator 
\begin{equation}
\langle \psi(t) |n_\alpha |\psi(t)\rangle = |\psi_a(t, x_l)|^2,
\end{equation}
where $\alpha= a  \;(\!\!\!\! \mod 4)$ and $\alpha$ is an integer-valued qubit index $\alpha\in[1,Q]$.  The multiple qubit number operators are reviewed in Appendix~\ref{qsm-unfolding-the-multiqubit-number-operator}. So the numbered state (\ref{numbered_position_spin_basis_states}) in the position-spin basis is encoded within the Hilbert space of the qubit system with the states
\begin{equation}
\label{numbered_qubit_basis_states}
\begin{split}
 |
 q_\uparrow(x_0) q_\downarrow(x_0)  
q_\uparrow(x_1) q_\downarrow(x_1)  
 \cdots q_\uparrow(x_{L-1}) q_\downarrow(x_{L-1}) \rangle.
\end{split}
\end{equation}

\subsection{Unitary stream and collide operators}

To simulate the dynamical behavior of a system of Dirac particles, we seek to specify a sequence of 2-qubit gate operations that will act on a  collection of qubits in a way that represents a particular quantum field theory.

Let us denote a ${\text{\scriptsize SWAP}}$ gate by $\chi$.
The stream operator, denoted $  S_a$ for $a=0,1$, causes a global shift to the right of the $a$th qubit on all the lattice nodes.  Therefore, $  S_a$ can be represented by a product of swaps acting on nearest neighbors
\begin{equation}
  S_a = \prod_{l=0}^{\frac{L-1}{2}}   \chi_{2l+a,2l+2+a}.
\end{equation}
Writing the full collide operator as a tensor product over all the points of the system, $  C=\bigotimes_{l=0}^{L-1} U_C$, the quantum algorithm we present for  the Dirac equation is a product of   collide  and  stream operators
\begin{equation}
\label{QLG_algorithm}
|\psi(t+\tau)\rangle =    S_1^{\hbox{\tiny T}} S_0^{} C |\psi(t)\rangle  = e^{-i\ell h_D/(\hbar c)}  |\psi(t)\rangle ,
\end{equation}
where $h_D$ is the Dirac Hamiltonian and $\ell$ and $\tau$ are the grid scale length and  time.
Here $  S_1^{\hbox{\tiny T}}$ denotes the transpose of $  S_1$ and is the inverse of $  S_1$.     Application of $  S_1^{\hbox{\tiny T}}$ causes a global shift to the left of the first qubit on all the lattice nodes.   
and where the streaming operator $  S_2$ causes a global shift to the right of the second qubit on all the lattice nodes. 
The collide operator is chosen so that the product decomposition in (\ref{QLG_algorithm}) is an exact representation of an evolution operator generated by $h_D$; that is, $S_1^{\hbox{\tiny T}} S_0^{} C |\psi(t)\rangle  = e^{-i\ell h_D/(\hbar c)}$ is not an approximation (in particular, both the Trotter's formula and the Baker-Campbell-Hausdorff formula are not needed so there are no error terms in the spatial derivatives).   The product of stream operators $S_1^{\hbox{\tiny T}} S_0^{}= e^{\sigma_z \ell  \partial_z}=e^{i p_z \ell/\hbar}$.

The efficiency of the quantum algorithm (\ref{QLG_algorithm}) becomes evident when it is used to simulate the dynamics of many quantum particles.  The case of multiple quantum particles is  handled by the same evolution operator used for the single particle case but generalized to handle Fermi-Dirac statistics. The collide operator may be expressed in terms of qubit creation and annihilation operators. The explicit form of this chirality breaking operator used in the quantum lattice gas algorithm for many-body quantum simulations is given in Appendix~\ref{Collide_operator_for_the_Dirac_equation}.  Yet, for simple test purposes, the quantum algorithm is implemented in Sec.~\ref{Quantum_simulation} in the one-body sector.  The particular sequence and number of quantum gate operations remains fixed, independent of the number of particles to be simulated.  The only difference is how the system's quantum state is initialized.

\section{Quantum lattice gas as composite rotations}
\label{Quantum_lattice_gas_as_composite_rotations}

In this section we present a quantum algorithm, constructed from the point of view of composite rotations, for modeling the dynamics of a Dirac particle in 1+1 dimensions.  The idea is to constrain the lattice-based quantum algorithmic representation of a Dirac particle in such a way that the hermitian generator of the unitary evolution is exactly the Dirac Hamiltonian. 
We give an explicit construction of the decomposition formula $e^{-i \arccos\! \sqrt{1-E^2\tau^2} \,(h_\circ + h')/E}= e^{ i \ell h_\circ} e^{-i  \arccos\!\sqrt{1-m^2\tau^2}\,{h'}/E}$ that is exactly computable \cite{PhysRevLett_2013_1},  where $\ell$ and $\tau$ denote the grid length and grid time,  and $E$ denotes the energy scale of particle dynamics.  This decomposition formula is exact in the sense that it does not require a limiting procedure as in Trotter decomposition $e^{-it(h_\circ + h')} = \lim_{n\rightarrow\infty} \big(e^{-it \,h_\circ/n}e^{-it \, h'/n}\big)^n$ \cite{Trotter_JSTOR_1959}.    A closed-form decomposition is possible when the modeled quantum system has a relativistic energy relation.   In the 1+1 dimensional case for free Dirac particle simulations with $h_D = h_\circ + h'$, the kinetic part of the Hamiltonian is $h_\circ = \sigma_z p_z c$ and the chiral breaking interaction part is $h'= \sigma_x m c^2$.  Since in 1+1 dimensions the quantum state is a 2-spinor, the unitary evolution generated by $h_D$, $h_\circ$, and $h'$ can each be viewed as a different rotation on the Bloch sphere.    So a rotation generated by $h_\circ$, followed by a rotation generated by $h'$, is equated to a single composite rotation generated by $h_D$.  This is the geometrical basis of the quantum lattice gas algorithm for the Dirac equation cast in the high-energy limit.

Consider a local evolution operator as a composition of qubit rotation operators $U_{\hat{\bm{n}}_2}=e^{-i \frac{\beta_2}{2} \hat{\bm{n}}_2\cdot \bm{\sigma}}$ and $U_{\hat{\bm{n}}_1}=e^{-i \frac{\beta_1}{2} \hat{\bm{n}}_1\cdot \bm{\sigma}}$, where $\bm{\sigma}=(\sigma_x,\sigma_y,\sigma_z)$ is a vector of Pauli matrices, $\hat{\bm{n}}_1$ and $\hat{\bm{n}}_2$ are unit vectors specifying the respective principal axes of rotation, and $\beta_1$ and $\beta_2$ are real-valued rotation angles.  The product of these rotations is
\begin{widetext}
\begin{subequations}
\label{double_rot_construction}
\begin{eqnarray}
\label{double_rot_construction_a}
U_{\hat{\bm{n}}_2}(\beta_2) U_{\hat{\bm{n}}_1}(\beta_1)
\!\!
&=&
\!\!
\left[
\cos \frac{\beta_2}{2} - i (\hat{\bm{n}}_2 \cdot \bm{\sigma}) \sin \frac{\beta_2}{2}
\right]
\left[
\cos \frac{\beta_1}{2} - i (\hat{\bm{n}}_1 \cdot \bm{\sigma}) \sin \frac{\beta_1}{2}
\right] 
\\
\nonumber
& = & 
\label{double_dot}
\cos \frac{\beta_1}{2} \cos \frac{\beta_2}{2} - \sin \frac{\beta_1}{2} \sin \frac{\beta_2}{2}
(\hat{\bm{n}}_1 \cdot \bm{\sigma}) (\hat{\bm{n}}_2 \cdot \bm{\sigma})
-
i \left[
\cos \frac{\beta_1}{2} \sin \frac{\beta_2}{2} (\hat{\bm{n}}_2 \cdot \bm{\sigma})
+ \sin \frac{\beta_1}{2} \cos \frac{\beta_2}{2} (\hat{\bm{n}}_1 \cdot \bm{\sigma})
\right]
\\
\nonumber
& = &
\cos \frac{\beta_1}{2} \cos \frac{\beta_2}{2}
- \sin \frac{\beta_1}{2} \sin \frac{\beta_2}{2} 
\hat{\bm{n}}_1 \cdot \hat{\bm{n}}_2 
-
\nonumber
i \Big[
\sin \frac{\beta_1}{2} \cos \frac{\beta_2}{2} \hat{\bm{n}}_1
+ \cos \frac{\beta_1}{2} \sin \frac{\beta_2}{2} \hat{\bm{n}}_2
\\
&&
\label{double_rot_construction_d}
- \sin \frac{\beta_1}{2} \sin \frac{\beta_2}{2} \hat{\bm{n}}_1 \times \hat{\bm{n}}_2
\Big] \cdot \bm{\sigma},
\end{eqnarray}
\end{subequations}
\end{widetext}
where in the last line we made use of the identity
\begin{equation}
\label{double_dot_identity}
(\hat{\bm{n}}_1 \cdot \bm{\sigma})\cdot (\hat{\bm{n}}_2 \cdot \bm{\sigma}) 
=
\hat{\bm{n}}_1 \cdot  \hat{\bm{n}}_2 + i \,\left( \hat{\bm{n}}_1\times \hat{\bm{n}}_2 \right)\cdot \bm{\sigma}.
\end{equation}
 Let us take $U_\text{\tiny S}^z=e^{-i \frac{\beta_2}{2} \hat{\bm{n}}_2\cdot \bm{\sigma}}$ as our stream operator and $U_\text{\tiny C}=e^{-i \frac{\beta_1}{2} \hat{\bm{n}}_1\cdot \bm{\sigma}}$ as our collision operator. 
Let us choose a reference frame where the particle motion occurs  along the $\hat{\bm{z}}$
\begin{subequations}
\label{unitary_stream_collide_operators}
\begin{equation}
\label{unitary_stream_operator_in_p_form}
U_\text{\tiny S}^z=e^{-i\frac{\beta_2}{2}\sigma_z}.
\end{equation} 
In this frame a general collision operator is 
\begin{equation}
\label{general_collision_operator}
U_\text{\tiny C}= e^{-i\frac{\beta_1}{2} (\alpha \sigma_x + \beta \sigma_y + \gamma \sigma_z )},
\end{equation} 
\end{subequations}
where $\alpha$, $\beta$, and $\gamma$ are real valued components subject to the constraint $\alpha^2 + \beta^2+ \gamma^2=1$.  
Furthermore, let us suppose that the unitary operators (\ref{unitary_stream_collide_operators}) are applied locally and homogeneously at all the points in the system. So, here we consider a construction whereby the two principal unit vectors specifying the axes of rotation are
\begin{eqnarray}
\label{axes_of_2_qubit_rotations}
 \hat{\bm{n}}_1 
&=& 
(\alpha, \beta, \gamma)
\qquad\qquad
 \hat{\bm{n}}_2
=
(0, 0, 1).
 \end{eqnarray}
With this choice, $\hat{\bm{n}}_1  \times  \hat{\bm{n}}_2 = (\beta, -\alpha,0)$ and $\hat{\bm{n}}_1  \cdot  \hat{\bm{n}}_2 =\gamma$, so (\ref{double_rot_construction}) is a quite general representation of a quantum lattice gas evolution operator
\begin{subequations}
\begin{eqnarray}
\label{double_rot_construction_explicit}
U_\text{\tiny S}^z \, U_\text{\tiny C}
& \stackrel{(\ref{axes_of_2_qubit_rotations})}{=} &
\cos \frac{\beta_1}{2} \cos \frac{\beta_2}{2}
- \gamma\sin \frac{\beta_1}{2} \sin \frac{\beta_2}{2} 
\\
\nonumber
& - &  
i \left(
\alpha\sin \frac{\beta_1}{2} \cos \frac{\beta_2}{2}
- \beta\sin \frac{\beta_1}{2} \sin \frac{\beta_2}{2}
\right)
 \sigma_x
\\
\nonumber
& - &  
i \left(
\beta\sin \frac{\beta_1}{2} \cos \frac{\beta_2}{2}
+\alpha\sin \frac{\beta_1}{2} \sin \frac{\beta_2}{2}
\right)
 \sigma_y
\\
\nonumber
& - &  
i \left(
 \gamma\sin \frac{\beta_1}{2} \cos \frac{\beta_2}{2} 
+ \cos \frac{\beta_1}{2} \sin \frac{\beta_2}{2}
\right)
\sigma_z
\\
\label{double_rot_construction_explicit_low_energy}
&\mapsto&
1 + \frac{i c\, p_z\tau}{\hbar} \sigma_z  - \frac{i {m} c^2\tau}{\hbar} \sigma_x,
\end{eqnarray}
\end{subequations}
where the last line is chosen as a construction.  The reason for choosing this construction is that  the quantum algorithm $\psi'=U_\text{\tiny S}^z \, U_\text{\tiny C}\psi$ is
\begin{equation}
\label{Dirac_equation_1_plus_1_dim_time_difference_form}
\psi'(z) =\left(1 + \frac{i c\, p_z\tau}{\hbar} \sigma_z  - \frac{i {m} c^2\tau}{\hbar} \sigma_x \right)\psi(z),
\end{equation}
which is a
time-difference representation of the equation of motion of a single free Dirac particle with a 2-spinor quantum state $\psi(z)=(\psi_\text{L}(z), \psi_\text{R}(z))^\text{T}$ defined over the set of points $\{z\}$ in a 1+1 dimensional spacetime.  That is, 
for small $\tau$ and for momentum operator $p_z = -i\hbar\partial_z$, (\ref{Dirac_equation_1_plus_1_dim_time_difference_form}) represents the Dirac equation for a relativistic quantum particle of mass ${m}$ 
\begin{equation}
i\hbar \partial_t \psi = -c\, p_z \sigma_z  \psi + {m} c^2 \sigma_x \psi.
\end{equation}

To establish a correspondence between (\ref{double_rot_construction_explicit}) and (\ref{double_rot_construction_explicit_low_energy}), we simply choose the real-valued components of $\hat{\bm{n}}_1$  to satisfy the following three conditions:
\begin{subequations}
\label{n_1_unit_vector_condition}
\begin{eqnarray}
\label{n_1_unit_vector_condition_a}
 \alpha\sin \frac{\beta_1}{2} \cos \frac{\beta_2}{2} 
&-&
\beta\sin \frac{\beta_1}{2} \sin \frac{\beta_2}{2}
 =   \frac{ {m} c^2\tau}{\hbar}  
\\
\label{n_1_unit_vector_condition_b}
\beta\sin \frac{\beta_1}{2} \cos \frac{\beta_2}{2}
&+&
\alpha\sin \frac{\beta_1}{2} \sin \frac{\beta_2}{2}  = 
0
\\
\label{n_1_unit_vector_condition_c}
\gamma\sin \frac{\beta_1}{2} \cos \frac{\beta_2}{2} 
&+&
 \ \ 
 \cos \frac{\beta_1}{2} \sin \frac{\beta_2}{2}
 = 
-\frac{c\, p_z\tau}{\hbar} .
\end{eqnarray}
Additionally, we should respect the reality condition that $\hat{\bm{n}}_1$ have unit norm%
\begin{equation}
\label{n_1_unit_vector_condition_d}
\alpha^2 + \beta^2+ \gamma^2=1
\end{equation}
\end{subequations}
that we established above with the collision operator (\ref{general_collision_operator}).
For the sake of simplicity, let us start with a specialized construction whereby $\hat{\bm{n}}_1$ is perpendicular to $\hat{\bm{n}}_2$. 
The solution of (\ref{n_1_unit_vector_condition}) in this special case is
\begin{equation}
\label{n1_perp_to_n2_solution}
\alpha=\cos\frac{\beta_2}{2} 
\qquad
\beta = -\sin\frac{\beta_2}{2}
\qquad
\gamma=0.
\end{equation}
Inserting (\ref{n1_perp_to_n2_solution}) into (\ref{n_1_unit_vector_condition_a}) gives
\begin{equation}
\label{beta_1_tau_formula}
\sin  \frac{\beta_1}{2} =  \frac{ {m} c^2\tau}{\hbar},  
\end{equation}
and in turn (\ref{n_1_unit_vector_condition_c}) is
\begin{equation}
\label{grid_equation_beta2_form}
 \sqrt{1-\left( \frac{ {m} c^2\tau}{\hbar}\right)^2} \sin \frac{\beta_2}{2}
 = 
-\frac{c\, p_z\tau}{\hbar} .
\end{equation}
In turn,
we have
\begin{subequations}
\begin{eqnarray}
\nonumber
\cos \frac{\beta_1}{2} \cos \frac{\beta_2}{2}
&=&
 \sqrt{1-\left( \frac{ {m} c^2\tau}{\hbar}\right)^2} 
  \sqrt{1-\frac{\left( \frac{ c\, p_z\tau}{\hbar}\right)^2}{1-\left( \frac{ {m} c^2\tau}{\hbar}\right)^2}} 
\\
\\
&=&
 \sqrt{1-\left( \frac{ E \tau}{\hbar}\right)^2} ,
\end{eqnarray}
\end{subequations}
with $E^2 = ({m} c^2)^2 + ( c\, p_z)^2$.
Therefore, the quantum lattice gas evolution operator (\ref{double_rot_construction_explicit}) is
\begin{eqnarray}
\nonumber
U_\text{\tiny S}^z \, U_\text{\tiny C}
&=&
 \sqrt{1-\left( \frac{ E \tau}{\hbar}\right)^2} 
 +
 \frac{ iE \tau}{\hbar}
 \left( \frac{ c\, p_z}{E} \sigma_z  - \frac{{m} c^2}{E} \sigma_x\right).
 \\
 \label{lattice_gas_operator_in_m_E_form_e}
\end{eqnarray}
This result leads us to define the rotation axis
\begin{equation}
\label{n_12_m_p_E_form}
\hat{\bm{n}}_{12} \equiv   - \frac{{m} c^2}{E}\, \hat{\bm{x}}+\frac{c\, p_z}{E}\, \hat{\bm{z}}.
\end{equation}
Since $(\hat{\bm{n}}_{12}\cdot\bm{\sigma})^2=\bm{1}$ (an involution), we can employ Euler's identity and the trigonometric identity $\sin (\cos^{-1} \sqrt{1-x^2})=x$, so we are free to write (\ref{lattice_gas_operator_in_m_E_form_e}) in a manifestly unitary form $e^{-i \frac{\beta_{12}}{2} \hat{\bm{n}}_{12}\cdot \bm{\sigma}}$ as follows:
\begin{subequations}
\label{lattice_gas_operator_in_exp_form}
\begin{eqnarray}
\nonumber
U_\text{\tiny S}^zU_\text{\tiny C}
 \!\!\!\!&=&\!\!\!\!
 \exp\left[
 i \cos^{-1}\!\left( \text{\scriptsize $\sqrt{1-\left( \frac{ E \tau}{\hbar}\right)^2}$}\right) \hat{\bm{n}}_{12}\cdot\bm{\sigma}
  \right]
\\
\label{lattice_gas_operator_in_exp_form_a}
\\
\nonumber
 &\stackrel{(\ref{n_12_m_p_E_form})}{=}&\!\!\!
 \exp\left[
 i \,\frac{\cos^{-1} \sqrt{1-\left( \frac{ E \tau}{\hbar}\right)^2}}{E} \left(\sigma_z  c\,p_z -  \sigma_x\, {{m}}c^2\right)
  \right].
\\
\label{lattice_gas_operator_in_exp_form_b}
\end{eqnarray}
The hermitian generator governing the dynamical behavior of the 2-spinor field $\psi$ is the Dirac Hamiltonian
\begin{equation}
\label{Dirac_hamiltonian_z_t_dimensions}
h_\text{D} = -\sigma_z  c\,p_z +  \sigma_x\, {{m}}c^2.
\end{equation}
This is a remarkable finding because nowhere in the derivation of (\ref{lattice_gas_operator_in_exp_form_b}) did we invoke the continuum limit where $\tau\rightarrow0$.  That is, $\tau$ may be taken to be a small but finite quantity, not necessarily infinitesimal.  Thus, because of the form of (\ref{Dirac_hamiltonian_z_t_dimensions}), Lorentz invariance would apply  to the quantum dynamics even though the spacetime is discrete, albeit there are unexpected departures from relativistic quantum mechanics, and we address the effect of these departures in Sec.~\ref{Unitary_finite_path_summation}. The rotation angle in (\ref{lattice_gas_operator_in_exp_form_b}) is  a real scalar quantity, so we may denote this as $\ell$ and write
\begin{eqnarray}
U_\text{\tiny S}^zU_\text{\tiny C} &=&
\label{lattice_gas_operator_in_exp_form_c}
 e^{
-i \,{\ell}\, h_\text{D}/(\hbar  c)
 },
\end{eqnarray}
\end{subequations}
where in the last line we made the identification
\(
\cos\!\left( \frac{E\ell}{\hbar c}\right) = \sqrt{1-\left( \frac{ E \tau}{\hbar}\right)^2},
\)
or expressing $\tau$ in terms of the grid size $\ell$, we find that the grid sizes must satisfy the transcendental equation
\begin{equation}
\label{tau_ell_transcendental_equation}
\frac{E \tau}{\hbar} = \sin\!\left( \frac{E\ell}{\hbar c}\right) .
\end{equation}

We know that
\begin{equation}
\label{beta_2_p_z_formula}
\frac{\beta_2}{2} = -\ell k_z
\end{equation}
because the stream operator 
\begin{equation}
\label{stream_operator}
U_\text{\tiny S}^z=e^{i\ell k_z\sigma_z} = e^{\sigma_z\ell \partial_z}
\end{equation}
 is just the shift operator that displaces the spin-up and spin-down components of the Dirac field by $\pm\ell$, respectively.

With the above results, we may rewrite $U_C$ in an analytical form that is useful for the  simulation of the quantum dynamics of a Dirac field. We begin by writing the collision operator (\ref{general_collision_operator}) as
\begin{subequations}
\begin{eqnarray}
U_C 
& = & 
e^{-i\frac{\beta_1}{2} \left[ \sigma_x \left(\cos\frac{\beta_2}{2}  - i \sigma_z\sin\frac{\beta_2}{2}   \right) \right]}
 \\
& = & 
e^{-i\frac{\beta_1}{2} \sigma_x e^{- i \sigma_z\frac{\beta_2}{2}}}
 \\
& = & 
\cos\frac{\beta_1}{2}-  \sigma_x e^{- i \sigma_z\frac{\beta_2}{2}}\sin\frac{\beta_1}{2}
 \\
& 
\stackrel{(\ref{beta_2_p_z_formula})}{\stackrel{(\ref{beta_1_tau_formula})}{=}} & 
\sqrt{1-\left(\frac{mc^2\tau}{\hbar}\right)^2}-  \sigma_x e^{ i \ell  p_z \sigma_z}\frac{mc^2\tau}{\hbar}.
\end{eqnarray}
\end{subequations}
Since 
\begin{equation}
\sigma_x e^{ i \ell  k_z \sigma_z} = 
\begin{pmatrix}
     0 &    1\\
    1  &  0
\end{pmatrix}
\begin{pmatrix}
   e^{ i \ell  k_z }   & 0   \\
0      &     e^{ -i \ell  k_z } 
\end{pmatrix}
=
\begin{pmatrix}
0&    e^{- i \ell  k_z }     \\
  e^{ i \ell  k_z } & 0
\end{pmatrix},
\end{equation}
the collision operator in matrix form is
\begin{equation}
\label{collide_operator}
U_C 
=
\begin{pmatrix}
  \sqrt{1-\left(\frac{mc^2\tau}{\hbar}\right)^2}    &
-i  e^{- i \ell  k_z }    \frac{mc^2\tau}{\hbar}
      \\
-i     e^{ i \ell  k_z }     \frac{mc^2\tau}{\hbar}
      & 
       \sqrt{1-\left(\frac{mc^2\tau}{\hbar}\right)^2}
\end{pmatrix}.
\end{equation}
For quantum simulation purposes, we wish to write $U_C = U_C(m,\ell, \gamma)$,  where $\gamma \equiv E/(mc^2)$.  Since $p_z= \sqrt{E^2-(mc^2)^2}$, we have $p_z = \hbar k_z = mc\sqrt{\gamma^2-1}$, and in turn the collision operator may be written as
\begin{equation}
\label{unitary_collision_operator}
U_C 
=
\begin{pmatrix}
  \sqrt{1-\left(\frac{mc^2\tau}{\hbar}\right)^2}    &
-i \, e^{- i \frac{mc\ell }{\hbar}\sqrt{\gamma^2-1} }    \frac{mc^2\tau}{\hbar}
      \\
-i \,    e^{ i \frac{mc\ell}{\hbar}\sqrt{\gamma^2-1}  }     \frac{mc^2\tau}{\hbar}
      & 
       \sqrt{1-\left(\frac{mc^2\tau}{\hbar}\right)^2}
\end{pmatrix}.
\end{equation}
Now we can rewrite (\ref{tau_ell_transcendental_equation}) as
\begin{equation}
\frac{m c^2 \tau}{\hbar} = \frac{1}{\gamma}\sin\!\left( \frac{\gamma mc\ell}{\hbar}\right) ,
\end{equation}
which allows us to eliminate the explicit $\tau$-dependence in the collide operator
\begin{widetext}
\begin{equation}
U_C 
=
\frac{1}{\gamma}
\begin{pmatrix}
  \sqrt{\gamma^2-\sin^2\!\left( \frac{\gamma mc\ell}{\hbar}\right)}    &
-i  \,e^{- i \frac{mc\ell }{\hbar}\sqrt{\gamma^2-1} }    \sin\!\left( \frac{\gamma mc\ell}{\hbar}\right)
      \\
-i  \,   e^{ i \frac{mc\ell}{\hbar}\sqrt{\gamma^2-1}  }    \sin\!\left( \frac{\gamma mc\ell}{\hbar}\right)
      & 
       \sqrt{\gamma^2-\sin^2\!\left( \frac{\gamma mc\ell}{\hbar}\right)}
\end{pmatrix}.
\end{equation}
\end{widetext}
So in natural lattice units ($\hbar =1$ and $c=1$), the quantum algorithm for the Dirac equation is represented by the following stream and collide operators:
\begin{subequations}
\label{Yepez_quantum_algorithm_Dirac_particles}
\begin{eqnarray}
 U_\text{\tiny S}^z & = & e^{\sigma_z\ell \partial_z} \\
 \nonumber
U_C & = & 
\scriptsize
\frac{1}{\gamma}
\begin{pmatrix}
  \sqrt{\gamma^2-\sin^2(\gamma m\ell)}    &
-i  \,e^{- i m\ell \sqrt{\gamma^2-1} }    \sin(\gamma m\ell)
      \\
-i  \,   e^{ i m\ell\sqrt{\gamma^2-1}  }    \sin(\gamma m\ell)
      & 
       \sqrt{\gamma^2-\sin^2(\gamma m\ell)}  
\end{pmatrix}.
\\
\end{eqnarray}
\end{subequations}
We will demonstrate the numerical performance of this quantum algorithm below in Sec.~\ref{Quantum_simulation}.

 In this Appendix~\ref{Collide_operator_for_the_Dirac_equation} we derive a quantum gate representation of the collide operator (\ref{unitary_collision_operator}) for a system of Dirac particles in 1+1 dimensions that is useful for representing the scattering of  $ \psi_\uparrow $ and $ \psi_\downarrow$ particles at a point
\begin{equation}
\begin{pmatrix}
      \psi'_\uparrow    \\
      \psi'_\downarrow
\end{pmatrix} 
=
\begin{pmatrix}
\sqrt{1-\epsilon^2}   &
-i  \epsilon\, e^{- i \frac{mc\ell }{\hbar}\sqrt{\gamma^2-1} }  
      \\
-i \,    \epsilon e^{ i \frac{mc\ell}{\hbar}\sqrt{\gamma^2-1}  }   
      & 
    \sqrt{1-\epsilon^2} 
\end{pmatrix}
\begin{pmatrix}
      \psi_\uparrow    \\
      \psi_\downarrow
\end{pmatrix},
\end{equation}
where $\epsilon  \equiv {mc^2\tau}/{\hbar}$.

\section{Quantum lattice gas  as a path summation}
\label{Quantum_lattice_gas_as_a_path_summation}

In this section, we will derive (\ref{Yepez_quantum_algorithm_Dirac_particles}) in a different way based on a path integral representation of relativistic quantum mechanics.

\subsection{Feynman path summation}

The probability amplitude that a quantum particle at position $z_a$ at time $t_a$ will transfer to a new position $z_b$ and time $t_b$ is given by the following path integral:
\begin{equation}
\label{feynman-path-integral}
K(z_a t_a ; z_b t_b) = \int {\cal{D}}[ z(t)] e^{i \frac{S[z(t)]}{\hbar}} ,
\end{equation}
where $\int {\cal D}[ z(t)]$ denotes integration over all trajectories $z(t)$ for which $z(t_a)=z_a$ and $z(t_b)=z_b$,
and where the increase of the action 
\begin{equation}
S = \int_{t_a}^{t_b} dt L[\dot z(t), z(t)],
\end{equation}
along a trajectory $z(t)$ is determined using the classical Lagrangian $L$. 
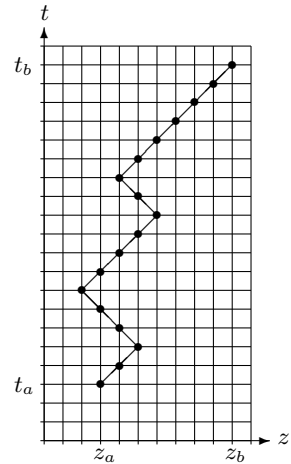
\begin{figure}[htbp]
\begin{center}
\setlength{\unitlength}{.5mm}
\begin{picture}(65,105)

\put(0,0){\vector(1,0){60}}
\put(62,-1){$z$}

\multiput(0,0)(5,0){12}{\line(0,-1){1}}
\put(13,-5){$z_{a}$}
\put(48,-5){$z_{b}$}

\put(0,0){\vector(0,1){110}}
\put(-1,112){$t$}

\multiput(0,0)(0,5){22}{\line(-1,0){1}}
\put(-8,13){$t_{a}$}
\put(-8,98){$t_{b}$}

\put(15,15){\circle*{2}}
\put(15,15){\line(1,1){10}}
\put(20,20){\circle*{2}}
\put(25,25){\circle*{2}}
\put(25,25){\line(-1,1){15}}
\put(20,30){\circle*{2}}
\put(15,35){\circle*{2}}
\put(10,40){\line(1,1){20}}
\put(10,40){\circle*{2}}
\put(15,45){\circle*{2}}
\put(20,50){\circle*{2}}
\put(25,55){\circle*{2}}
\put(30,60){\line(-1,1){10}}
\put(30,60){\circle*{2}}
\put(25,65){\circle*{2}}
\put(20,70){\line(1,1){30}}
\put(20,70){\circle*{2}}
\put(25,75){\circle*{2}}
\put(30,80){\circle*{2}}
\put(35,85){\circle*{2}}
\put(40,90){\circle*{2}}
\put(45,95){\circle*{2}}
\put(50,100){\circle*{2}}

\multiput(0,0)(5,0){12}{\line(0,1){105}}

\multiput(0,0)(0,5){22}{\line(1,0){55}}

\end{picture}
\end{center}
\caption{\label{discretized-relativistic-trajectory}\footnotesize Example trajectory of a massive relativistic particle starting at location $z_{a}$ at time $t_{a}$  and ending at $z_{b}$ at time $t_{b}$.  The total number of steps is $N=17$, so the elapsed time is $t=17{\tau}$. The number of steps to the right minus the number to the left is $M=7$, so the net distance traversed is $z=7{\ell}$. The relativistic particle moves at the speed of light $c\equiv{\ell}/{\tau}$. The number of bends is $R=4$.}
\end{figure}
Feynman established a discrete representation of the path integral in 1+1 dimensions  \cite{feynman-65-1st-qlga}, using a infinite square lattice to compute (\ref{feynman-path-integral}) for a relativistic quantum particle
\begin{equation}
\label{riazanov-feynman-path-summation}
K_{\alpha\beta}(z_a t_a ; z_b t_b) = \lim_{\stackrel{N\rightarrow\infty}{\tau\rightarrow 0}} \sum_{R\ge 0}\Phi_{\alpha\beta}(R) \left(i  \frac{mc^{2}{\tau}}{\hbar}\right)^R,
\end{equation}
where ${\tau}\equiv({t_b-t_a})/{N}$, where $\alpha$ and $\beta$ are the $\pm$ components of the spinor amplitude field (spin-up or spin-down), $\Phi_{\alpha\beta}(R)$ is the number of paths with $N$ steps and $R$ bends, where the length of each step is ${\ell} \equiv {(z_b-z_a)}/{M}\equiv c \,{\tau}$, where $c$ is the speed of light, and where $m$ is the mass of the quantum particle.  An example relativistic trajectory with 4 bends along $\hat z$ is depicted in Figure~\ref{discretized-relativistic-trajectory}.

\begin{figure*}[htbp]
\begin{center}
\xy
(-25,-38)*{z_b};
(-36,-77)*{z_a};
(-30,-57)*{\includegraphics[width=1.5in]{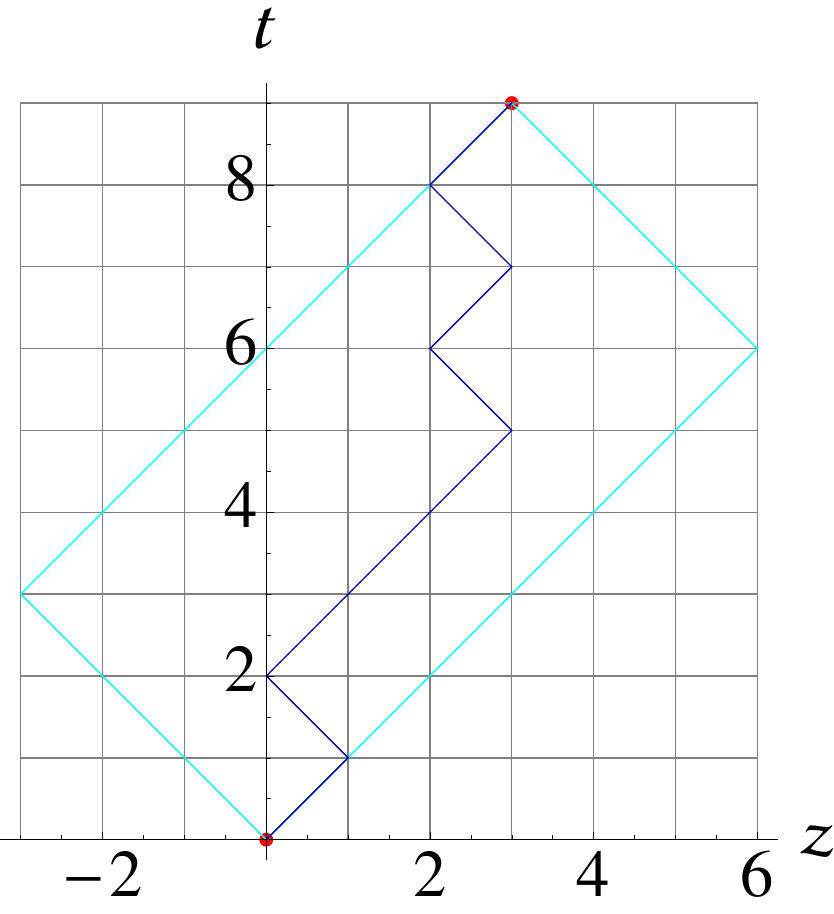}};
(65,-55)*{\includegraphics[width=5.25in]{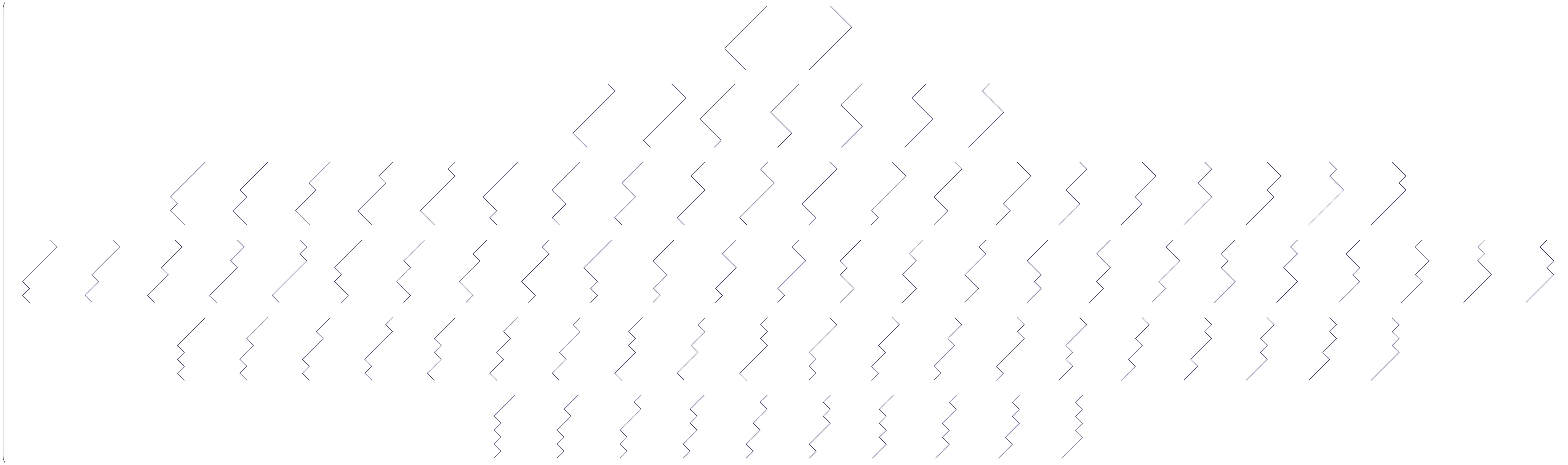}};
(-38,-30)*{t_b-t_a=9};
(-39.5,-34)*{|\bm{z}_b-\bm{z}_a|=3};
(-5,-30)*{\text{Bends}};
(-5,-38)*{\text{\normalsize 1}};
(-5,-45)*{\text{\normalsize 2}};
(-5,-52)*{\text{\normalsize 3}};
(-5,-59)*{\text{\normalsize 4}};
(-5,-66)*{\text{\normalsize 5}};
(-5,-73)*{\text{\normalsize 6}};
\endxy
\end{center}
\caption{\label{pathsExample} \footnotesize Example problem in 1+1 dimensions with $N=9$ and $M=3$. Left: Allowable rectangular region of the square lattice within the light cone (cyan), with sides of length $Q=\frac{N+M}{2}=6$ and $P=\frac{N-M}{2}=3$, and path (blue) $(1,-1,1,1,1,-1,1,-1,1)$ with 6 bends.  Right: Enumeration of all possible paths, the $84$ permutations of the set $(1,1,1,1,1,1, -1, -1,-1)$. }
\end{figure*}
The allowable region of the lattice is bounded by the intersection of two light cones, with boundaries $z=\pm c (t-t_{a})+z_{a}$ and $z=\pm c(t-t_b)+z_b$ for $t_a\le t\le t_b$ and   $z_a\le z\le z_b$.  One light cone originates at the spacetime point $(z_{a}, t_{a})$ and an inverted light zone terminates at $(z_{b}, t_{b})$, see Figure~\ref{pathsExample}.  With $z_{b}-z_{a}=M{\ell}$ and $t_{b}-t_{a}=N{\tau}$,
the edges of the allowable rectangular region are given by $P \equiv \lfloor\frac{N-M}{2}\rfloor$ and $Q\equiv\lceil\frac{N+M}{2}\rceil$, for $N\ge M\ge 0$.  Hence, the paths are the permutations of the set with $N=P+Q$ members $\pm1$:
\begin{equation}
\label{ }
(\!\!\!\!\underbrace{1,1,\dots, 1}_{Q \; \text{number of 1's}}\underbrace{-1,-1,\cdots, -1}_{P\; \text{number of -1's}}).
\end{equation}
The number of permutations is the binomial coefficient:
\begin{equation}
\label{ }
\text{number of paths} = 
\left(
\begin{matrix}P + Q \cr 
P
\end{matrix}
\right) =  \left(
\begin{matrix}P + Q \cr 
Q
\end{matrix}
\right)  .
\end{equation}

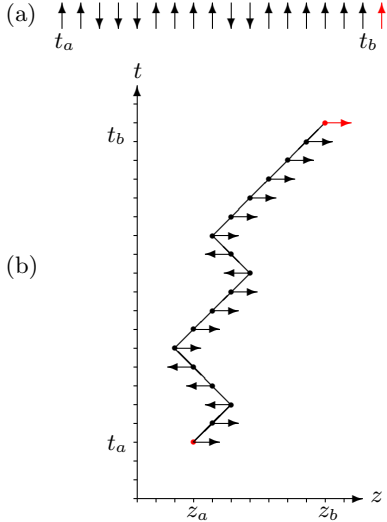
\begin{figure}[htbp]
\begin{center}
\setlength{\unitlength}{.5mm}
\begin{picture}(65,145)

\put(0,0){\vector(1,0){60}}
\put(62,-1){$z$}

\multiput(0,0)(5,0){12}{\line(0,-1){1}}
\put(13,-5){$z_{a}$}
\put(48,-5){$z_{b}$}

\put(0,0){\vector(0,1){110}}
\put(-1,112){$t$}

\multiput(0,0)(0,5){22}{\line(-1,0){1}}
\put(-8,13){$t_{a}$}
\put(-8,95){$t_{b}$}

\put(15,15){\line(1,1){10}}
\textcolor{red}{\put(15,15){\circle*{1.5}}}
\put(20,20){\circle*{1.5}}
\put(25,25){\line(-1,1){15}}
\put(25,25){\circle*{1.5}}
\put(20,30){\circle*{1.5}}
\put(15,35){\circle*{1.5}}
\put(10,40){\line(1,1){20}}
\put(10,40){\circle*{1.5}}
\put(15,45){\circle*{1.5}}
\put(20,50){\circle*{1.5}}
\put(25,55){\circle*{1.5}}
\put(30,60){\line(-1,1){10}}
\put(30,60){\circle*{1.5}}
\put(25,65){\circle*{1.5}}
\put(20,70){\line(1,1){30}}
\put(20,70){\circle*{1.5}}
\put(25,75){\circle*{1.5}}
\put(30,80){\circle*{1.5}}
\put(35,85){\circle*{1.5}}
\put(40,90){\circle*{1.5}}
\put(45,95){\circle*{1.5}}
\textcolor{red}{\put(50,100){\circle*{1.5}}}

\put(15,15){\line(1,1){10}}
\put(15,15){\vector(1,0){7}}
\put(20,20){\vector(1,0){7}}
\put(25,25){\vector(-1,0){7}}
\put(20,30){\vector(-1,0){7}}
\put(15,35){\vector(-1,0){7}}
\put(10,40){\vector(1,0){7}}
\put(15,45){\vector(1,0){7}}
\put(20,50){\vector(1,0){7}}
\put(25,55){\vector(1,0){7}}
\put(30,60){\vector(-1,0){7}}
\put(25,65){\vector(-1,0){7}}
\put(20,70){\vector(1,0){7}}
\put(25,75){\vector(1,0){7}}
\put(30,80){\vector(1,0){7}}
\put(35,85){\vector(1,0){7}}
\put(40,90){\vector(1,0){7}}
\put(45,95){\vector(1,0){7}}
\textcolor{red}{\put(50,100){\vector(1,0){7}}}

\put(-20,125){\vector(0,1){7}}
\put(-15,125){\vector(0,1){7}}
\put(-10,132){\vector(0,-1){7}}
\put(-5,132){\vector(0,-1){7}}
\put(-0,132){\vector(0,-1){7}}
\put(5,125){\vector(0,1){7}}
\put(10,125){\vector(0,1){7}}
\put(15,125){\vector(0,1){7}}
\put(20,125){\vector(0,1){7}}
\put(25,132){\vector(0,-1){7}}
\put(30,132){\vector(0,-1){7}}
\put(35,125){\vector(0,1){7}}
\put(40,125){\vector(0,1){7}}
\put(45,125){\vector(0,1){7}}
\put(50,125){\vector(0,1){7}}
\put(55,125){\vector(0,1){7}}
\put(60,125){\vector(0,1){7}}
\textcolor{red}{\put(65,125){\vector(0,1){7}}}

\put(-22,120){$t_{a}$}
\put(59,120){$t_{b}$}

\put(-35,127){(a)}
\put(-35,60){(b)}

\end{picture}
\end{center}
\caption{\label{spin-representation-of-relativistic-trajectory}\footnotesize (a) Spin representation of the trajectory of a massive relativistic particle starting at time $t_{a}$ and ending at time $t_{b}$ for $N=7$ and $M=17$.  (b) Quantum lattice  gas representation of the same trajectory where the particle is stream-plus (or ``spin-up'') as it moves to the right and stream-minus (or ``spin-down'') as it moves to the left. Post-collision spin orientations are shown. The last spin variable (red) does not determine the path. The final spin $s_N=1$ (red) in this  path is shown in an example post-collisional state.}
\end{figure}
Because the summation (\ref{riazanov-feynman-path-summation}) occurs on a discrete spacetime lattice, in 1+1 dimensions it is possible to enumerate all the paths originating at point $a$ and ending at point $b$ using $N$ spin variables $s_i$, for $i=0,1,2,\dots,N-1$.  This is depicted in Figure~\ref{spin-representation-of-relativistic-trajectory}a for the example relativistic trajectory. The last spin variable, shown in red, is not included in the sum because it is not determinative of the path.  Identifying $\alpha$ with $s_{0}$ and $\beta$ with $s_N$ (which are considered  initial and final spin conditions), the summation (\ref{riazanov-feynman-path-summation}) is equivalent to:
\begin{equation}
\label{riazanov-feynman-path-summation-spin-form1}
K_{s_{0}s_{N}}(z_a t_a ; z_b t_b) = 
\lim_{\stackrel{N\rightarrow\infty}{\tau\rightarrow 0}}\sum_{\{ s_0, \dots , s_{N-1}\}|\text{M}} \left(i  \frac{mc^{2}{\tau}}{\hbar}\right)^R,
\end{equation}
where for now we consider $N$ fixed with the understanding that representation (\ref{riazanov-feynman-path-summation-spin-form1}) is unitary only in the continuum limit  $N\rightarrow \infty$ and ${\tau}\rightarrow 0$ (which we will take at the end of the derivation), and where the set $\{ s_0,\dots, s_{N-1}\}$ specifies a discrete trajectory with a path length constrained by the condition
\begin{equation}
\label{fixed-distance-requirement}
\sum_{i=0}^{N-1}  s_i = \frac{z_b - z_a}{{\ell}}.
\end{equation}
The starting and ending points are fixed, yet the sum on the R.H.S. of (\ref{riazanov-feynman-path-summation-spin-form1}) includes a summing over the initial spin $s_0$ and not over $s_N$.  The condition (\ref{fixed-distance-requirement}) is equivalent to fixing the spin magnetization $M=\sum_{i=0}^{N-1}  s_i$ of a system of $N$ spins.

\subsection{Revised path summation}
\label{Unitary_finite_path_summation}

Here we establish a discrete representation of the path integral in 1+1 dimensions that is an improvement over the Feynman representation (\ref{riazanov-feynman-path-summation}).  We  need only a finite-sized square grid to compute the probability amplitude for a Dirac particle that obeys relativistic quantum mechanics:
\begin{equation}
\label{Yepez1D_riazanov-feynman-path-summation}
K_{\alpha\beta}(a , b) 
=\sum_{R\ge 0}\Phi_{\alpha\beta}(R) 
\left( 
\text{
\scriptsize
$\sqrt{1-\left(\frac{mc^{2}\tau}{\hbar}\right)^2}$
}
\right)^{\overline{R}} 
\!\!
\left(i \, \frac{mc^{2}\tau}{\hbar}
\right)^R
\!\!,
\end{equation}
where the  initial point is $a = (t_a, z_a)$ and the final point is $b=(t_b, z_b)$, where the grid scales are  ${\tau}\equiv({t_b-t_a})/{N}$ and ${\ell} \equiv {(z_b-z_a)}/{M}\equiv c \,{\tau}$ for integers $N$ (number of time steps) and $M$ (spin chain magnetization),  where $\alpha$ and $\beta$ are the $\pm$ components of the spinor amplitude field (spin-up or spin-down), $\Phi_{\alpha\beta}(R)$ is the number of paths with $N$ steps and $R$ bends, where $\overline{R} \equiv N - R$, where $c$ is the speed of light, and where $m$ is the mass of the quantum particle.

Because the summation (\ref{Yepez1D_riazanov-feynman-path-summation}) occurs on a discrete spacetime lattice, in 1+1 dimensions it is possible to enumerate all the paths originating at point $a$ and ending at point $b$ using $N$ spin variables $s_i$, for $i=0,1,2,\dots,N-1$.  
  Identifying $\alpha$ with $s_{0}$ and $\beta$ with $s_N$, 
  the summation (\ref{Yepez1D_riazanov-feynman-path-summation}) is equivalent to:
\begin{equation}
\label{Yepez1D_riazanov-feynman-path-summation-spin-form1}
K_{s_{0}s_{N}}(a,b) = 
\hspace{-0.25in}
\sum_{\{ s_0, \dots , s_{N-1}\}|\text{M}}
\!\!
\left( 
\text{
\scriptsize
$\sqrt{1-\left(\frac{mc^{2}\tau}{\hbar}\right)^2}$
}
\right)^{\overline{R}} 
\!\!
\left(i \, \frac{mc^{2}\tau}{\hbar}
\right)^R
\!\!,
\end{equation}
where the set $\{ s_0,\dots, s_{N-1}\}$ specifies a discrete trajectory with a path length constrained by the condition
\begin{equation}
\label{Yepez1D_fixed-distance-requirement}
M\equiv \sum_{i=0}^{N-1}  s_i = \frac{z_b - z_a}{{\ell}}.
\end{equation}
Condition (\ref{Yepez1D_fixed-distance-requirement}) is equivalent to fixing the spin magnetization of a spin chain consisting of $N$ spins.  

At the $i$th step, the particle continues to move straight when $ s_{i}= s_{i+1}$, and it changes direction when $ s_{i}=- s_{i+1}$.  As the particle moves (or { streams}) to the right, its spin orientation is ``spin-up'' and as it moves to the left it is ``spin-down.''
 Therefore, the following binary value counts the occurrence of a bend at the $i$th step:
\begin{equation}
\frac{1}{2}(1- s_{i} s_{i+1})=\left\{
\begin{matrix}
0, & \qquad \hbox{no bend} \cr
1, & \qquad \hbox{bend}.
\end{matrix}
\right.
\end{equation}
Hence, the following sum counts the total number of bends and nonbends, respectively, in a path:
\begin{subequations}
\begin{eqnarray}
\label{Yepez1D_number-of-bends}
R&=&\frac{1}{2}\sum_{i=0}^{N-1}(1- s_{i} s_{i+1})
\\
\overline{R}&=&\frac{1}{2}\sum_{i=0}^{N-1}(1+ s_{i} s_{i+1}) = N-R.
\end{eqnarray}
\end{subequations}

With the change of variables
\begin{subequations}
\begin{eqnarray}
\label{Yepez1D_coupling_constant_over_KT}
\mu 
&\equiv &
-\frac{1}{2}\log
\sqrt{1-\left(\frac{mc^{2}\tau}{\hbar}\right)^2}
\\
\nu 
&\equiv &
-\frac{1}{2}\log\left(i \, \frac{mc^{2}\tau}{\hbar}\right),
\end{eqnarray}
\end{subequations}
 the kernel (\ref{Yepez1D_riazanov-feynman-path-summation-spin-form1}) can be written as the partition function of an ensemble of spins with nearest neighbor coupling and with fixed total magnetization
\begin{widetext}
\begin{equation}
\label{Yepez1D_riazanov-feynman-path-summation-spin-form2}
K_{ s_{0} s_{N}}
=
\sum_{\{ s_0, \dots , s_{N-1}\}}\delta\left(M,\sum_{i=0}^{N-1} s_{i}\right)\;
e^{-\mu\sum_{i=0}^{N-1}(1+ s_{i} s_{i+1})}e^{-\nu\sum_{i=0}^{N-1}(1- s_{i} s_{i+1})},
\end{equation}
where the Kronecker delta $\delta(a,b)=1$ for $a=b$ and $\delta(a,b)=0$ for $a\ne b$.  We may write the Kronecker delta as follows:
\begin{equation}
\label{Yepez1D_kronecker-delta-for-fixed-magnetization}
\delta\left(M,\sum_{i=0}^{N-1} s_{i}\right) = \frac{1}{2N}\sum_{n=-N}^{N-1}e^{i\frac{2\pi n}{N}\left(M-\sum_{i} s_{i}\right)},
\end{equation}
since $M$ and $\sum_{i=0}^{N-1} s_{i}$ are integers.
Then inserting (\ref{Yepez1D_kronecker-delta-for-fixed-magnetization}) into (\ref{Yepez1D_riazanov-feynman-path-summation-spin-form2}) gives
\begin{equation}
\label{Yepez1D_riazanov-feynman-path-summation-spin-form3}
K_{ s_{0} s_{N}}= 
\frac{1}{2N}\sum_{n=-N}^{N-1}e^{ i  \left(\frac{2\pi n}{N}\right)M}
\sum_{\{ s_0, \dots , s_{N-1}\}} 
e^{- i \left(\frac{2\pi n}{N}\right)\sum_{i=0}^{N-1} s_{i}-\mu\sum_{i=0}^{N-1}(1+ s_{i} s_{i+1})-\nu\sum_{i=0}^{N-1}(1- s_{i} s_{i+1})}.
\end{equation}
We may write the sum as $\sum_{i=0}^{N-1} s_{i}=\frac{1}{2}( s_0- s_{N})+ \frac{1}{2}\sum_{i=0}^{N-1}( s_{i}
+ s_{i+1})$,  and in turn we pull down the summation in the argument of the exponential to form the following product: 
\begin{equation}
\label{Yepez1D_riazanov-feynman-path-summation-spin-form4}
K_{ s_{0} s_{N}}=
\frac{1}{2N}\sum_{n=-N}^{N-1}e^{i\left(\frac{2\pi n}{N}\right)M}
\sum_{\{ s_0, \dots , s_{N-1}\}} 
e^{-i\pi \left(\frac{n}{N}\right)( s_0- s_{N})}
\prod_{i=0}^{N-1}e^{-i\pi \left(\frac{n}{N}\right)( s_{i}+ s_{i+1})-\mu(1+ s_{i} s_{i+1})-\nu(1- s_{i} s_{i+1})}.
\end{equation}
We define a unitary transfer operator  $\mathscr{U}$ as
\begin{equation}
\label{Yepez1D_hamiltonian_transfer_matrix-components}
\mathscr{U}_{ s_{i}, s_{i+1}}\equiv e^{-\mu(1+ s_{i} s_{i+1})-\nu(1- s_{i} s_{i+1})-i\pi\left(\frac{n}{N}\right)( s_{i}+ s_{i+1})},
\end{equation}
so that (\ref{Yepez1D_riazanov-feynman-path-summation-spin-form4}) becomes
\begin{equation}
\label{Yepez1D_riazanov-feynman-path-summation-transfer-form}
K_{ s_{0} s_{N}}=
\frac{1}{2N}\sum_{n=-N}^{N-1}e^{i\left(\frac{2\pi n}{N}\right)M}
\sum_{s_0=\pm1}
e^{-i\pi \left(\frac{n}{N}\right)( s_0- s_{N})}
\mathscr{Z}_{ s_0 s_{N}},
\end{equation}
where we have defined
\begin{equation}
\label{Yepez1D_gersch-transfer-form}  
\mathscr{Z}_{ s_{0} s_{N}}
\equiv
\sum_{\{ s_1, \dots , s_{N-1}\}} 
\prod_{i=0}^{N-1}
\mathscr{U}_{ s_{i}, s_{i+1}}
=
\sum_{ s_1=\pm1} \cdots\sum_{ s_{N-1}=\pm1} 
\mathscr{U}_{ s_0, s_1}\mathscr{U}_{ s_1, s_{2}}\cdots\mathscr{U}_{ s_{N-1}, s_{N}}.
\end{equation}
\end{widetext}
The matrix form of (\ref{Yepez1D_hamiltonian_transfer_matrix-components}) is
\begin{equation}
\label{Yepez1D_hamiltonian_transfer_matrix}
\mathscr{U} =
\left(
\begin{matrix}
\mathscr{U}_{1,1} & \mathscr{U}_{-1,1} \cr
\mathscr{U}_{1,-1} & \mathscr{U}_{-1,-1}
\end{matrix}
\right)
=
\left(
\begin{matrix}
e^{-i 2\pi n/N}  e^{-2\mu}& e^{-2\nu} \cr
e^{-2\nu}  & e^{i2\pi n/N}  e^{-2\mu}
\end{matrix}
\right)
\end{equation}
and so (\ref{Yepez1D_gersch-transfer-form}) becomes simply an $N$ fold matrix multiplication of $\mathscr{U}$:
\begin{equation}
\label{Yepez1D_gersch-transfer-form-matrix-form}
\mathscr{Z} 
\!\!
=
\!\!
\left(
\begin{matrix}
\mathscr{Z}_{1,1} & \mathscr{Z}_{-1,1} \cr
\mathscr{Z}_{1,-1} & \mathscr{Z}_{-1,-1}
\end{matrix}
\right)
\!\!
=
\!\!
\left(
\begin{matrix}
e^{-i 2\pi n/N}  e^{-2\mu}& e^{-2\nu} \cr
e^{-2\nu}  & e^{i2\pi n/N} e^{-2\mu}
\end{matrix}
\right)^{N},
\end{equation}
which is independent of the spin variables.

We can rewrite (\ref{Yepez1D_hamiltonian_transfer_matrix}) as
\begin{equation}
\label{Yepez1D_hamiltonian_transfer_matrix_SC_form}
\mathscr{U} =
\begin{pmatrix}
e^{-i 2\pi n/N}  & 0 \cr
0 & e^{i2\pi n/N} 
\end{pmatrix}
\begin{pmatrix}
  e^{-2\mu}& e^{i 2\pi n/N}e^{-2\nu} \cr
e^{-i2\pi n/N} e^{-2\nu}  &  e^{-2\mu}
\end{pmatrix},
\end{equation}
which has the form of a quantum lattice gas evolution operator for the Dirac equation
\begin{equation}
\label{Yepez1D_transfer_matrix_collide_stream_form}
\mathscr{U} = \mathscr{CS}.
\end{equation}
That is,  the stream and collide operators are respectively 
\begin{subequations}
\begin{eqnarray}
\mathscr{S} 
&=&
\begin{pmatrix}
e^{-i 2\pi n/N}  & 0 \cr
0 & e^{i2\pi n/N} 
\end{pmatrix}
\\
\mathscr{C} 
&=&
\begin{pmatrix}
  e^{-2\mu}& e^{i 2\pi n/N}e^{-2\nu} \cr
e^{-i2\pi n/N} e^{-2\nu}  &  e^{-2\mu}
\end{pmatrix}.
\end{eqnarray}
\end{subequations}
Now with wave number $k_z$ defined as
\begin{equation}
\ell k_z \equiv \frac{2\pi n}{N}
\end{equation}
and using (\ref{Yepez1D_coupling_constant_over_KT}) to revert back to the original variables
\begin{eqnarray}
\label{Yepez1D_coupling_constant_over_KT_exp_form}
e^{-2\mu} 
=
\sqrt{1-\left(\frac{mc^{2}\tau}{\hbar}\right)^2}
\qquad
e^{-2\nu} 
=
i \, \frac{mc^{2}\tau}{\hbar},
\qquad
\end{eqnarray}
the stream and collide operators may  be written as
\begin{subequations}
\begin{eqnarray}
\mathscr{S} 
&=&
e^{- \sigma_z i \ell k_z }  
\\
\mathscr{C} 
&=&
\begin{pmatrix}
\sqrt{1-\left(\frac{mc^{2}\tau}{\hbar}\right)^2}& i \frac{mc^{2}\tau}{\hbar}  e^{i \ell k_z }\cr
i\frac{mc^{2}\tau}{\hbar}e^{-i\ell k_z } & \sqrt{1-\left(\frac{mc^{2}\tau}{\hbar}\right)^2}
\end{pmatrix},
\end{eqnarray}
\end{subequations}
which have the identical analytical form of (\ref{stream_operator}) and (\ref{collide_operator}), respectively. 
The manifestly unitary transfer matrix (\ref{Yepez1D_transfer_matrix_collide_stream_form}) is
\begin{subequations}
\label{Yepez1D_unitary_transfer_matrix}
\begin{eqnarray}
\nonumber
\mathscr{U} 
&=&
\begin{pmatrix}
\sqrt{1-\left(\frac{mc^{2}\tau}{\hbar}\right)^2} \, e^{-i  \ell k_z }& i \frac{mc^{2}\tau}{\hbar} \cr
i\frac{mc^{2}\tau}{\hbar}&  \sqrt{1-\left(\frac{mc^{2}\tau}{\hbar}\right)^2 }\, e^{i \ell k_z }
\end{pmatrix}
\\
\label{Yepez1D_unitary_transfer_matrix_a}
\\
\label{Yepez1D_unitary_transfer_matrix_b}
&=&
\sqrt{1-\left(\frac{mc^{2}\tau}{\hbar}\right)^2} \, e^{-i  \ell k_z \sigma_z}
- i \,\sigma_x  \frac{mc^{2}\tau}{\hbar} 
\\
&=&
\label{Yepez1D_unitary_transfer_matrix_c}
\sqrt{1-\left(\frac{mc^{2}\tau}{\hbar}\right)^2} \, \cos \ell k_z 
\\
\nonumber
& - & 
i \, \sqrt{1-\left(\frac{mc^{2}\tau}{\hbar}\right)^2} \,  \sigma_z\sin \ell k_z
 -
  i \,\sigma_x  \frac{mc^{2}\tau}{\hbar} .
\end{eqnarray}
\end{subequations}
This leads us to make the following identification for the momentum of  the quantum particle
\begin{equation}
\label{Yepez1D_ansatz}
\frac{p_z c \tau}{\hbar}
=
 \sqrt{1-\left(\frac{mc^{2}\tau}{\hbar}\right)^2} \, \sin \ell k_z ,
\end{equation}
and this represents an ansatz to resolve the representation,
 where the momentum defined this way contributes to the  energy  as $E^2 = p^2 c^2 + (mc^2)^2$. 
Squaring (\ref{Yepez1D_ansatz}) gives
\begin{equation}
p^2 c^2 =
 \left[
 \frac{\hbar^2}{\tau^2}
 -\left({mc^{2}}\right)^2
 \right] \sin^2  \!\ell k_z ,
\end{equation}
and then adding $(mc^2)^2$ to both sides gives
\begin{equation}
\label{grid_level_relativistic_energy_relation}
E^2 = 
 \frac{\hbar^2}{\tau^2} \sin^2  \!\ell k_z + mc^2\cos^2  \!\ell k_z,
 \end{equation}
a novel grid-level form of the relativistic energy relation, which we will discuss below in more detail.  For now, let us rewrite (\ref{grid_level_relativistic_energy_relation}) as
\begin{equation}
\label{grid_level_relativistic_energy_relation_form_2}
\sqrt{1-\left(\frac{E\tau}{\hbar}\right)^2} = \sqrt{1-\left(\frac{mc^{2}\tau}{\hbar}\right)^2} \, \cos \ell k_z.
\end{equation}
Inserting (\ref{Yepez1D_ansatz}) and (\ref{grid_level_relativistic_energy_relation_form_2}) into (\ref{Yepez1D_unitary_transfer_matrix_c}) allows us to rewrite the unitary transfer function in a form 
\begin{subequations}
\label{Yepez1D_unitary_transfer_operator}
\begin{eqnarray}
\mathscr{U} 
& = &
\sqrt{1-\left(\frac{E\tau}{\hbar}\right)^2} 
\,
- 
\left.
i \frac{E\tau}{\hbar}
\middle(
\frac{p_z c }{E}\sigma_z
+ \frac{mc^2}{E} \sigma_x
\right)
\qquad
\\
\nonumber
&=&
 \exp\left[
- i \,\frac{\cos^{-1} \sqrt{1-\left( \frac{ E \tau}{\hbar}\right)^2}}{E} \left(\sigma_z  c\,p_z +  \sigma_x\, {m}c^2\right)
  \right]
  \\
  &=&
  e^{-i \tau \left(\sigma_z  c\,p_z +  \sigma_x\, {m}c^2\right)/\hbar}+ {\cal O}({\tau}^{3}),
\end{eqnarray}
\end{subequations}
since $\cos^{-1}\sqrt{1-\epsilon^2}=\epsilon + \epsilon^3/6+\cdots$.
In turn (\ref{Yepez1D_gersch-transfer-form-matrix-form}) becomes
\begin{subequations}
\begin{eqnarray}
\mathscr{Z} & = & (\mathscr{C}\mathscr{S})^{N} \\
\label{Yepez1D_quantum-lattice-gas-N-fold-transfer-matrix}
& = & e^{i(\sigma_{x}  mc^{2}-\sigma_{z} p_{n} c) N {\tau}/\hbar}+{\cal O}({\tau}^{3}).
\end{eqnarray}
\end{subequations}
Making the following change of variables
$z=z_{b}-z_{a}= {\ell} M$  and 
$t = t_{b}-t_{a}={\tau} N$,
 and inserting (\ref{Yepez1D_quantum-lattice-gas-N-fold-transfer-matrix}) into (\ref{Yepez1D_riazanov-feynman-path-summation-transfer-form}), the kernel becomes
\begin{widetext}
\begin{equation}
\label{Yepez1D_quantum-lattice-gas-kernel}
K_{ s_{0} s_{N}}=
\frac{1}{2N}\sum_{n=-N}^{N-1}
e^{i k_z z}
\sum_{s_0=\pm1}
e^{-i \frac{k_z{\ell}}{2}( s_0- s_{N})}
\left[
e^{i(\sigma_{x}  mc^{2}-\sigma_{z} p_{n} c) N {\tau}/\hbar}
\right]_{ s_{-1} s_{N}}.
\end{equation}
\end{widetext}
For large $N$  (in the low-energy limit), we can neglect the high-momentum grid scale  ${\ell}$ term compared with the low-momentum length scale   $z$ terms, so we make the approximation
\begin{equation}
\sum_{s_0=\pm1}
e^{-i \frac{k_z{\ell}}{2}( s_0- s_{N})}
\approxeq
\sum_{s_0=\pm1}
1
=2.
\end{equation}
So, in turn, we  write the kernel as 
\begin{equation}
\label{Yepez1D_quantum-lattice-gas-kernel_form2}
K_{ s_{0} s_{N}}=
\frac{1}{N}\sum_{n=-N}^{N-1} e^{i k_z z}
\left[
e^{i(\sigma_{x}  mc^{2}-\sigma_{z} p_{n} c) N {\tau}/\hbar}
\right]_{ s_{-1} s_{N}}.
\end{equation}
In the continuum limit, the summation goes over to an integral ($\frac{1}{N}\sum_{n}\rightarrow\frac{\ell}{h}
\int dp$) and so we have
\begin{subequations}
\label{Yepez1D_quantum-lattice-gas-kernel-for-a-continuum}
\begin{eqnarray}
K_{\alpha\beta}(z,t)& \equiv & \lim_{N\rightarrow\infty}K_{ s_{0} s_{N}}\\
& =&
\frac{\ell}{ \hbar}
\int_{-\infty}^{\infty}\frac{dp}{2\pi}\;
e^{i \frac{p z}{\hbar}}
\left[
e^{i(\sigma_{x}  mc^{2}-\sigma_{z} p c) t/\hbar}
\right]_{\alpha\beta}.
\qquad
\end{eqnarray}
\end{subequations}

\begin{figure}[!h!b!t]
\begin{center}
\includegraphics[width=3.4in]{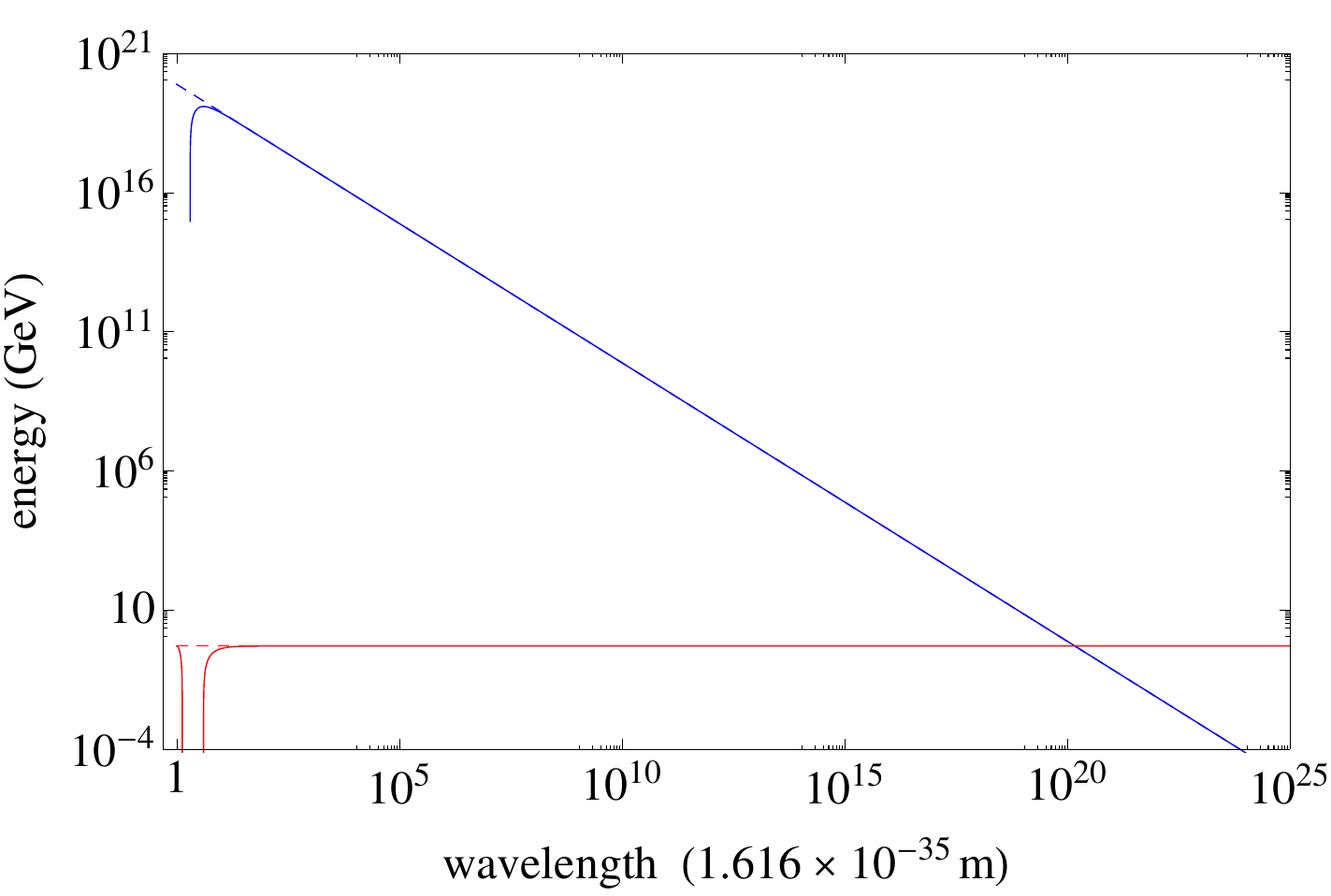}
\caption{\label{ModifiedDeBroglieRelation} \footnotesize Log-log plot of (\ref{Yepez1D_modified_QM_relations}) for mass (red) and momentum (blue) in GeV of a single proton versus its wavelength measured in units of the Planck length, $\ell=1.616 \times 10^{-35}$m. The straight lines are the de Broglie relation  of quantum mechanics, $p=h/\lambda$ (blue dashed line), and the invariant mass of special relativity, 
${m}=0.511$ MeV 
(red dashed line).  
Respectively, the slopes are $-1$ and $0$ for the standard theories. 
The intersection of the mass and momentum lines occurs at 
the Compton wavelength of the Dirac particle.
}
\end{center}
\end{figure}

The form of the relativistic energy relation (\ref{grid_level_relativistic_energy_relation}) leads us to define high-energy relations for momentum and mass
\begin{subequations}
\label{Yepez1D_modified_QM_relations}
\begin{eqnarray}
\label{Yepez1D_modified_QM_relations_a}
p_\text{grid}
& \equiv&
 \frac{\hbar}{c\tau } \sin  \ell k_z
=  \frac{\hbar}{c\tau} \sin  \!\left(\frac{2 \pi \ell}{\lambda}\right)
\\
\label{Yepez1D_modified_QM_relations_b}
m_\text{grid}
& \equiv&
m \cos  \ell k_z
=
m \cos  \!\left(\frac{2 \pi \ell}{\lambda}\right),
\qquad
\end{eqnarray}
\end{subequations}
where we use the Compton wave length (i.e. $k_z = 2\pi/\lambda$). Equation (\ref{Yepez1D_modified_QM_relations_a}) is a modified de Broglie relation.
Plots of (\ref{Yepez1D_modified_QM_relations}) are given in Fig.~\ref{ModifiedDeBroglieRelation}.  The modified de Broglie relation (\ref{Yepez1D_modified_QM_relations_a}) has the effect of reducing the vacuum energy that arises from a chiral field.   

\section{Quantum simulation}
\label{Quantum_simulation}

\subsection{Continuity relation for a spinor field}

If we multiply the relativistic wave equation for a Dirac particle by the adjoint 4-spinor, then we have
\begin{subequations}
\begin{equation}
\psi^\dagger \left( i\hbar c\,\gamma^\mu\partial_\mu  -mc^2 \right) \psi=0,
\end{equation}
and the adjoint of this equation is
\begin{equation}
\psi^\dagger \left( -i\hbar c  \overleftarrow{\partial_\mu}  \gamma^\mu  -mc^2\right) \psi=0.
\end{equation}
\end{subequations}
Taking the difference of these two equations, and dividing the result through by $\hbar$, gives a continuity relation for the 4-spinor particle 
\begin{equation}
\partial_\mu( i c\, \psi^\dagger \gamma^\mu \psi) 
= 0 ,
\end{equation}
where one identifies the   probability flux density 
as
\begin{equation}
\label{Dirac_4_flux_density}
u^\mu \equiv c\,\psi^\dagger\gamma^\mu\psi,
\end{equation}
 a 4-vector governed by a flux continuity relation. 
The reason for dropping the overall phase factor of $i$ is to have the temporal component $u_0= c \, \overline{\psi}\psi$ be real valued.
In contrast, the  4-current density, which may be defined as
\begin{equation}
\label{Dirac_4_current_density}
j^\mu \equiv i c\,\overline{\psi}\gamma^\mu\psi,
\end{equation}
satisfies the  continuity relation
\begin{equation}
\label{covariant_continuity_relation}
\partial_\mu j^\mu = 0,
\end{equation}
where the conserved 4-current is $j^\mu \equiv (\rho c, \bm{j})$.  Thus, the conservation of probability (or more precisely the conservation of particle number) is manifestly covariant\footnote{The  common definition  $j^\mu \equiv c\,\overline{\psi}\gamma^\mu\psi$ ensures that $j^0=\psi^\dagger\psi$ is a real-valued and positive-definite scalar.}.

\subsection{Dirac particle in a square well potential in 1+1 dimensions}

Let us consider a  quantum particle  
 in a confining one-dimensional lattice with grid points $z = n \ell$, for integer $n$ ($0\le n\le L$), with stream and collide operators given by (\ref{Yepez_quantum_algorithm_Dirac_particles}) and with the similarity transformation given by $R=(\sigma_x+\sigma_z)/\sqrt{2}$.  The grid-level equation of motion for this finite quantum system is
\begin{subequations}
\begin{equation}
\label{Yepez_qlg_1_1_dimensions}
\eta'= R\, U_s^z U_C R^\dagger\eta.
\end{equation}
 We will use (\ref{Yepez_qlg_1_1_dimensions}) for the purpose of modeling the dynamical behavior of a Dirac particle in a square well potential. The low-energy effective field theory (in the rotating frame) of (\ref{Yepez_qlg_1_1_dimensions}) is the Dirac equation
\begin{equation}
\label{EFT_QLG_1_1_dimensions}
i\hbar \partial_t\eta(z,t) = p c\,\sigma_z \eta(z,t) + m c^2 \sigma_x  \eta(z,t),
\end{equation}
where the momentum operator is identified as $p = -i \hbar \partial_z$.\footnote{Here we are writing the Dirac operator $\alpha=\sigma_z$, whereas earlier we had $\alpha=-\sigma_z$. The overall sign is specified by the direction of the spatial displacement caused by the stream operator $U_s^z$.}

We will work in the nonrotating frame $\psi \equiv R \,\eta$ where the effective equation of motion is
\begin{equation}
\label{Dirac_equation_1D_rotating_frame}
i\hbar \partial_t\psi(z,t) = p c\,\sigma_x \psi(z,t) + m c^2 \sigma_z  \psi(z,t).
\end{equation}
\end{subequations}
Using separation of variables, the 2-spinor field is
\begin{equation}
\psi(z,t) = 
\begin{pmatrix}
      \phi(z)    \\
      \xi(z)
\end{pmatrix}
e^{-i (p z - E t)/\hbar} ,
\end{equation}
and in turn the eigenequation is
\begin{equation}
E\psi = \begin{pmatrix}
m c^2      &  p c   \\
p c      &   -mc^2
\end{pmatrix}
\psi.
\end{equation}
Explicitly writing out the coupled component equations
\begin{subequations}
\begin{eqnarray}
E\phi
 & = & 
 mc^2 \phi + p c\, \xi
  \\
E\xi & = & 
pc \, \phi - mc^2 \xi,
\end{eqnarray}
\end{subequations}
one immediately identifies  two solutions types
\begin{subequations}
\label{phi_xi_solutions}
\begin{eqnarray}
\phi
 & = & 
\frac{p c}{E-mc^2}\, \xi
  \\
\xi & = & 
\frac{p c}{E+mc^2} \, \phi.
\end{eqnarray}
\end{subequations}
That is, the eigensolutions for a free Dirac particle have the form
\begin{eqnarray}
\psi(z,0)
=
\begin{cases}
\phi(z) \begin{pmatrix}
      1    \\
      \frac{p c}{E+mc^2}   
\end{pmatrix}  
e^{-i p z /\hbar} ,   
& \text{for } E>0
 \\
\xi(z) \begin{pmatrix}
      \frac{p c}{E-mc^2}   \\
      1    
\end{pmatrix}  
e^{-i p z /\hbar}   , 
      & \text{for } E<0.
\end{cases}
\end{eqnarray}
Let us consider the positive-energy solution for a plane wave with momentum eigenvalue $p=\hbar k$.  Normalizing such that $\langle \psi_k|\psi_{k'}\rangle \equiv \int dz \,\psi_k^\dagger(z,0) \psi_{k'}(z,0) = \delta(k-k')$, we find that $\langle \psi_k|\psi_k\rangle = 2E/({E+mc^2})$, so in turn we may write the plane-wave solution as
\begin{subequations}
\begin{equation}
\psi_k(z)
=
\sqrt{\frac{E+mc^2}{2E}}
 \begin{pmatrix}
      1    \\
     \sqrt{ \frac{E-mc^2}{E+mc^2}}   
\end{pmatrix}  
e^{-i k z },
\end{equation}
where we made use of the relativistic energy relation $E^2 = (pc)^2 + (mc^2)^2$.  We can rewrite this solution by splitting it into its right-going (spin-up) and left-going (spin-down) components\footnote{Right and left-going chirality and spin-up and spin-down properties of a Dirac particle are the physically the same properties in one spatial dimension. This is not the case in higher dimensions.}
\begin{equation}
\psi_k
=
\sqrt{\frac{1}{2E}}
\left[
 \underbrace{
 \sqrt{ {E+mc^2}}  
 \begin{pmatrix}
      1    \\
    0 
\end{pmatrix}  
}_{\text{right-goer (spin up)}}
+
 \underbrace{
 \sqrt{ {E-mc^2}}  
 \begin{pmatrix}
      0    \\
    1 
\end{pmatrix}  
}_{\text{left-goer (spin down)}}
\right]
e^{-i k z }.
\end{equation}
\end{subequations}

As a way to avoid the Klein paradox, we can model the external confining square well barrier as regions in space where the mass of the Dirac particle is large.  This is depicted as
\begin{equation}
\label{square_well_diagram}
\xy
{\ar (0,0)*{}; (0,35)*++{V(z)\cong m(z)}};
{\ar (0,0)*{}; (35,0)*++{z}};
(-10,25)*{}; (0,25)*{} **@{-};
(15,25)*{}; (28,25)*+{M} **@{-};
(0,5)*{}; (17.5,5)*+{m} **@{-}; 
(15,5)*{}; (15,25)*{} **@{-}; 
(-5,15)*{}; (20,15)*+{E} **@{--}; 
(25,10)*{\text{I}};
(7.5,10)*{\text{II}};
(-7.5,10)*{\text{III}};
(0,-4)*{0};
(15,-4)*{L};
\endxy
\end{equation}
In (\ref{square_well_diagram}) the mass of the Dirac particle is $m$ for $0\le z \le L$ (region II) and its mass is $M> m$ for $z>L$ (region I) and for $z<0$ (region III).
So the plane-wave solutions in (\ref{square_well_diagram}) are
\begin{subequations}
\begin{eqnarray}
\psi_k^{\text{I}} (z)
& = & 
A\,
 e^{i k' z}
\begin{pmatrix}
      1    \\
      \frac{\hbar k' c}{E+M c^2}  
\end{pmatrix}
 \\
\psi_k^{\text{II}}  (z)
& = & 
B\,
 e^{i k z}
\begin{pmatrix}
      1    \\
      \frac{\hbar k c}{E+m c^2}  
\end{pmatrix}
+
C\,
 e^{-i k z}
\begin{pmatrix}
      1    \\
      \frac{-\hbar k c}{E+m c^2}  
\end{pmatrix}
\qquad
 \\
\psi_k^{\text{III}}  (z)
& = & 
D\,
 e^{-i k' z}
\begin{pmatrix}
      1    \\
      \frac{-\hbar k' c}{E+M c^2}  
\end{pmatrix}.
\end{eqnarray}
\end{subequations}
For convenience, let us write the the 2-spinor field in region II as
\begin{equation}
\label{psi_in_region_II}
\psi_k^{\text{II}} (z)=
\begin{pmatrix}
      B\,
 e^{i k z}    +
C\,
 e^{-i k z}\\
       \left(
       B\,
 e^{i k z}    -
C\,
 e^{-i k z}
 \right) P
\end{pmatrix},
\end{equation}
where 
\begin{equation}
\label{P_definition}
P \equiv \frac{\hbar k c}{(E+m c^2)}.
\end{equation}%

Choosing  boundary conditions to ensure that the Dirac particle is appropriately confined within the square well in region II is a rather subtle matter.  We will choose boundary conditions such that the probability flux density vanishes at $z=0$ and $z=L$ \cite{EurJPhys17_1996_p19}. It is a remarkable property of the spinor structure of a Dirac particle that we do not have to choose $\psi(0)=0$ and $\psi(L)=0$ even when $M = \infty$, nor do the components of the 2-spinor field individually have to be continuous functions of position across the boundaries.\footnote{Albeit the probability density and probability flux density are both continuous across the boundaries.}  Boundary conditions with continuous probability flux density $\overline{\psi}\psi\equiv\psi^\dagger \gamma^0\psi$ vanishing at the container walls were originally introduced in the MIT bag model of bound hadrons \cite{PhysRevD.9.3471,PhysRevD.10.2599}.

Let us write the Dirac equation (\ref{Dirac_equation_1D_rotating_frame}) as
\begin{subequations}
\begin{equation}
\label{Dirac_equation_1D_rotating_frame_alpha_beta_form}
i\hbar \partial_t\psi(z,t) = \left( - i \hbar c \alpha_z \partial_z + m(z) c^2 \beta\right)  \psi(z,t),
\end{equation}
where $\alpha_z=\sigma_x$ and $\beta=\sigma_z$ and where $m(z)=m$ for $0\le z\le L$ and $m(z)=M$ otherwise. In the chiral representation, this equation may be written in manifestly covariant form
\begin{equation}
\label{Dirac_equation_1D_rotating_frame_gamma_form}
i\hbar c ( \gamma^0 \partial_{0}+ \gamma^z \partial_z)\psi - m(z) c^2   \psi = 0,
\end{equation}
\end{subequations}
where $\partial_0 \equiv \partial_{ct}$, $\gamma^0 = \beta$, and $\gamma^z= \beta \alpha_z = -i \sigma_y$.
In this representation, the probability flux density is the difference of the particle's probability occupancy of its spin-up and spin-down states
\begin{equation}
\label{probability_density_flux_chiral_rep_1D}
\overline{\psi}\psi=
\begin{array}{l}
\begin{pmatrix}
      \psi_R^\dagger    
      &  
      \psi_L^\dagger 
      \end{pmatrix}
          \\
          {}
\end{array}
 \begin{pmatrix}
 1     & 0   \\
   0   &  -1
\end{pmatrix}
\begin{pmatrix}
      \psi_R    
      \\  
      \psi_L 
      \end{pmatrix}
      =
       \psi_R^\dagger   \psi_R    
-
       \psi_L^\dagger  \psi_L .        
\end{equation}

Equating the 4-current density  (\ref{Dirac_4_current_density}) along the $-\hat{\bm{z}}$-direction to the probability flux density (\ref{Dirac_4_flux_density}) at the left wall, the boundary condition at $z=0$ is 
\begin{subequations}
\label{left_wall_boundary_condition}
\begin{equation}
j_z (0)= -i c\,\overline{\psi}_k(0)\gamma_z \psi_k(0) =  c\,\overline{\psi}_k(0)\psi_k(0) =  u(0)
\end{equation}
or
\begin{equation}
\label{left_wall_boundary_condition_matrix_form}
 \begin{pmatrix}
  0    &  -1  \\
1      &  0
\end{pmatrix}
 \begin{pmatrix}
       \psi_R^{\text{II}}(0) 
    \\
       \psi_L^{\text{II}}(0) 
\end{pmatrix}
 = 
 i
 \begin{pmatrix}
       \psi_R^{\text{II}}(0) 
    \\
       \psi_L^{\text{II}}(0) 
\end{pmatrix}.
\end{equation}
\end{subequations}
 One chooses the boundary condition (\ref{left_wall_boundary_condition}) to force both $j_z(0)=0$ and $u(0)=0$, which represents both vanishing  probability current escaping the square well and vanishing probability flux at the left wall.  We will now verify that this is indeed the case.

From (\ref{left_wall_boundary_condition_matrix_form}) we see that the components are constrained by the relation $\psi_L^{\text{II}}(0) = -i\, \psi_R^{\text{II}}(0)$.  Using the plane-wave solution (\ref{psi_in_region_II}), the left wall boundary condition implies
\begin{subequations}
\label{B_C_P_equation}
\begin{equation}
\label{B_C_P_equation_a}
i (B-C) P = B+C
\end{equation}
or
\begin{equation}
\label{B_C_P_equation_b}
C = B \,\frac{i P -1}{i P + 1}.
\end{equation}
\end{subequations}
Inserting (\ref{psi_in_region_II}) into  (\ref{probability_density_flux_chiral_rep_1D}), we verify that the probability flux density vanishes at the left wall
\begin{equation}
\overline{\psi}(0)\psi(0)=
|B+C|^2 - |B-C|^2 P^2\stackrel{(\ref{B_C_P_equation_a})}{=}0.
\quad \checkmark
\end{equation}

\begin{figure}[!h!t!b!p]
\begin{center}
\includegraphics[width=3.35in]{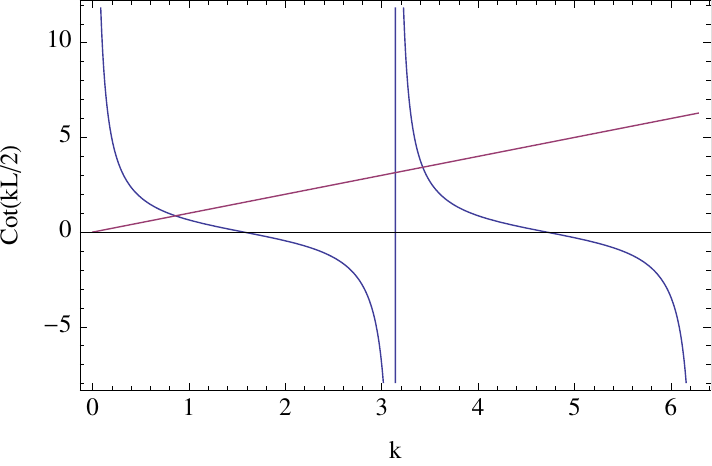}
\caption{\label{Dirac_SquareWell_TranscendentalEq}  \footnotesize Solution to transcendental equation (\ref{B_C_P_equation_right_c_2}) for a square well of size $L=2$ for a Dirac particle of mass $m=1/2$ in lattice units $\hbar=c=1$. The first crossing occurs at $k=0.860334$.}
\end{center}
\end{figure}

Now we can apply similar boundary conditions at the right wall.
The boundary condition at $z=L$ is 
\begin{subequations}
\label{right_wall_boundary_condition}
\begin{equation}
j_z (L)= i c\, \overline{\psi}_k(L)\gamma_z \psi_k(L) = c\, \overline{\psi}_k(L)\psi_k(L) = u(L)
\end{equation}
or
\begin{equation}
 \begin{pmatrix}
  0    &  -1  \\
1      &  0
\end{pmatrix}
 \begin{pmatrix}
       \psi_R^{\text{II}}(L) 
    \\
       \psi_L^{\text{II}}(L) 
\end{pmatrix}
 = 
 -i
 \begin{pmatrix}
       \psi_R^{\text{II}}(L) 
    \\
       \psi_L^{\text{II}}(L) 
\end{pmatrix}.
\end{equation}
\end{subequations}
 The boundary condition (\ref{right_wall_boundary_condition}) forces both $j_z(L)=0$ and $u(L)=0$, which represents both vanishing probability current escaping the square well and vanishing probability flux at the right wall.
So the components are constrained by the relation $\psi_L^{\text{II}}(L) = i\, \psi_R^{\text{II}}(L)$.  Using the plane-wave solution (\ref{psi_in_region_II}), the right wall boundary condition implies
\begin{subequations}
\label{B_C_P_equation_right}
\begin{equation}
\label{B_C_P_equation_right_a}
-i \left(B \,e^{i k L}-C\,e^{-i k L}\right) P = B\,e^{i k L}+C\,e^{-i k L}.
\end{equation}
Defining $e^{i\theta}\equiv C/B$ and multiplying  through by  $e^{-i\theta/2}$, (\ref{B_C_P_equation_right_a}) becomes
\begin{equation}
\label{B_C_P_equation_right_b}
-i \left(e^{i (k L-\theta/2)}-e^{-i (k L-\theta/2)}\right) P = e^{i (k L-\theta/2)}+e^{-i (k L-\theta/2)}
\end{equation}
or
\begin{equation}
\label{B_C_P_equation_right_c}
\cot\left(k L-\frac{\theta}{2} \right) = P.
\end{equation}
\end{subequations}
From (\ref{B_C_P_equation_b}) we have
\begin{equation}
\label{e_i_theta_identity}
e^{i \theta}=\frac{i P -1}{i P + 1} = \frac{P^2-1}{P^2+1} + i \frac{2P}{P^2+1},
\end{equation}
so the phase angle is determined by $\tan\theta = 2P/(P^2-1)$.
Additionally, we can write (\ref{B_C_P_equation_right_b}) as
\begin{equation}
-(i P+1) e^{i k L} = (-i P + 1) e^{i \theta} e^{-i k L}
\end{equation}
or
\begin{equation}
e^{2 i k L} = \frac{i P - 1}{i P+1 } e^{i \theta} \stackrel{(\ref{e_i_theta_identity})}{=} e^{2i \theta}.
\end{equation}
This implies that $\theta=k L$, so we can write (\ref{B_C_P_equation_right_c}) as
\begin{equation}
\label{B_C_P_equation_right_c_2}
\cot\left(\frac{k L}{2} \right)  \stackrel{(\ref{P_definition})}{=}  \frac{\hbar k c}{(E+m c^2)}.
\end{equation}
This is a transcendental equation whose solution for wave number $k$ ensures both vanishing probability current escaping the square well and vanishing probability flux at the right wall.  The solution of (\ref{B_C_P_equation_right_c_2}) is shown graphically in Fig.~\ref{Dirac_SquareWell_TranscendentalEq} for a relativistic case where $\hbar k > m c^2$.
With this $k$ value, we can determine the value of $P$ and in turn determine the value of the coefficient $C$ in terms of $B$ by using (\ref{B_C_P_equation_b}).

\begin{figure}[!h!t!b!p]
\begin{center}
\includegraphics[width=3.35in]{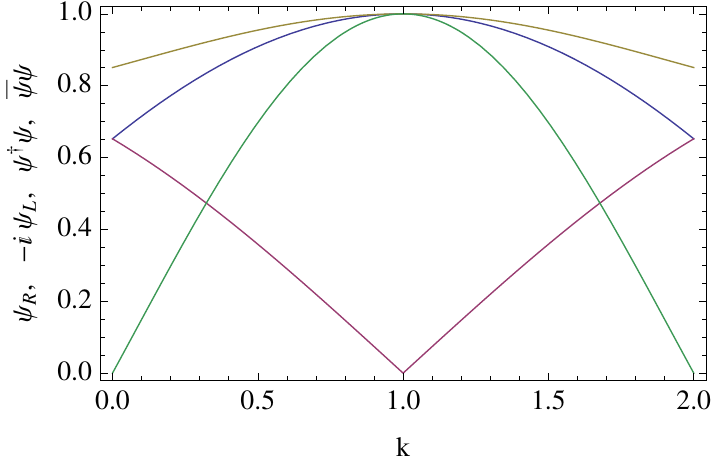}
\caption{\label{PsiRegionII} \footnotesize Plane-wave solution for a relativistic case with $k=0.860334$, square well size $L=2$, and Dirac particle mass $m=1/2$ in lattice units $\hbar=c=1$. The spin components $\psi_R$ (blue) and $\psi_L$ (red) are matched but in the relativistic regime ($k>m$) do not vanish at the $z=0$ and $z=2$ walls, nor does the probability density $\psi^\dagger\psi$ (gold) vanish at the walls.  However, the 4-flux 0-component $\overline{\psi}\psi$ (green) does indeed vanish at the boundary walls.}
\end{center}
\end{figure}

Plots of the 2-spinor components $\Re\{\psi_R\}$ and $\Im\{\psi_L\}$ and plots of the probability density $\rho\equiv\psi^\dagger\psi=\overline{\psi}\gamma^0\psi$ and particle flux density's time component $\psi^\dagger\gamma^0\psi=\overline{\psi}\psi$ are shown in Fig.~\ref{PsiRegionII} for a relativistic case. The physical interpretation of the particle dynamics is that our plane-wave solution represents a perfectly matched  situation where the Dirac particle is trapped because of total internal reflection occurring with $\rho\ne 0$ and $|\psi_R| = |\psi_L|$ at the boundaries.  At $z=0$ the spin-down (left-going) state scatters off the wall into the spin-up (right-going) state, which is the bounce-back collision $\psi_L(0)\rightarrow\psi_R(0)$.  Likewise, at $z=L$ the spin-up (right-going) state scatters off the wall into the spin-down (left-going) state, which is the bounce-back collision $\psi_R(L)\rightarrow\psi_L(L)$.  This is an example where one can interpret the spin state dynamics in terms of kinetic particle motion in position space at the grid level.

\begin{figure}[!h!t!b!p]
\begin{center}
\includegraphics[width=3.35in]{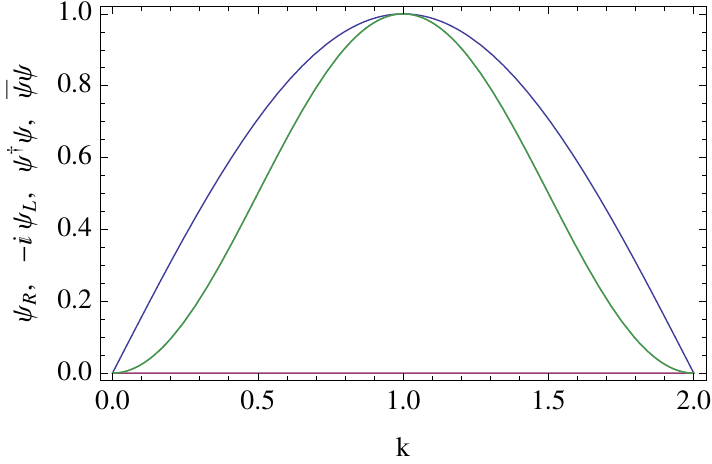}
\caption{\label{PsiRegionIINR} \footnotesize Plane-wave solution for a nonrelativistic case with $k=1.5704$, square well size $L=2$, and Dirac particle mass $m=2000$ in lattice units $\hbar=c=1$. The spin components $\psi_R$ (blue) and $\psi_L\approx 0$ (red) are matched near zero at the $z=0$ and $z=2$ walls.  That is, in the norrelativistic regime ($k\ll m$) these components nearly vanish at the $z=0$ and $z=2$ walls. The probability density $\psi^\dagger\psi$ (gold) also vanishes at the walls.  The 4-flux 0-component $\overline{\psi}\psi$ (green) still vanishes at the boundary walls and  $\overline{\psi}\psi\approx \psi^\dagger\psi$ because $\psi_L\approx 0$, so the green curve overlaps the gold curve.}
\end{center}
\end{figure}

In the nonrelativistic regime where $\hbar k c \lll m c^2$, then the transcendental equation (\ref{B_C_P_equation_right_c_2}) simplifies to 
\begin{equation}
\cot\left(\frac{k L}{2} \right)  \approx  0,
\end{equation}
which is analytically solvable:  the wave number solution is $k \simeq \pi/L$.  In  this limit, $P\approx 0$ and $B= e^{i\pi}C=-C$, so the spinor field in region  II reduces to
\begin{equation}
\label{psi_in_region_II_NR}
\psi_k^{\text{II}} (z)
\approx
2i B\begin{pmatrix}
\sin{k z}   
 \\
 0
 \end{pmatrix}.
\end{equation}
The lower (fast) component vanishes while the upper (slow) component is just the usual ground state solution of the Schroendinger wave equation for a square well potential, and these components are shown in Fig.~\ref{PsiRegionIINR}.

\begin{figure}[!h!t!b!p]
\begin{equation*}
\hspace{-0.25in}
\xy
(-10,0)*{\includegraphics[width=2in]{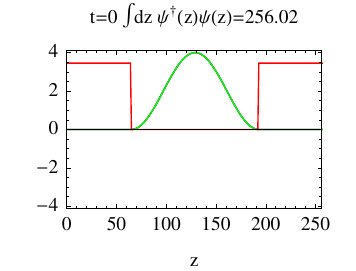}};
(35,0)*{\includegraphics[width=2in]{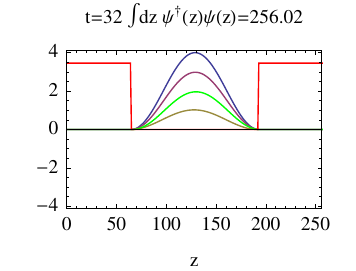}};
(-10,-40)*{\includegraphics[width=2in]{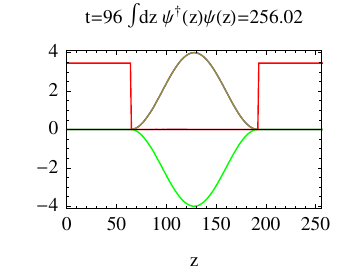}};
(35,-40)*{\includegraphics[width=2in]{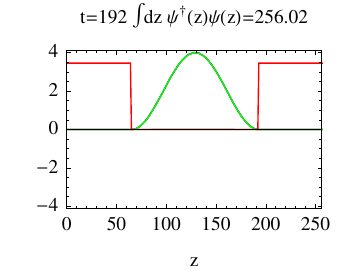}};
\endxy
\end{equation*}
\vspace{-0.15in}
\caption{\label{DiracSquareWell_Simulation} \footnotesize Numerical prediction of the dynamical behavior of the lowest energy  plane-wave eigenstate for a nonrelativistic case for a Dirac particle with wave number $k=1.5704\approx \pi/2$ confined to a square well.  The grid size is $L=256$.  The Dirac particle's mass is $m=1.94707+ 2000\pi$ and in lattice units $\hbar=c=1$. The barrier (red) is modeled with $M=3.45218$.  The spin components $\psi_R$ (blue) and $\psi_L$ (purple) nearly vanish at the walls, and the probability density $\psi^\dagger\psi$ (gold) vanishes at the walls too.  The 4-flux 0-component $\overline{\psi}\psi$ (green) becomes negative  at $t=92$.  The green curve overlaps the gold curve at $t=0$ and at $t=192$.}
\end{figure}

\begin{figure}[!h!t!b!p]
\includegraphics[width=2.75in]{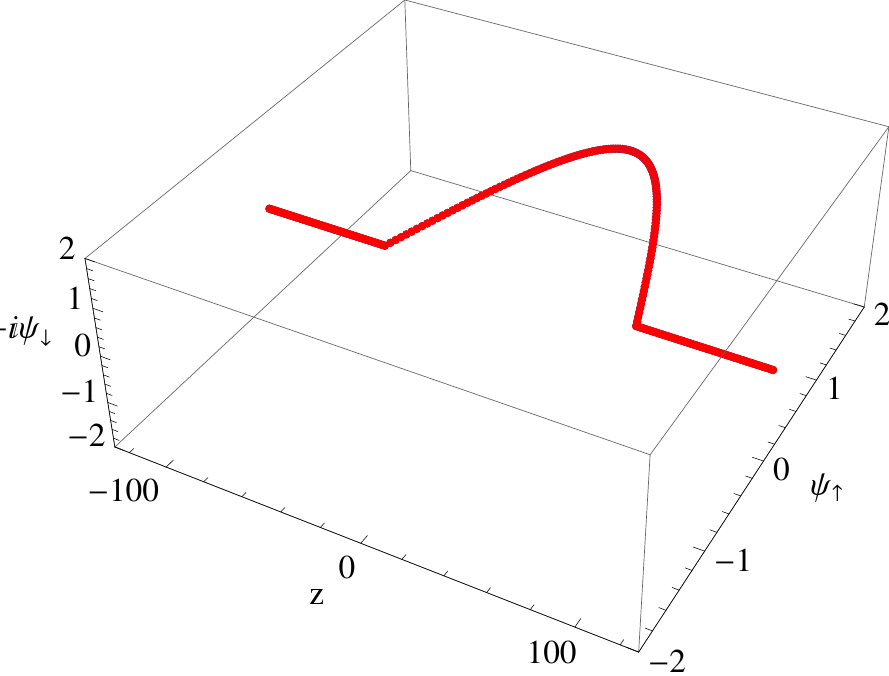}
\includegraphics[width=2.75in]{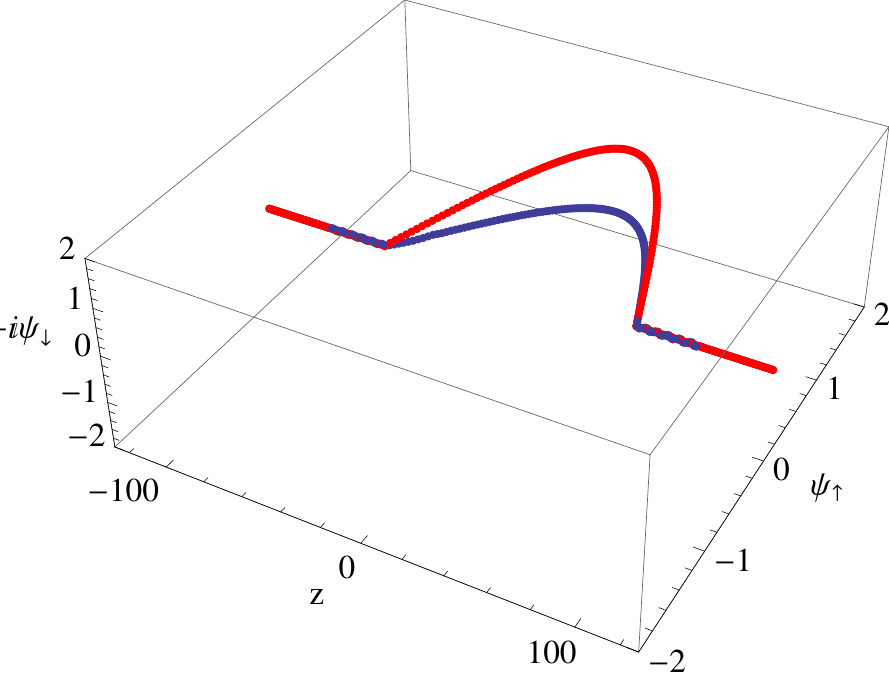}
\includegraphics[width=2.75in]{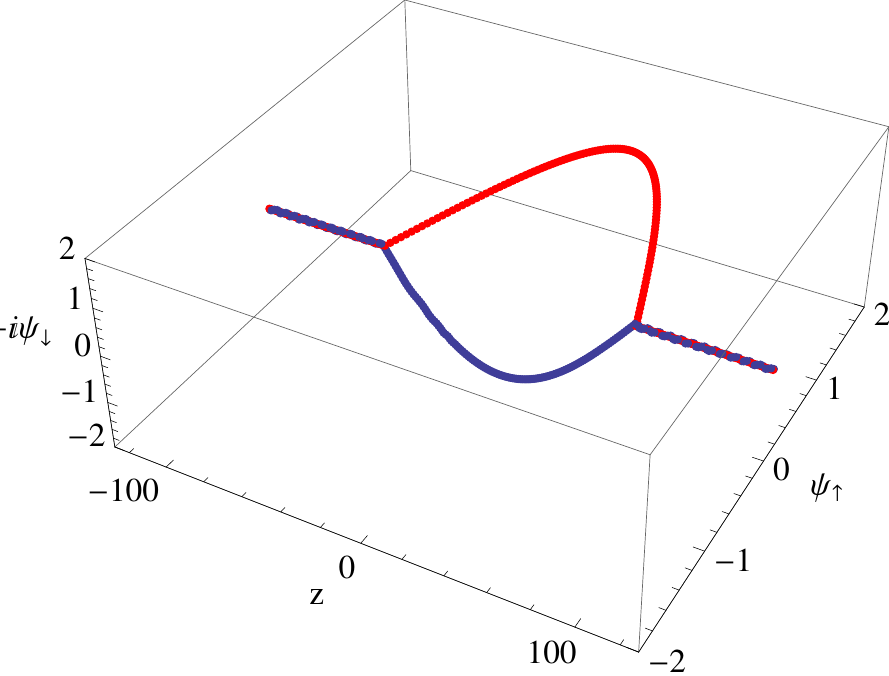}
\includegraphics[width=2.75in]{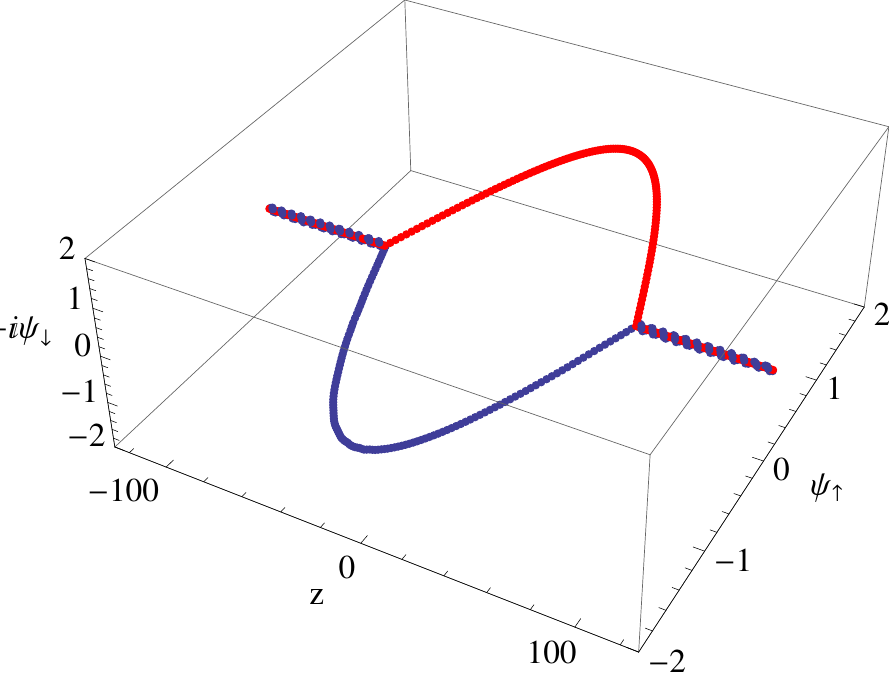}
\caption{\label{DiracSquareWell_Parametric} \footnotesize Parametric plot (blue curve) of $\Re[\psi_R(z)]$ and $\Im[\psi_L(z)]$ versus $z$ for the same simulation as shown in Fig.~\ref{DiracSquareWell_Simulation} for $t=0, 32, 96, 192$.  The numerical solution is seen to wrap around the initial state (red curve) as time progresses in regions I and III inside the potential barrier. }
\end{figure}

A quantum simulation of a Dirac particle confined to a square well is shown in Fig.~\ref{DiracSquareWell_Simulation}. The quantum simulation is in excellent agreement with theory.
Another way to see the behavior of a Dirac particle confined to a square well potential is shown in Fig.~\ref{DiracSquareWell_Parametric} as a parametric plot.  The 2-spinor field is not a perfect approximation of a nonrelativistic scalar field because the 2-spinor field twists inside of the potential barrier. Despite this twisting (a relativistic  effect), the shape of the wave function within the square well (region II) remains sinusoidal.

\begin{figure}[!h!t!b!p]
\begin{equation*}
\hspace{-0.25in}
\xy
(-10,0)*{\includegraphics[width=2in]{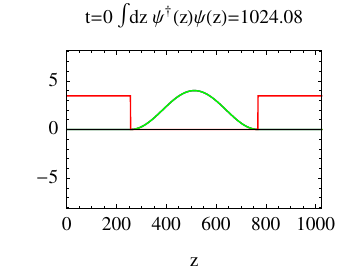}};
(35,0)*{\includegraphics[width=2in]{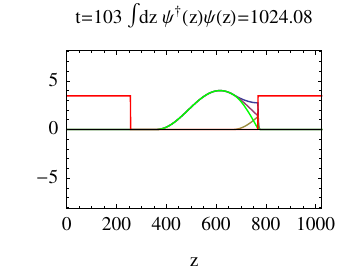}};
(-10,-40)*{\includegraphics[width=2in]{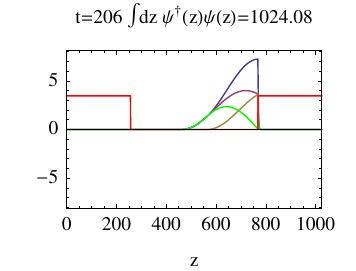}};
(35,-40)*{\includegraphics[width=2in]{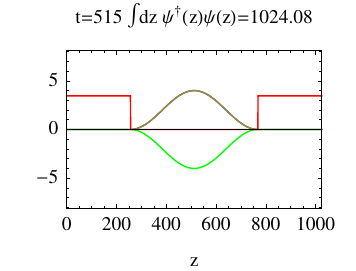}};
(-10,-80)*{\includegraphics[width=2in]{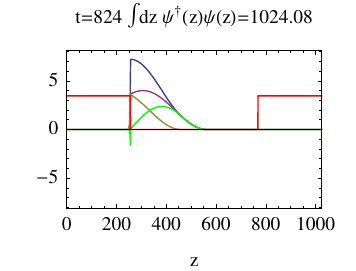}};
(35,-80)*{\includegraphics[width=2in]{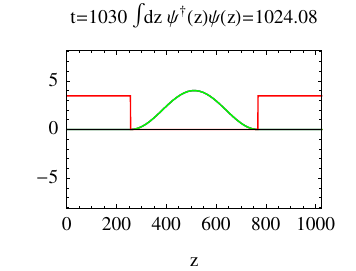}};
\endxy
\end{equation*}
\vspace{-0.15in}
\caption{\label{WeylSquareWell_Simulation} \footnotesize Numerical prediction of the dynamical behavior of the lowest energy  plane-wave eigenstate for the case of a Weyl particle (m=0) with wave number $k=1.5704\approx \pi/2$ confined to a square well.  The grid size is $L=1024$.   The barrier (red) is modeled with $M=3.45218$.  The quantum particle initially moves to the right as the upper component of the 2-spinor field is nonzero while the lower component is zero.  The Weyl particle remains confined to the square well, reflecting off of boundary walls.}
\end{figure}

A quantum simulation of a Weyl particle confined to a square well is shown in Fig.~\ref{WeylSquareWell_Simulation}. The quantum simulation is in excellent agreement with theory.
%

\subsection{Dirac particle in a harmonic potential in 1+1 dimensions}

Here we will use the quantum algorithm (\ref{Yepez_quantum_algorithm_Dirac_particles}) to simulate a nonrelativistic scalar particle in an external parabolic potential. We will employ the quantum lattice gas algorithm for a relativistic Dirac particle in the parameter regime where $mc^2 \gg pc$. 

Let us begin with the effective nonrelativistic particle dynamics in 1+1 dimensions as governed by the Schroedinger wave equation
\begin{equation}
i\hbar \partial_t\phi(z,t) = -\frac{\hbar^2}{2m} \partial_{zz} \phi(z,t) + \frac{1}{2} \kappa z^2 \phi(z,t).
\end{equation}
Using separation of variables
\begin{equation}
\phi(z,t) = f(z)\,e^{-i E t/\hbar} 
\end{equation}
gives
\begin{subequations}
\begin{equation}
\label{eigenequation}
-\frac{\hbar^2}{2m} d_{zz} f(z) + \left(\frac{1}{2} \kappa z^2 -E\right) f(z)=0
\end{equation}
or
\begin{equation}
d_{zz} f(z) + \left( \frac{2m E}{\hbar^2}-   \frac{m \kappa}{\hbar^2} z^2\right) f(z)=0.
\end{equation}
Defining $b\equiv \sqrt{m \kappa /(4\hbar^2)}$, the eigenequation may be written as
\begin{equation}
\label{eigenequation_form2}
d_{zz} f(z) + \left(4b E \sqrt{\frac{m}{\kappa \hbar^2}}-  4 b^2 z^2\right) f(z)=0.
\end{equation}
\end{subequations}
Consider a solution of the form
\begin{subequations}
\label{f_solution}
\begin{equation}
f(z) = e^{-b \,z^2} h(\varsigma\, z) ,
\end{equation}
where $\varsigma \equiv (m\kappa/\hbar^2)^{\frac{1}{4}}$.  Then, 
\begin{eqnarray}
d_z f 
&=&
e^{-b \,z^2} d_z h(\varsigma\, z) - 2 b\, z\, e^{-b \,z^2}  h(\varsigma\, z) 
\\
\nonumber
d_{zz} f
&=&
e^{-b \,z^2} d_{zz} h(\varsigma\, z) - 4 b\, z\, e^{-b \,z^2} d_z h(\varsigma\, z) 
\\
&+&\left( 4 b^2 z^2-2b\right) e^{-b \,z^2}  h(\varsigma\, z) .
\end{eqnarray}
\end{subequations}
Inserting (\ref{f_solution}) in the eigenequation (\ref{eigenequation_form2}) leads to 
\begin{equation}
d_{zz} h( \varsigma\, z) - 4 b\, z\, d_z h(\varsigma\, z) +  \left( 4b E \sqrt{\frac{m}{\kappa \hbar^2}}-2 b\right) h(\varsigma\, z) =0,
\end{equation}
or since $\varsigma^2=2b$ this is
\begin{equation}
d_{\varsigma z,\varsigma z} h( \varsigma\, z) - 2 \varsigma\, z\, d_{\varsigma z} h(\varsigma\, z) + 2 \left( E \sqrt{\frac{m}{\kappa\hbar^2}}-\frac{1}{2}\right) h(\varsigma\, z) =0,
\end{equation}
which is Hermite's ordinary differential equation
\begin{equation}
d_{z'z'} h( z') - 2 z'\, d_{z'} h( z') + 2 n\, h( z') =0,
\end{equation}
for $z'\equiv \varsigma\, z$ and 
\begin{equation}
n = \left( E \sqrt{\frac{m}{\kappa\hbar^2}}-\frac{1}{2}\right) .
\end{equation}

\begin{figure}[!h!b!t!p]
\begin{center}
\includegraphics[width=2.5in]{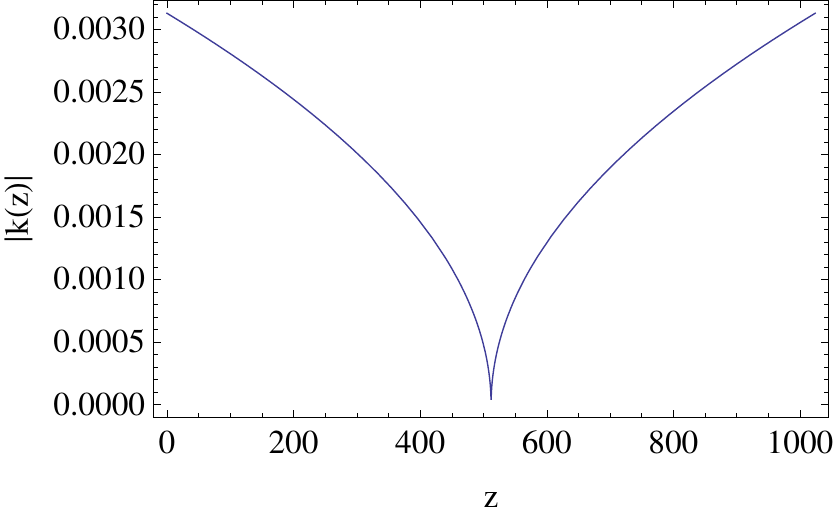}
\caption{\label{DiracParabolicPotential_WaveNumber} \footnotesize Wave number variation for parameters $L=1024$ with $\kappa= 0.01/L^2$ for a particle of mass $m=1/2$.  }
\end{center}
\end{figure}

\begin{figure*}[!h!b!t!p]
\begin{center}
\includegraphics[width=2.3in]{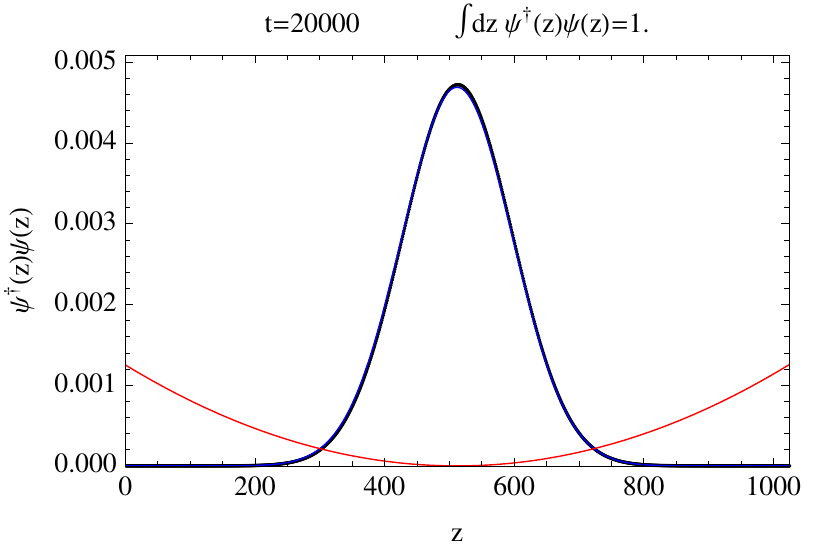}
\includegraphics[width=2.3in]{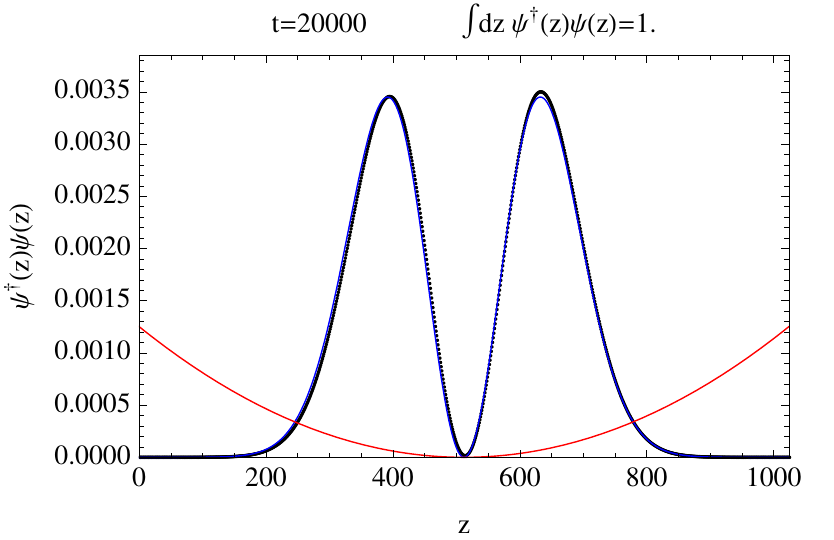}
\includegraphics[width=2.3in]{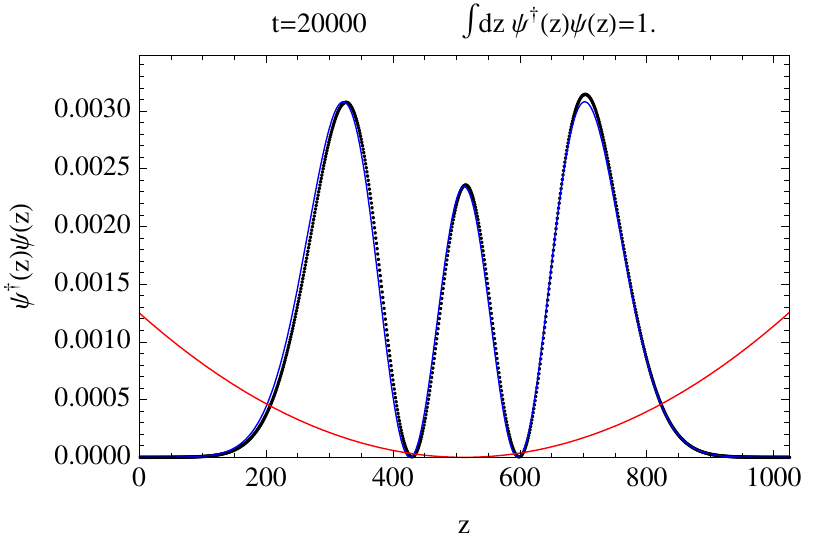}
\includegraphics[width=2.3in]{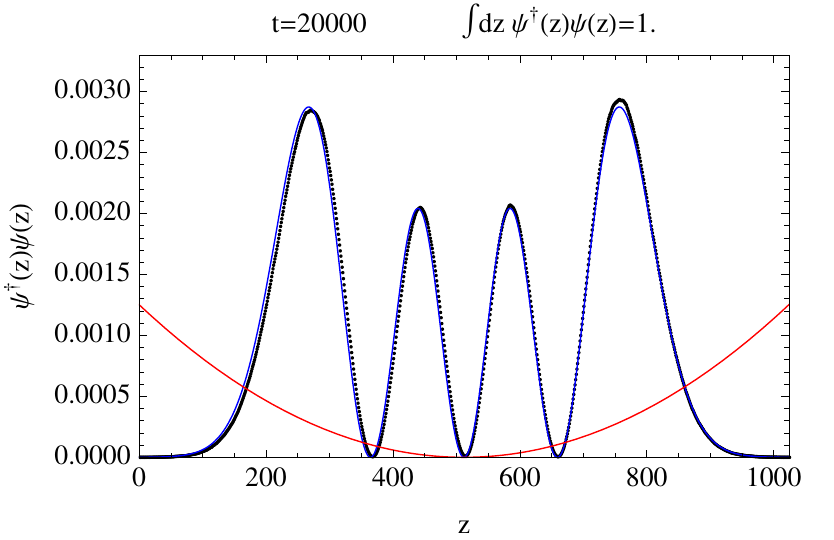}
\includegraphics[width=2.3in]{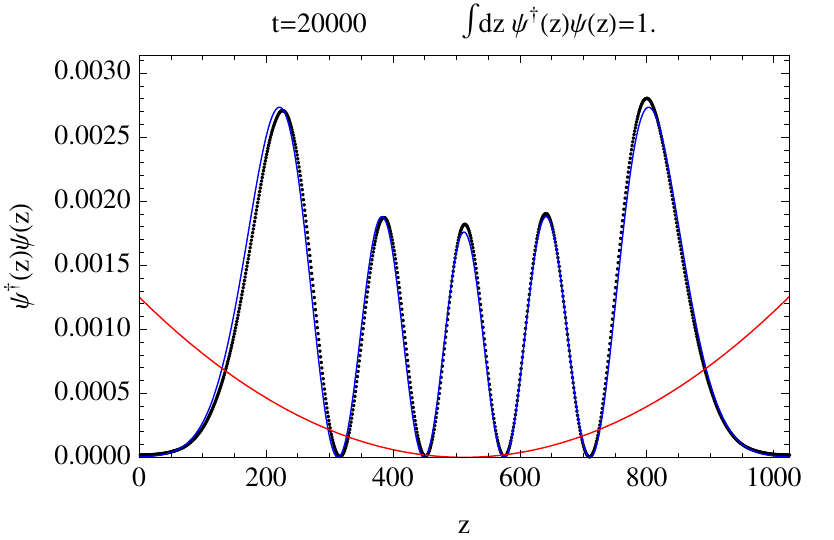}
\includegraphics[width=2.3in]{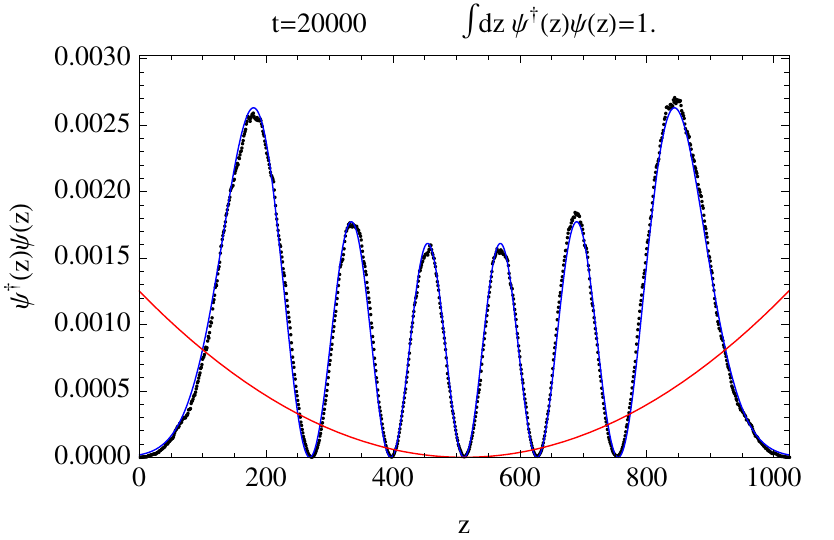}
\caption{\label{DiracParabolicPotentialMultipleEnergyLevels} \footnotesize Hermite polynomial solutions (blue curve) for energy levels $n=0,1,2,3,4,5$  and the numerically predicted the ground state solutions (black dots) of the harmonic oscillator with external potential $V(z) = \kappa z^2/2$ (red curve) for parameters $L=1024$ with $\kappa= 0.01/L^2$ for a particle of mass $m=1/2$.  During the numerical simulation, the numerical predictions oscillates about the exact ground state solution over time. The numerical solutions are shown at $t=20000$, near a recurrence point where the L2 norm error is minimal.  The numerical prediction begins to breakdown at the $n=5$ energy level because the small $L=1024$ grid cannot fully resolve five peaks.}
\end{center}
\end{figure*}

We  now demonstrate the quantum algorithm (\ref{Yepez_quantum_algorithm_Dirac_particles}) by applying it to the harmonic oscillator problem.  Since $\varsigma=\sqrt{2b}= \left({m \kappa }/{\hbar}\right)^{\frac{1}{4}}$, the quantum simulation is possible to do by employing the analytical solution
\begin{equation}
\label{harmonic_oscillator_solution}
\phi(z)  = H_n\left[\left( \frac{m \kappa }{\hbar}\middle)^{\frac{1}{4}} \middle( z - \frac{L}{2}\right) \right] e^{- \sqrt{\frac{m \kappa }{4\hbar^2}}   \left( z - \frac{L}{2}\right)^2 },
\end{equation}
so long as we run the quantum algorithm in the nonrelativistic regime with $\hbar k \ll mc^2$.  What we need do is let the Dirac particle's mass be position dependent, using it to encode the confining potential 
\begin{equation}
\label{m_dependence_harmonic_oscillator_ground_state}
m(z) = {m}+ \frac{\kappa}{2} \,z^2,
\end{equation}
 and we also need to use (\ref{harmonic_oscillator_solution}) as the first component of the initial  2-spinor field that has a vanishing second component ($\xi=0$)
\begin{equation}
\label{nonrelativistic_2_spinor}
\psi(z,0)
=
\begin{pmatrix}
      \phi (z) \\
      0
\end{pmatrix} .
\end{equation}
Adding together the coupled equations in (\ref{phi_xi_solutions}), we have
\begin{equation}
\label{sum_of_phi_and_xi}
\phi + \xi = \frac{- 2 i  E(z) \hbar c}{E(z)^2-(m(z) c^2)^2} \partial_z\left( \phi+\xi\right).
\end{equation}
Let us work out the ground state solution to the harmonic oscillator problem, which has the form
\begin{subequations}
\begin{equation}
\varphi(z) \equiv \phi(z) + \xi(z) = \varphi_\circ e^{-b z^2},
\end{equation}
and which upon inserting into (\ref{sum_of_phi_and_xi}) gives
\begin{equation}
\varphi  = 
 \frac{4 i  E(z) \hbar c\, b z}{E(z)^2-(m(z) c^2)^2} 
 \varphi
 = 
 \frac{4 i  E(z)  c\, b z}{\hbar\, k(z)^2} 
 \varphi
\end{equation}
or
\begin{equation}
 k(z)^2  = 
4 i   c^2\, b z
\sqrt{ k(z)^2 + \left(\frac{m(z) c}{\hbar}\right)^2}.
\end{equation}
Therefore, in lattice units $\hbar=1$ and $c=1$, the wave number must satisfy the fourth-order polynomial equation
\begin{equation}
k(z)^4 + 16 b^2 z^2 \left(k(z)^2 + m(z)^2\right) = 0.
\end{equation}
\end{subequations}
A physical solution to this is
\begin{equation}
k(z) = 2 b\sqrt{-2z^2 +\sqrt{4z^4 - m(z)^2 z^2/b^2}}.
\end{equation}
A plot of $|k(z)|$ is given in Fig.~\ref{DiracParabolicPotential_WaveNumber}. 
The relativistic energy is $E(z) = \sqrt{|k(z)|^2 + m(z)^2}$, so in turn the Lorentz factor that we need for the quantum simulation is
\begin{equation}
\label{Lorentz_factor_harmonic_oscillator_ground_state}
\gamma(z) = \frac{\sqrt{|k(z)|^2 + m(z)^2}}{m(z)}.
\end{equation}
The particular quantum algorithm obtained by inserting (\ref{m_dependence_harmonic_oscillator_ground_state}) and  (\ref{Lorentz_factor_harmonic_oscillator_ground_state})  into (\ref{Yepez_quantum_algorithm_Dirac_particles}) allows us to perform a numerical quantum simulation that unitarily evolves the 2-spinor initial state (\ref{nonrelativistic_2_spinor}).  The observed numerical behavior is that $|\psi(z,t)|\approxeq |\psi(z,0)|$ for all time, which is the expected behavior of any energy eigenstate.  The quantum simulation  agrees with theory, as shown in Fig.~\ref{DiracParabolicPotentialMultipleEnergyLevels}.

\section{Conclusion}

A novel quantum informational representation of Dirac particle dynamics was presented.  The treatment was restricted to 1+1 spacetime dimensions for the sake of pedagogy.  Hopefully, the treatment in low dimensions reduced the complexity so that the model is more readily understood by readers not already familiar with the quantum lattice gas method.  The representation of relativistic quantum mechanics presented here constitutes a quantum algorithm useful for quantum simulations of systems with one or more Dirac particles.  Regardless of the number of particles in any  particular quantum simulation, the quantum gate protocol remains fixed.  When implemented on a large-scale quantum computer,  the quantum algorithm is  efficient---handling many-body Fermi simulations with an equal number of quantum gate operations that scales only as the volume of the grid.  The quantum algorithm exploits the relativistic energy relation to decompose the unitary evolution generated by the Dirac Hamiltonian into two distinct unitary evolution steps, one representing pure kinetic transport and the other representing a chiral symmetry breaking particle-particle interaction.  The quantum lattice gas model is a unitary and finite version of the Feynman chessboard model of quantum mechanical path integration.  

Several quantum simulations were performed and the quantum lattice gas model was shown to be useful for  quantum simulations. We demonstrated that it is useful for  nonrelativistic quantum simulations when the rest energy of the modeled Dirac particle is much greater than its kinetic energy.  Some test simulations of a Dirac particle confined to a square well potential were carried out.  To avoid the Klein  paradox, the external scalar potential was treated as arising from the effective particle mass, which was parametrized as varying in space.  Thus, a discontinuity in the effective mass of the Dirac particle serves as a boundary of a step barrier.   The numerical results obtained from the quantum simulations were in excellent agreement with the analytical predictions.  Subsequent articles explaining the quantum lattice gas algorithm in more  situations will focus on quantum simulations: (1) in the relativistic regime, (2) in higher spatial dimensions, and (3) with nonlinear particle-particle interactions.

\section{Acknowledgements}

This work was supported by grant no. AFOSR 11RV13COR RDSM and the DoD HPCMP  Quantum Computing Program at the MHPCC and conducted under an AFRL-UH Educational Partnership Agreement (2010-AFRL/RD-EPA-03 2013 Amendment 1).

\appendix

\section{Qubit}
\label{Qubit_review}

\subsection{Qubit representations}

\begin{figure}[!h!b!p!t]
\xy
(0,0)*{\includegraphics[width=1.625in]{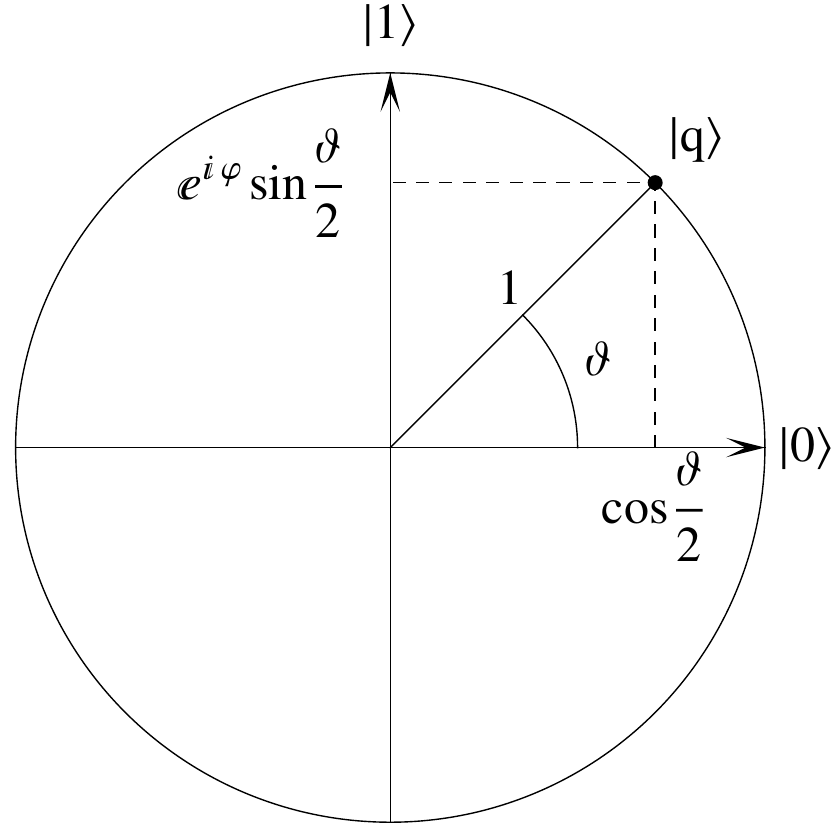}};
(45,0)*{\includegraphics[width=1.65in]{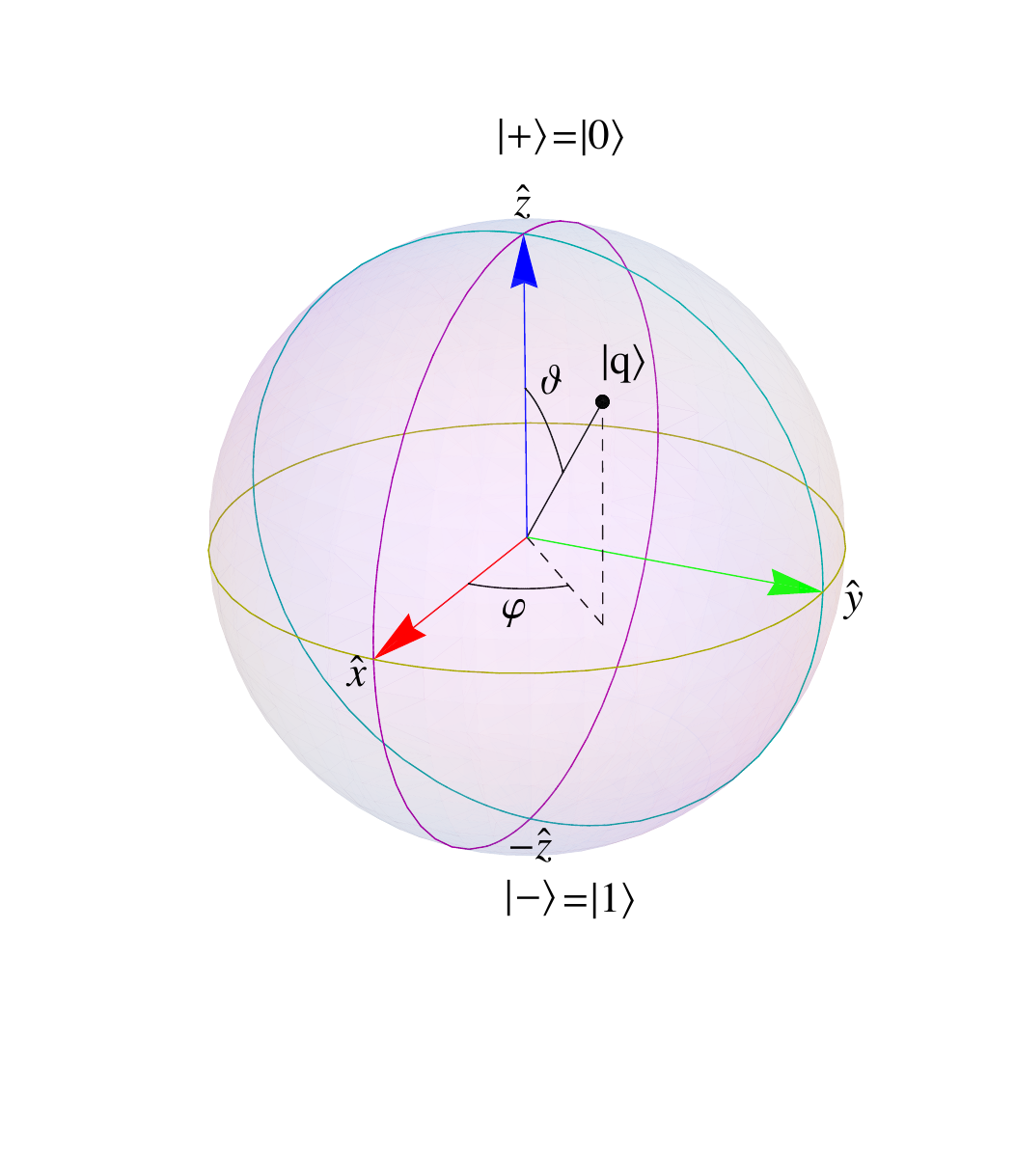}};
\endxy
\caption{\label{qubit_depiction}A qubit in Hilbert space in its SU(2) representation  (left), and the same qubit on the Bloch sphere in its O(3) representation (right).  SU(2) and O(3) are homomorphic.}
\end{figure}

A qubit is the two-level quantum state representing a unit of  information  (one bit of classical information) accessible by measurement.  The quantum logic states {}``one''
 and  {}``zero''  (called ``minus'' and ``plus" on the Bloch sphere) are denoted
\begin{subequations}
\begin{eqnarray}
|1\rangle
&=&
\begin{pmatrix}0\\
1\end{pmatrix}\quad\text{or alternate symbol  $|-\rangle$}
\\
|0\rangle
&=&
\begin{pmatrix}1\\
0\end{pmatrix}\quad\text{or alternate symbol  $|+\rangle$}.
\end{eqnarray}
\end{subequations}
The alternate symbols $|+\rangle$ and $|-\rangle$ are used to denote logical states because the names {}``up'' and {}``down,'' and the respective
symbols $\mid\uparrow\rangle$ and $\mid\downarrow\rangle$, are reserved to denote the spin states of 
spin-$\frac{1}{2}$ particles.  A qubit, as an abstraction of a two-state quantum object, represents the superposition state 
\begin{equation}
\label{qubit_superposition_state}
|q\rangle=\alpha|0\rangle+\beta|1\rangle=\alpha\begin{pmatrix}1\\
0\end{pmatrix}+\beta\begin{pmatrix}0\\
1\end{pmatrix},
\end{equation}
where $\alpha$ and $\beta$ are complex numbers.  These complex numbers are called probability amplitudes.  The basis states are orthonormal
 \begin{subequations} 
\begin{eqnarray}
\langle0|0\rangle & = & \langle1|1\rangle =1 \\
\langle0|1\rangle & = & \langle1|0\rangle=0.
\end{eqnarray}
 \end{subequations} 
It might seem that a qubit should have four free real-valued parameters (two magnitudes and two phases):
\begin{equation}
|q\rangle=\begin{pmatrix}\alpha\\
\beta\end{pmatrix} = 
\begin{pmatrix}\phi_0\; e^{i\theta_0}\\
\phi_1\; e^{i\theta_1}\end{pmatrix}.
\end{equation}
Yet, for a qubit to contain only one classical bit of information, the qubit need only be unimodular (normalized to unity)
\(
\alpha^{*}\alpha+\beta^{*}\beta=1.
\)
Hence it lives on the complex unit circle, depicted on the top of Figure~\ref{qubit_depiction}.  This normalization  
constrains the value of the magnitudes, so we can write a qubit as
\begin{equation}
\label{qubit_f_superposition_state}
|q\rangle= 
\begin{pmatrix}\sqrt{1-f}\\
\sqrt{f}\; e^{i\varphi}\end{pmatrix},
\end{equation}
where $0\le f\le 1$ and where an irrelevant overall phase is factored out.  The length (or norm) of the qubit is thus an invariant quantity
\begin{equation}
\langle q| q\rangle = |\alpha|^2 + |\beta|^2 = |\sqrt{1-f}|^2 + |\sqrt{f}|^2=1.
\end{equation}
 The quantum property of measurement  follows from identifying the moduli squared of the amplitude as an occupation probability $f$ and $1-f$ for the qubit to occupy its logical states $|1\rangle$ and $|0\rangle$, respectively, as follows:
\begin{eqnarray}
 f & = & |\beta|^2 
 \qquad\qquad
 1-f  =  |\alpha|^2  .
\end{eqnarray}
There are only two relevant free parameters to specify the state of a qubit, but upon measurement, the qubit originally in the superposition state (\ref{qubit_f_superposition_state}) is found to occupy only one of its logical states 
\begin{equation}
\label{ }
|q\rangle \xrightarrow{\text{measure}}
\begin{cases}
      & |1\rangle, \text{ with probability } f, \\
      & |0\rangle,  \text{ with probability }1-f.
\end{cases}
\end{equation}
Thus, upon a single measurement, $|q\rangle$ is found to be in either the state $|0\rangle$ or $|1\rangle$, an outcome that is said to be specified by a single classical bit $\in\{0,1\}$.   Thus in actual experiments, the occupation probability $f$ equals the frequency of occurrence of the result $1$ obtained from many repeated measurements.

The state $|q(t)\rangle$ of a time-dependent qubit, as a two-energy level quantum mechanical entity, is governed by the Schroedinger wave equation
\begin{equation}
i\hbar\frac{\partial}{\partial t}|q(t)\rangle =\frac{\hbar \omega}{2} \sigma_z |q(t)\rangle.
\end{equation}
The energy eigenvalues are $\pm\hbar \omega/2$ and energy eigenstates are
\begin{equation}
|0\rangle\equiv \left(
\begin{matrix}
1\cr 0
\end{matrix}
\right) \hspace{0.25in}
|1\rangle\equiv\left(
\begin{matrix}
0\cr 1
\end{matrix}
\right),
\end{equation}
where $|0\rangle$ is the ground state and $|1\rangle$ is the excited state of the qubit.
In terms of the angular frequency $\omega$ (e.g. Rabi frequency), the time-dependent qubit state is
\begin{equation}
|q(t)\rangle = {\cal A}_0 e^{-i\frac{\omega}{2} t}|0\rangle + {\cal A}_1 e^{i\frac{\omega}{2} t}|1\rangle,
\end{equation}
where the complex probability amplitudes satisfy $|{\cal A}_0|^2 +|{\cal A}_1|^2=1$ since the qubit resides on the complex circle in Hilbert space (or the Bloch sphere in spin space).   

The qubit  eigenstates may be expressed on the Bloch sphere with $\hat u = (\sin\theta\cos\varphi, \sin\theta\sin\varphi, \cos\theta)$  as
\begin{subequations}
\begin{eqnarray}
\nonumber
| + \rangle_u &=&
\begin{pmatrix}
 \cos \frac{\theta}{2} e^{-i \frac{\varphi}{2}}    \\
  \sin \frac{\theta}{2} e^{i \frac{\varphi}{2}} 
\end{pmatrix}
=
\cos \frac{\theta}{2} e^{-i \frac{\varphi}{2}} |0\rangle
+
  \sin \frac{\theta}{2} e^{i \frac{\varphi}{2}} |1\rangle,
\\
\\
\nonumber
| - \rangle_u &=&
\begin{pmatrix}
 -\sin \frac{\theta}{2} e^{-i \frac{\varphi}{2}}    \\
  \cos \frac{\theta}{2} e^{i \frac{\varphi}{2}} 
\end{pmatrix}
=
 -\sin \frac{\theta}{2} e^{-i \frac{\varphi}{2}}    |0\rangle
 +
  \cos \frac{\theta}{2} e^{i \frac{\varphi}{2}} |1\rangle . 
  \\
\end{eqnarray}
\end{subequations}
Writing the 2-spinor basis states in terms of qubit states, we have 
\begin{subequations}
\begin{eqnarray}
\nonumber
\xi(\uparrow) 
& \equiv &
e^{i \frac{\varphi}{2}}
| + \rangle 
= 
\begin{pmatrix}
 \cos \frac{\theta}{2}  \\
 e^{i \varphi} \sin \frac{\theta}{2}  
\end{pmatrix}
=\cos \frac{\theta}{2}|0\rangle
+
  \sin \frac{\theta}{2} e^{i \varphi} |1\rangle,
\\
\\
\nonumber
\xi(\downarrow) 
& \equiv &
e^{-i \frac{\varphi}{2}}
| - \rangle
=
\begin{pmatrix}
 -  e^{-i \varphi} \sin \frac{\theta}{2}    \\
  \cos \frac{\theta}{2}  
\end{pmatrix}
\\
&=&
 -\sin \frac{\theta}{2} e^{-i \varphi}    |0\rangle
 +
  \cos \frac{\theta}{2}  |1\rangle .
\end{eqnarray}
\end{subequations}

The space of all possible orientations of $|q\rangle$ on the complex unit circle is called the Hilbert space. 
 In the logical basis, the two degrees of freedom of the qubit is often expressed as two angles $\theta$ and $\varphi$, where $f=\sin^2\left(\frac{\theta}{2}\right)$. So without any loss of generality the Hilbert space representation of  a qubit (\ref{qubit_superposition_state}) can  be written as
\begin{equation}
\label{qubit_Hilbert_space_Euler_angles}
|q\rangle=\cos\!\left(\frac{\theta}{2}\right)|0\rangle+ \sin\!\left(\frac{\theta}{2}\right) e^{i\varphi}|1\rangle.
\end{equation}
These angles have a well known geometrical interpretation as Euler angles.   

To understand the geometrical interpretation of a qubit, consider a three-dimensional space with ``unit vectors'' $\sigma_x$, $\sigma_y$, and $\sigma_z$ chosen as an orthonormal basis.   In quantum information theory, one represents each basis element by a $2\times 2$ matrix, a traceless hermitian generators of two-dimensional special unitary group, SU(2).  To do so, one defines the symmetric product (dot product) as
 \begin{subequations} 
 \label{dot_and_cross_products}
\begin{equation}
\sigma_i\cdot \sigma_j \equiv 
\left.
\frac{1}{2}
\middle(
\sigma_{i}\cdot\sigma_{j}
+
\sigma_{j} \cdot\sigma_{i}
\right).
\end{equation}
Furthermore, one defines the anti-symmetric product (cross product) as
\begin{equation}
\sigma_i\times \sigma_j \equiv 
\left.
-\frac{i}{2}
\middle(
\sigma_{i} \cdot\sigma_{j}
-
\sigma_{j} \cdot\sigma_{i}
\right).
\end{equation}
 \end{subequations} 
Note that the centered dot symbol on the R.H.S. of (\ref{dot_and_cross_products}) denotes matrix multiplication.
Thus, a basis that is orthonormal satisfies the following conditions
 \begin{subequations} 
 \label{orthonormality_conditions}
\begin{eqnarray}
\sigma_i\cdot \sigma_j 
&=&
\begin{cases}
      & 1, \text{ for } i=j \text{ (normal)}, \\
      & 0, \text{ otherwise} \text{ (orthogonal)},
\end{cases}
\\
\label{SU2_structure_equation}
\sigma_i\times \sigma_j 
&=&
\begin{cases}
      & 0, \text{ for } i=j, \\
      & \sigma_k, \text{ for cyclic indices}.
\end{cases}
\end{eqnarray}
 \end{subequations} 
A fundamental matrix representation that satisfies (\ref{orthonormality_conditions}) is the well-known Pauli basis
\begin{equation}
\label{Pauli_matrices}
\sigma_1 = 
\begin{pmatrix}
   0   &   1 \\
    1  &  0
\end{pmatrix},
\qquad
\sigma_2 = 
\begin{pmatrix}
   0   &   -i \\
    i  &  0
\end{pmatrix},
\qquad
\sigma_3 = 
\begin{pmatrix}
   1  &   0 \\
    0  &  -1
\end{pmatrix}.
\end{equation}
The Pauli matrices (\ref{Pauli_matrices}) satisfy the orthonormality conditions (\ref{orthonormality_conditions}) which is just the  structure equation for the SU(2) group
\(
\left[S_i,S_j\right] = i \,\epsilon_{ijk} \,S_k, 
\)
where $S_i\equiv\frac{\sigma_i}{2}$ and the structure constant $\epsilon_{ijk}$ is the anti-symmetric Levi-Civita symbol. Now we can express the qubit (\ref{qubit_Hilbert_space_Euler_angles}) in vector form (i.e. with three real components) as follows:
\begin{equation}
\label{qubit_Bloch_sphere_Euler_angles}
 \vec q=(\sin\theta\cos\varphi, \,\sin\theta\sin\varphi,\,\cos \theta).
\end{equation} %
(\ref{qubit_Bloch_sphere_Euler_angles}) is a representation of a qubit on the Bloch sphere where $\theta$ is the elevation angle and $\varphi$ is the azimuthal angle.  In this representation, depicted on the bottom of Fig.~\ref{qubit_depiction}, the qubit is considered as a vector element of the three-dimensional orthogonal group, O(3).
Defining the Pauli spin vector (which has matrix components)
\(
\vec \sigma \equiv (\sigma_1, \sigma_2, \sigma_3),
\)
a qubit can also be expressed in matrix form
 \begin{subequations} 
\begin{eqnarray}
\label{M_q_definition}
M_q 
& \equiv& \vec q \cdot \vec \sigma
\\
&=&
\sin\theta\cos\varphi\,\sigma_1+\sin\theta\sin\varphi\,\sigma_2+\cos \theta\,\sigma_3
\qquad
\\
&\stackrel{(\ref{Pauli_matrices})}{=} &
\begin{pmatrix}
   \cos\theta   &  e^{-i \varphi}\sin\theta   \\
     e^{i \varphi}\sin\theta   &  -\cos\theta
\end{pmatrix}.
\end{eqnarray}
 \end{subequations} 
In this representation, the qubit is expressed as a matrix element of the SU(2) group.  In quantum information, usually $2\times 2$ unitary matrices are considered single-qubit quantum gates, but such matrices can themselves represent qubits too.  Table~\ref{qubit_representations} gives a summary of the three qubit representations
\begin{table}[hbtp!]
  \centering 
  \begin{tabular}{c|cl}
  Representations &\multicolumn{2}{c}{\sc Qubit} \\
\hline
Hilbert space & $|q\rangle$ & $=\cos\left(\frac{\theta}{2}\right)|0\rangle+ \sin\left(\frac{\theta}{2}\right) e^{i\varphi}|1\rangle$       \\
O(3) group & $ \vec q$ &$=(\sin\theta\cos\varphi, \,\sin\theta\sin\varphi,\,\cos \theta)$    \\
 SU(2) group & $M_q$ &
 $
 =
 \begin{pmatrix}
   \cos\theta   &  e^{-i \varphi}\sin\theta   \\
     e^{i \varphi}\sin\theta   &  -\cos\theta
\end{pmatrix}$     \\
\hline
\end{tabular}
  \caption{\label{qubit_representations} Qubit representations.}
\end{table}

\subsection{Rotation by similarity transformation}

Now that we see a qubit as simply a unit vector on the complex circle (Hilbert space representation) or a unit vector on the Bloch sphere (O(3) representation), we can consider rotations of the qubit's state that keep its length (or norm) invariant.
Remarkably, such a rotation of a qubit is conveniently accomplished by employing its SU(2) representation as a $2\times 2$ unitary matrix. Then, the qubit rotation is induced by a similarity transformation, which is to say a double-sided transformation acting from the left and the right side.  The unitary matrix (acting from the left) along with its matrix inverse (acting from the right) that is customarily employed for such rotations, about the $i$th principle axis say, is
\begin{equation}
\label{unitary_rotation_ith_axis}
U_i(\theta)\equiv e^{-i \frac{\theta}{2} \sigma_i}
=\sigma_0 \cos \left(\frac{\theta}{2}\right) - i \sigma_i \sin \left(\frac{\theta}{2}\right) ,
\end{equation}
where the identity matrix is $\sigma_0=\begin{pmatrix}
   1   & 0   \\
     0 & 1 
\end{pmatrix}$.
Explicitly, the unitary matrices for the principles directions are
\begin{subequations}
\begin{eqnarray}
U_1(\theta)
&=&
\begin{pmatrix}
   \cos \left(\frac{\theta}{2}\right)    & -i \sin \left(\frac{\theta}{2}\right)    \\
   -i \sin \left(\frac{\theta}{2}\right)    & \cos \left(\frac{\theta}{2}\right)  
\end{pmatrix}
\\
U_2(\theta)
&=&
\begin{pmatrix}
   \cos \left(\frac{\theta}{2}\right)    & -\sin \left(\frac{\theta}{2}\right)    \\
   \sin \left(\frac{\theta}{2}\right)    & \cos \left(\frac{\theta}{2}\right)  
\end{pmatrix}
\\
U_3(\theta)
&=&
\begin{pmatrix}
   e^{-i \frac{\theta}{2}}    & 0    \\
   0    &   e^{i \frac{\theta}{2}}
\end{pmatrix}.
\end{eqnarray}
\end{subequations}
A general rotation of a qubit about  axis $\hat{n} = (n_x, n_y, n_z)$ is built using the following unitary matrix (along with its inverse)
\begin{equation}
U_{\hat{n}}(\theta) = e^{-i \frac{\theta}{2} \hat{n} \cdot \bm{\sigma}}
=\cos \left(\frac{\theta}{2}\right) - i (\hat{n}\cdot \bm{\sigma}) \sin \left(\frac{\theta}{2}\right) ,
\end{equation}
where the identity on the R.H.S. follows since $(\hat{n}\cdot \bm{\sigma})^2=\mathbf{1}$.

Now a qubit rotation by angle $\theta$ about the arbitrary axis $\hat n$ is expressed as the similarity transformation mentioned above
\begin{equation}
\label{rotated_qubit}
M_{q'}  = U_{\hat n}(\theta) \cdot M_q\cdot U_{\hat n}^\dagger(\theta).
\end{equation}
Here again the centered dot symbol represents matrix multiplication.  The $\dagger$ symbol denotes the matrix adjoint, {\it i.e.} complex conjugate of the components of the matrix combined with matrix transposition.   Since 
\begin{equation}
U_{\hat n}^\dagger(\theta) = U_{\hat n}^{-1}(\theta)=U_{\hat n}(-\theta), 
\end{equation}
we simply compute the rotated qubit (\ref{rotated_qubit}) as follows
\begin{eqnarray}
M_{q'} &=&
U_{\hat n}(\theta) \cdot M_q\cdot U_{\hat n}(-\theta).
\end{eqnarray}
Of course using $M_q = \vec q \cdot \vec \sigma$ we can write this similarity transformation directly in terms of the 3-vector $\vec q$ and the resulting 3-vector $\vec {q'}$ 
\begin{eqnarray}
\label{rotated_q_similarity_transformation}
\vec {q'} \cdot \vec \sigma&=&
U_{\hat n}(\theta) \cdot (\vec {q} \cdot \vec \sigma)\cdot U_{\hat n}(-\theta),
\end{eqnarray}
which determines $\vec {q'}$ in terms of the original vector $\vec q$, the axis of rotation $\hat n$, and the angular rotation amount $\theta$.\footnote{This reduces to the useful  formula
\begin{equation*}
\label{Rodriques_rotation_formula}
\vec {q'}
=
\cos\theta \;\vec {q} 
+
\left(1-\cos\theta\right)
\hat n (\hat n \cdot \vec q)
+
\sin\theta\;\hat n \times \vec q,
\end{equation*}
known as Rodrigues' rotation formula.  As a check of the rotation formula (\ref{Rodriques_rotation_formula}), consider the two cases when $\hat n \perp \vec q$ and $\hat n \parallel \vec q$ :
\begin{equation*}
\vec{q'}
=
\begin{cases}
  \cos\theta\, \vec q + \sin\theta \, \hat n\times \vec q    & \text{ for } \hat n \perp \vec q, \\
  \cos\theta\, \vec q  +    \left(1-\cos\theta\right)
\hat n (\hat n \cdot \vec q)=\vec q   & \text{ for } \hat n \parallel \vec q,
\end{cases}
\end{equation*}
which is correct by inspection.}

\section{Quantum gates as matrices and  operators}

\subsection{\label{Singleton_ladder_operators}Singleton ladder operators}

There are two basic operators from which all other quantum operator are constructed.  These operators are 
${\scriptsize
\begin{pmatrix}0 & 0\\
1 & 0\end{pmatrix}
}$ and
${\scriptsize
\begin{pmatrix}0 & 1\\
0 & 0\end{pmatrix}
}$,  which by matrix multiplication generate
${\scriptsize
\begin{pmatrix}0 & 0\\
0 & 1\end{pmatrix}
}$ and
${\scriptsize
\begin{pmatrix}1 & 0\\
0 & 0\end{pmatrix}
}$.
A one in each slot---what could be simpler?  Each of these four operators carries physical significance.  They are named for their function.
\\\\
 \begin{subequations} 
\noindent Raising ladder operator: 
\begin{equation}
a^{\dagger}=
{\scriptsize
\begin{pmatrix}0 & 0\\
1 & 0\end{pmatrix}
}
=\frac{1}{2}
\left(
\sigma_{1}- i \sigma_{2}
\right) 
\end{equation}

\noindent Lowering ladder operator: 
\begin{equation}
a=
{\scriptsize
\begin{pmatrix}0 & 1\\
0 & 0\end{pmatrix}
}
 =
\frac{1}{2}
\left(
\sigma_{1}+ i \sigma_{2}
\right).
\end{equation}

\noindent 1 number (particle) operator: 
\begin{equation}
n=
{\scriptsize
\begin{pmatrix}0 & 0\\
0 & 1  
\end{pmatrix}
}
= a^{\dagger} a
=
\frac{1}{2}
\left(
1-  \sigma_{3}
\right).
\end{equation}

\noindent 0 number (hole) operator: 
\begin{equation}
h=\bar{n}=
{\scriptsize
\begin{pmatrix}1 & 0\\
0 & 0\end{pmatrix}
}
= a\, a^{\dagger}
=
\frac{1}{2}
\left(
1+  \sigma_{3}
\right).
\end{equation}
 \end{subequations} 
Operating on logical  states (qubit basis states), the singleton ladder operators give 
 \begin{subequations} 
\begin{eqnarray}
a^{\dagger}|0\rangle & = & |1\rangle\qquad\quad\text{Raise 0 to 1}\\
a^{\dagger}|1\rangle & = & 
{\scriptsize
\begin{pmatrix}0\\
0\end{pmatrix}
}
\qquad\text{Exclusion of 1's}\\
a|0\rangle & = & 
{\scriptsize
\begin{pmatrix}0\\
0\end{pmatrix}
}
\qquad\text{Exclusion of 0's}\\
a|1\rangle & = & |0\rangle\qquad\quad\text{Lower 1 to 0},
\end{eqnarray}
 \end{subequations} 
where the state 
${\scriptsize
\begin{pmatrix}0\\
0\end{pmatrix}
}$ is called {\bf oblivion}.
Furthermore, operating on the logical states, the singleton number operators give
 \begin{subequations} 
\begin{eqnarray}
n|0\rangle & = & 
{\scriptsize
\begin{pmatrix}0\\
0\end{pmatrix}
}
\qquad\text{Exclusion of 0's}\\
n|1\rangle & = & |1\rangle\qquad\quad\text{Counts 1's} \\
h|0\rangle & = & |0\rangle\qquad\quad\text{Counts 0's}\\
h|1\rangle & = & 
{\scriptsize
\begin{pmatrix}0\\
0\end{pmatrix}
}
\qquad\text{Exclusion of 1's}.
\end{eqnarray}
 \end{subequations} 
From the simple identity 
\begin{equation}
\label{basic_number_identity}
n+h=\bm{1}
\end{equation}
follows the anticommutation relation algebraically expressing the local exclusion principle
\begin{equation}
a^{\dagger}a+a\, a^{\dagger}=\bm{1}.
\end{equation}
In this lecture we use $\bm{1}\equiv \bm{1}_2$.
Finally, the observable number {}``1\char`\"{} (a bit of information) is implicitly defined as the
eigenvalue of $n$: \begin{equation}
n|1\rangle=
{\scriptsize
\begin{pmatrix}0 & 0\\
0 & 1\end{pmatrix}
\begin{pmatrix}0\\
1\end{pmatrix}
}
=1
{\scriptsize
\begin{pmatrix}0\\
1\end{pmatrix}
}
=1|1\rangle.
\end{equation}

\subsection{Multiple objects}

\subsubsection{Qubits}

\noindent Tensor product state--the state of independent qubits: 
\begin{eqnarray*}
\bigotimes_{i=1}^{Q}|q_{i}\rangle & = & |q_{1}\rangle\otimes|q_{2}\rangle\otimes\cdots\otimes|q_{Q}\rangle\\
 & = & |q_{1}\rangle|q_{2}\rangle\cdots|q_{Q}\rangle\qquad\text{used for a few qubits, $Q\lessapprox3$}\\
 & = & |q_{1}q_{2}\cdots q_{Q}\rangle\qquad\text{numbered state, $|q_{i}\rangle=|0\rangle$ or $|1\rangle$,}
 \end{eqnarray*}
for all $i=1,\dots,Q$.

\subsubsection{Qubit number operators}
\label{qsm-unfolding-the-multiqubit-number-operator}

Using the singleton number operator ($Q=1$)
\begin{equation}
  n = \left(
\begin{matrix}
0 & 0 \cr 0 & 1
\end{matrix}
\right),
\end{equation}
we can generate the multiple qubit number operators.  So, the two qubit number operators ($Q=2$) are expressed as the following tensor products of $  n$ with identity
\begin{eqnarray}
  n^{(2)}_1 
  &\equiv&
  n \otimes {\bf 1}
=
\left(
\begin{matrix}
0 & 0 \cr 0 & 1
\end{matrix}
\right)\otimes \left(
\begin{matrix}
1 & 0 \cr 0 & 1
\end{matrix}
\right)
=
\left(
\begin{matrix}
0 & 0 & 0 & 0 \cr
0 & 0 & 0 & 0 \cr
0 & 0 & 1 & 0 \cr
0 & 0 & 0 & 1
\end{matrix}
\right)
\qquad
\end{eqnarray}
and
\begin{eqnarray}
  n^{(2)}_2 
 &\equiv&
{\bf 1} \otimes   n
= 
\left(
\begin{matrix}
1 & 0 \cr 0 & 1
\end{matrix}
\right)\otimes \left(
\begin{matrix}
0 & 0 \cr 0 & 1
\end{matrix}
\right)
=\left(
\begin{matrix}
0 & 0 & 0 & 0 \cr
0 & 1 & 0 & 0 \cr
0 & 0 & 0 & 0 \cr
0 & 0 & 0 & 1
\end{matrix}
\right) ,
\qquad
\end{eqnarray}
where $\bf 1$ denotes the $2\times 2$ identity matrix.
Similarly, the three qubit number operators ($Q=3$) are expressed as the following tensor products 
\begin{eqnarray}
\nonumber
  n^{(3)}_1 \!\!& = &    n \otimes {\bf 1}\otimes {\bf 1},
  \quad
  n^{(3)}_2  \!\!=   {\bf 1} \otimes   n \otimes {\bf 1},
  \quad
  n^{(3)}_3 \!\!=   {\bf 1}  \otimes {\bf 1}\otimes   n. 
  \\
\end{eqnarray}
For any system with  $Q$ qubits, the $\alpha^{\hbox{\tiny th}}$ number operator, $  n_\alpha$ can be expressed in a way that depends on a single $  n$ placed at the  $\alpha^{\hbox{\tiny th}}$ position within the following tensor product:
\begin{eqnarray}
  n_\alpha & = &\overbrace{{\bf 1} \otimes {\bf 1} \otimes \cdots  \underbrace{\otimes\;   n \;\otimes}_{\alpha^{\hbox{\tiny th}}-\hbox{term}}  \cdots \otimes {\bf 1}}^{Q-\hbox{terms}}
=
\bm{1}^{\otimes\alpha} \otimes n.
\qquad
\end{eqnarray}
This identity represents the unfolding of the $Q$-qubit system number operator as a tensor product.

\subsection{Fermionic ladder operators}

All quantum gate operations can be represented in terms of the fermionic {\it qubit creation} and {\it qubit annihilation} operators in the number representation, denoted $  a^\dagger_\alpha$ and $  a_\alpha$ respectively.  This approach serves as  a general computational formulation applicable to any quantum algorithm.  Acting on a system of $Q$ qubits, $  a^\dagger_\alpha$ and $  a_\alpha$ create and destroy a fermionic number variable at the $\alpha\hbox{th}$ qubit
\begin{eqnarray}
\nonumber
  a^\dagger_\alpha |n_1  \dots n_\alpha \dots n_Q\rangle & = & 
\left\{
\begin{matrix}
0&, &  n_\alpha=1\cr
\epsilon\; |n_1 \dots 1 \dots n_Q\rangle&, &   n_\alpha=0\cr
\end{matrix}
\right. 
\\
\\
\nonumber
  a_\alpha |n_1  \dots n_\alpha \dots n_Q\rangle & = & 
\left\{
\begin{matrix}
\epsilon\; |n_1  \dots 0 \dots n_Q\rangle&, & n_\alpha=1\cr
0&, & n_\alpha=0\cr
\end{matrix}
\right. ,\\
\end{eqnarray}
where the phase factor is 
\begin{equation}
\epsilon = (-1)^{\sum_{i=1}^{\alpha-1}n_i}.
\end{equation}
See page 17 of Ref.~\cite{fetter-71} for this way of determining $\epsilon$ used by condensed matter theorists.
The fermionic ladder operators satisfy the anticommutation relations
\begin{eqnarray}
\label{fermionic_anti_commutation_relations}
\{   a_\alpha,   a^\dagger_\beta\} & =&  \delta_{\alpha\beta}
\qquad
\{   a_\alpha,   a_\beta\}  =   0
\qquad
\{   a^\dagger_\alpha,   a^\dagger_\beta\}  =  0.
\qquad
\end{eqnarray}
The number operator $  n_\alpha \equiv   a^\dagger_\alpha   a_\alpha$ has eigenvalues of 1 or 0  in the number representation when acting on a pure state, corresponding to the $\alpha\hbox{th}$ qubit being in state $|1\rangle$ or $|0\rangle$ respectively.

\subsubsection{Jordan-Wigner transformation}

With the logical one state of a qubit 
\(
{\scriptsize
|1\rangle = \begin{pmatrix}
     0    \\
      1  
\end{pmatrix}
},
\)
notice that 
\(
\sigma_z |1\rangle = -|1\rangle,
\)
so one can count the number of preceding bits that contribute to the overall phase shift due to fermionic bit exchange involving the $i$th qubit with tensor product operator, 
 \(
 \sigma_z^{\otimes i-1} |\psi \rangle = (-1)^{N_i}| \psi\rangle.
\)
The phase factor is determined by the number of bit crossings 
\(
N_i = \sum_{k=1}^{i-1}n_k
\)
in the state $|\psi\rangle$
and where the Boolean number variables are $n_k\in [ 0,1]$.
Hence, an annihilation operator is decomposed into a tensor product known as the Jordan-Wigner transformation \cite{JordanWigner1928}
\begin{equation}
\label{new_annihilation_operator_gamma}
a_i =\sigma_z^{\otimes i-1} \otimes\, a\otimes \bm{1}^{\otimes Q-i}
\end{equation}
for integer $i \in [1,Q]$.

That is,  begin with the single annihilation operator $a=
{\footnotesize
\begin{pmatrix}
   0   &  0  \\
   1   &  0
\end{pmatrix}
}
$. Then, the $i$th fermionic annihilation operator is a system of $Q$ qubits has a matrix representation that is expressible as the tensor product of $i-1$ number of Pauli $\sigma_3$ matrices, one single $a$, followed by $Q-i$ number of ones as follows:
\begin{equation}
\label{new_annihilation_operator_i}
a_i = \left(\bigotimes_{k=1}^{i-1} \sigma_3\middle) \otimes\, a \,\otimes \middle(\bigotimes_{k'=i+1}^{Q} \bm{1}\right).
\end{equation}
Since (\ref{new_annihilation_operator_i}) is the tensor product of $Q$ elements, each one a $2\times 2$ matrix, the resulting representation of $a_i$ is a matrix of size $2^Q \times 2^Q$, as expected.  Since all the components of $a_i$ are real ({\it i.e.} 0, 1, or $-1$), the $i$th creation operator is simple enough to compute by just transposing (\ref{new_annihilation_operator_i}), $a^\dagger_i = a^\text{T}$.   That (\ref{new_annihilation_operator_i}) satisfies the usual anticommutation relations is straightforward to prove.

First, using (\ref{new_annihilation_operator_i}) and since $\sigma_3^2 = \bm{1}$ and $\{a, a^\dagger\}  = \bm{1}$, we know that
\begin{subequations}
\label{anti_commutator_prove_1}
\begin{eqnarray}
\{a_i, a^\dagger_i\} & = & \left(\bigotimes_{k=1}^{i-1} \bm{1}\middle) \otimes\, \{a, a^\dagger\}  \,\otimes \middle(\bigotimes_{k'=i+1}^{Q} \bm{1}\right) \\
& = & \bigotimes_{k=1}^{Q} \bm{1} \\
& = & \mathbf{1}_{2^Q}. 
\end{eqnarray}
\end{subequations}
Similarly, $\{a_i, a_i\}=0$ and $\{a^\dagger_i, a^\dagger_i\}=0$ follow from the singleton anticommutators $\{a,a\}=0$ and $\{a^\dagger,a^\dagger\}=0$, respectively.
Second, and without loss of generality, for the case of $i<j$, we have
%
\begin{widetext}
\begin{subequations}
\label{anti_commutator_prove_2}
\begin{eqnarray}
\nonumber
\{a_i, a^\dagger_j\} 
& = & 
\left(\bigotimes_{k=1}^{i-1} \sigma_3^2\middle) \otimes\, a  \sigma_3  
\,\otimes
\middle(\bigotimes_{k'=i+1}^{j-1} \sigma_3\middle)
\otimes\,
 a^\dagger  \,\otimes \middle(\bigotimes_{k''=j+1}^{Q} \bm{1}\right)
 +
\left(\bigotimes_{k=1}^{i-1} \sigma_3^2\middle) \otimes\,  \sigma_3  a
\,\otimes
\middle(\bigotimes_{k'=i+1}^{j-1} \sigma_3\middle)
\otimes\,
 a^\dagger  \,\otimes \middle(\bigotimes_{k''=j+1}^{Q} \bm{1}\right)
 \\
 \\
 \nonumber
& = & 
\left(\bigotimes_{k=1}^{i-1} \bm{1}\middle) \otimes\, a  \sigma_3  
\,\otimes
\middle(\bigotimes_{k'=i+1}^{j-1} \sigma_3\middle)
\otimes\,
 a^\dagger  \,\otimes \middle(\bigotimes_{k''=j+1}^{Q} \bm{1}\right)
 +
\left(\bigotimes_{k=1}^{i-1} \bm{1}\middle) \otimes\,  \sigma_3  a
\,\otimes
\middle(\bigotimes_{k'=i+1}^{j-1} \sigma_3\middle)
\otimes\,
 a^\dagger  \,\otimes \middle(\bigotimes_{k''=j+1}^{Q} \bm{1}\right)
 \\
 \\
& = &\left(\bigotimes_{k=1}^{i-1} \bm{1}\middle) \otimes\, \{a, \sigma_3\} 
\,\otimes
\middle(\bigotimes_{k'=i+1}^{j-1} \sigma_3\middle)
\otimes\,
 a^\dagger  \,\otimes \middle(\bigotimes_{k''=j+1}^{Q} \bm{1}\right)
  \\
& = & 0 ,
\end{eqnarray}
\end{subequations}
\end{widetext}
%
since $\{a, \sigma_3\}= 0$. Similarly, we know $\{a_i, a_j\}=0$ and $\{a^\dagger_i, a^\dagger_j\}=0$.  Thus, we arrive at the end of the proof by combining what we have learned from (\ref{anti_commutator_prove_1}) and (\ref{anti_commutator_prove_2})
\begin{equation*}
\{a_i, a^\dagger_j\} =\delta_{ij} \qquad  \{a_i, a_j\}=0 \qquad \{a^\dagger_i, a^\dagger_j\}=0,
\end{equation*}
for any $i$ and $j$.

\subsubsection{Matrix representation}

In the basis where qubits $|q_1\rangle$ and $|q_2\rangle$ are ordered left to right $|q_1 q_2\rangle$,  the creation operators are
\begin{equation}
\label{qsm-creation-operators-representation}
\begin{split}
  a^\dagger_1  
  &= a^\dagger \otimes \bm{1}
  \\
&=
\left(
\begin{matrix}
0 & 0 \cr 1 & 0
\end{matrix}
\right)\otimes \left(
\begin{matrix}
1 & 0 \cr 0 & 1
\end{matrix}
\right)
  \\ 
  &=
  {\scriptsize
   \left(
\begin{matrix}
0 & 0 & 0 & 0 \cr 
0 & 0 & 0 & 0 \cr 
1 & 0 & 0 & 0 \cr 
0 & 1 & 0 & 0 
\end{matrix}
 \right)
 },
 \end{split}
\hspace{0.5in}
\begin{split}
  a^\dagger_2 
  &= \sigma_3\otimes a^\dagger
  \\
&=
\left(
\begin{matrix}
1 & 0 \cr 0 & -1
\end{matrix}
\right)\otimes \left(
\begin{matrix}
0 & 0 \cr 1 & 0
\end{matrix}
\right)
  \\
  &= 
    {\scriptsize
\left(
\begin{matrix}
0 & 0 & 0 & 0 \cr 
1 & 0 & 0 & 0 \cr 
0 & 0 & 0 & 0 \cr 
0 & 0 & -1 & 0 
\end{matrix}
\right)
}.
\end{split}
\end{equation}

Since $  a^\dagger_1$ and $  a^\dagger_2$ have real components, the annihilation operators are the transposes of the matrices given in (\ref{qsm-creation-operators-representation}), $  a_1 = (  a_1^\dagger)^T$ and $  a_1 = (  a_1^\dagger)^T$:
\begin{equation}
\label{qsm-destruction-operators-representation}
\begin{split}
  a_1
  &= a \otimes \bm{1} 
    \\
&=
\left(
\begin{matrix}
0 & 1 \cr 0 & 0
\end{matrix}
\right)\otimes \left(
\begin{matrix}
1 & 0 \cr 0 & 1
\end{matrix}
\right)
  \\
  &=
    {\scriptsize
 \left(
\begin{matrix}
0 & 0 & 1 & 0 \cr 
0 & 0 & 0 & 1 \cr 
0 & 0 & 0 & 0 \cr 
0 & 0 & 0 & 0 
\end{matrix}
 \right)
 },
 \end{split}
\hspace{0.5in}
\begin{split}
  a_2
    &= \sigma_3\otimes a
      \\
&=
\left(
\begin{matrix}
1 & 0 \cr 0 & -1
\end{matrix}
\right)\otimes \left(
\begin{matrix}
0 & 1 \cr 0 & 0
\end{matrix}
\right)
   \\
  &= 
    {\scriptsize
\left(
\begin{matrix}
0 & 1 & 0 & 0 \cr 
0 & 0 & 0 & 0 \cr 
0 & 0 & 0 & -1 \cr 
0 & 0 & 0 & 0 
\end{matrix}
\right)
}.
 \end{split}
\end{equation}

\subsection{\label{Conservative_quantum_logic_gates} Representations of perpendicular quantum gates}

 A type of quantum logic gate useful for casting quantum algorithms in various computational physics applications is a conservative quantum gate.  It is a 2-qubit universal quantum gate associated with perpendicular pairwise entanglement.
   A conservative quantum gate conserves the ``bit count'' in the number representation  of the qubit system (i.e. the total spin magnetization of a spin-$\frac{1}{2}$ system).
    If conservative quantum gates are used to model basic qubit-qubit interactions in a large qubit system, then the large scale dynamics of the qubit system is ultimately constrained by a number continuity equation, as was mentioned earlier.
   
   In the most general situation, it is sufficient to consider only a block diagonal matrix that has a $2\times 2$ sub-block, which causes entanglement and is a member of the special unitary group SU(2).  We can neglect the overall phase factor because this does not affect the quantum dynamics and therefore our sub-block need not be a member of the more general unitary group U(2).   If $  U$ is a member of SU(2), it can be parameterized using three real numbers, $\xi$, $\zeta$, and $\vartheta$, as follows
\begin{equation}
\label{qsm-su2-matrix}
  U \equiv \left(
\begin{matrix}
e^{i \xi}\cos \vartheta & -e^{i \zeta}\sin \vartheta\cr
-e^{-i \zeta}\sin \vartheta & -e^{-i\xi}\cos \vartheta 
\end{matrix}
\right)
=
\begin{pmatrix}
A      & B   \\
     C & D 
\end{pmatrix}.
\end{equation}

We can represent a general conservative quantum logical gate by the $4\times 4$ unitary matrix 
\begin{equation}
\label{idempotent-conservative-q-logic-gate-matrix}
 \Upsilon=
 {\scriptsize
 \left(
\begin{matrix}
1 & 0 & 0 & 0 \cr 
0 & A & B & 0 \cr 
0 & C & D & 0 \cr 
0 & 0 & 0 & E 
\end{matrix}
\right)
} .
\end{equation}
We choose this form for $  \Upsilon$ because we want to entangle only two of the basis states, $|01\rangle$ with $|10\rangle$, so as to conserve particle number, and that is why we call $  \Upsilon$ a {\it conservative} quantum gate.  The component in the top-left corner is set to unity because we do not want $  \Upsilon$ to alter the vacuum state $|00\rangle$ in any way.  However, we may allow the component in the bottom-right corner to be arbitrary.  We will see that the value of this component will depend on the particle statistics, reflecting whether quantum logic gates are used to model quantum gases with particles obeying Fermi statistics or not.

\subsubsection{Operator representation}

It is instructive to work out the ladder operators in the $Q=2$ case, where it is simple to write down the matrix representation. Remarkably, all the results carry over to the arbitrary size qubit systems with $Q\ge 2$. Consider the following five quadratic operators:
\begin{equation}
\label{qsm-flip-operators-representation}
  a^\dagger_1   a_2= 
   {\scriptsize
\left(
\begin{matrix}
0 & 0 & 0 & 0 \cr
0 & 0 & 0& 0 \cr 
0 & 1 & 0 & 0 \cr 
0 & 0 & 0 & 0 
\end{matrix}
 \right)
}
\hspace{0.5in}
  a^\dagger_2   a_1= 
   {\scriptsize
\left(
\begin{matrix}
0 & 0 & 0 & 0 \cr 
0 & 0 & 1 & 0 \cr 
0 & 0 & 0 & 0 \cr 
0 & 0 & 0 & 0
 \end{matrix}
 \right)},
\end{equation}
including the compound number operators
\begin{equation}
\label{compound-number-operators-representation}
\begin{matrix}
  n_1 ({\bf 1}-  n_2)= 
   {\scriptsize
\left(
\begin{matrix}
0 & 0 & 0 & 0 \cr 
0 & 0 & 0 & 0 \cr 
0 & 0 & 1 & 0 \cr 
0 & 0 & 0 & 0
 \end{matrix}
 \right)}
&\hspace{0.1in}&
({\bf 1}-  n_1)  n_2= 
 {\scriptsize
\left(
\begin{matrix}
0 & 0 & 0 & 0 \cr 
0 & 1 & 0 & 0 \cr 
0 & 0 & 0 & 0 \cr 
0 & 0 & 0 & 0 
\end{matrix}
\right)}
\cr
  n_1  n_2= 
   {\scriptsize
\left(
\begin{matrix}
0 & 0 & 0 & 0 \cr 
0 & 0 & 0 & 0 \cr 
0 & 0 & 0 & 0 \cr 
0 & 0 & 0 & 1 
\end{matrix}
\right)}.
\end{matrix}
\end{equation}
The conservative quantum gate (\ref{idempotent-conservative-q-logic-gate-matrix}) can be expressed in terms of the operators  (\ref{qsm-flip-operators-representation}) and (\ref{compound-number-operators-representation}) given above:
%
\begin{subequations}
\begin{eqnarray}
\nonumber
  \Upsilon & =&  {\bf 1} +(A-1) ({\bf 1}-  n_1)  n_2 + B  a^\dagger_2   a_1 + C  a^\dagger_1   a_2 
  \\
  &+& (D-1)  n_1({\bf 1}-  n_2) +(E-1)  n_1  n_2\\
\nonumber
& =&  {\bf 1} +(A-1)   n_2 + B  a^\dagger_2   a_1 + C  a^\dagger_1   a_2 \\
\label{idempotent-conservative-q-logic-gate-operator}
&+& (D-1)  n_1 -(A+D-E-1)  n_1  n_2 .
\end{eqnarray}
\end{subequations}
%
We would like to find the Hamiltonian, $H$ say, associated with $ \Upsilon$.  Letting $z$ denote a complex parameter, we begin by parametrizing (\ref{idempotent-conservative-q-logic-gate-operator}) in terms of $z$
\begin{equation}
\label{upsilon-series-expansion}
  \Upsilon(z) = e^{z  H},
\end{equation}
and then we solve for $  H$.  To do this, we series expand in the parameter $z$:
\begin{equation}
\label{upsilon_series_expansion_expansion}
  \Upsilon(z) =  {\bf 1} + z  H + \frac{z^2}{2}  H^2 + \cdots.
\end{equation}
There are two cases of interest: first when the Hamiltonian is idempotent, $  H^2 =   H$, then (\ref{upsilon_series_expansion_expansion}) reduces to 
\begin{equation}
\label{HsqEqualHcase}
  \Upsilon(z) = {\bf 1} + (e^z -1)   H,
\end{equation}
and second when $ H^2 \neq  H$ but $H^3 =   H$ and $  H^4 =   H^2$, then (\ref{upsilon-series-expansion}) reduces to 
\begin{equation}
\label{HcubedEqualHcase}
  \Upsilon(z) = {\bf 1} + \sinh z \,  H + (\cosh z -1)   H^2.
\end{equation}
These cases are worked out below.  A remarkable feature of this approach to deriving is that the imposition of the idempotent or tri-idempotent constraint will gives us a novel way to derive the exchange properties associated with Fermi statistics.

\subsubsection{$  H^2 =   H$ case}
\label{representations-of-conservative-quantum-gates}

From (\ref{idempotent-conservative-q-logic-gate-matrix}) and (\ref{HsqEqualHcase}),  we can solve for $  H$:
\begin{equation}
\label{idempotent-hamiltonian-matrix}
  H = \frac{1}{e^z-1} (\Upsilon - {\bf 1}) 
=
 \frac{1}{e^z-1}
 {\scriptsize
 \left(
 \begin{matrix}
0 & 0 & 0 & 0 \cr 
0 & A-1 & B & 0 \cr 
0 & C & D-1 & 0 \cr 
0 & 0 & 0 & E-1 
\end{matrix}
\right)} .
\end{equation}

Let us pick a new set of variables to simplify matters:
\begin{subequations}
\label{variable-substitution-1}
\begin{eqnarray}
{\cal A} & = & \frac{A-1}{e^z -1} \hspace{0.25in} {\cal B}  =  \frac{B}{e^z -1}\\
\label{variable-substitution-2}
{\cal C} & = & \frac{C}{e^z -1} \hspace{0.25in} {\cal D}  =  \frac{D-1}{e^z -1}\\
\delta & = &  \frac{E-1}{e^z -1}.
\end{eqnarray}
\end{subequations}
Then inserting (\ref{variable-substitution-1}) into (\ref{idempotent-hamiltonian-matrix}), the Hamiltonian has the simple matrix and operator representation
\begin{equation}
\label{idempotent-hamiltonian-matrix-form2}
  H 
=  
  {\scriptsize
\left(
 \begin{matrix}
0 & 0 & 0 & 0 \cr 
0 & {\cal A} & {\cal B} & 0 \cr 
0 & {\cal C} & {\cal D} & 0 \cr 
0 & 0 & 0 & \delta 
\end{matrix}
\right)},
\end{equation}
and from this we deduce the operator form of the idempotent Hamiltonian
\begin{equation}
\label{idempotent-hamiltonian-operator-form2}
  H =   
 {\cal B}  a^\dagger_2   a_1 
+ {\cal C}  a^\dagger_1   a_2 
+ {\cal D}  n_1({\bf 1}-  n_2) +{\cal A}({\bf 1}-  n_1)  n_2 
+\delta   n_1  n_2 .
\end{equation}

Next, inserting the new variables (\ref{variable-substitution-1}) into (\ref{idempotent-conservative-q-logic-gate-matrix}) and (\ref{idempotent-conservative-q-logic-gate-operator}), the matrix and operator representations for the conservative quantum logic gate become
\begin{subequations}
\begin{eqnarray}
\label{idempotent-conservative-q-logic-gate-form2}
 \Upsilon(z) & = & e^{z   H} 
 \\
 \nonumber
& = &  
{\scriptsize
\left(
\begin{matrix}
1 & 0 & 0 & 0 \cr 
0 &(e^z-1){\cal A}+1 & (e^z-1){\cal B} & 0 \cr 
0 & (e^z-1){\cal B}^\dagger  & (e^z-1){\cal D}+1 & 0 \cr 
0 & 0 & 0 & (e^z-1)\delta +1 
\end{matrix}
\right) 
}
\qquad
\\
\\
\nonumber
  & = &  {\bf 1} 
+(e^z-1)\Big[{\cal B}  a^\dagger_2   a_1 
+ {\cal C}  a^\dagger_1   a_2 
\\
&+& {\cal D}  n_1({\bf 1}-  n_2) +{\cal A}({\bf 1}-  n_1)  n_2 
+\delta   n_1  n_2\Big] .
\end{eqnarray}
Since the Hamiltonian must be Hermitian, $  H=  H^\dagger$, we know that ${\cal C} ={\cal B}^\dagger$ and $\delta=\delta^\dagger$, so $\delta$ must be a real valued number.   Also, since the Hamiltonian is idempotent, $  H^2=  H$, we get the additional constraint equations on the components:
\begin{eqnarray}
{\cal A}^2 - {\cal A} + |{\cal B}|^2 & = & 0\\
{\cal A} + {\cal D}& = &1\\
{\cal D}^2 - {\cal D} + |{\cal B}|^2 &=& 0,
 \end{eqnarray}
 which admit the solutions:
 \begin{eqnarray}
 \label{idempotent-solution1}
{\cal A} & =& \frac{1}{2}\left(1 \pm \sqrt{1-4|{\cal B}|^2 } \right)\\
 \label{idempotent-solution2}
{\cal D} & =& \frac{1}{2} \left(1 \mp  \sqrt{1-4|{\cal B}|^2 } \right).
\end{eqnarray}
\end{subequations}
Then inserting (\ref{idempotent-solution1}) and (\ref{idempotent-solution2}) into (\ref{idempotent-hamiltonian-matrix-form2}) and (\ref{idempotent-hamiltonian-operator-form2}), we can specify the idempotent Hamiltonian with only one free complex parameter:
\begin{subequations}
\begin{eqnarray}
  H 
& = & 
  {\scriptsize
\left(
 \begin{matrix}
0 & 0 & 0 & 0 \cr 
0 & \frac{1}{2} \pm  \frac{1}{2}\sqrt{1-4|{\cal B}|^2 }  & {\cal B} & 0 \cr 
0 & {\cal B}^\dagger &\frac{1}{2} \mp  \frac{1}{2}\sqrt{1-4|{\cal B}|^2 }  & 0 \cr 
0 & 0 & 0 & \delta 
\end{matrix}
\right)} \\
\nonumber
&  = &  
 {\cal B}  a^\dagger_2   a_1 
+ {\cal B}^\dagger  a^\dagger_1   a_2 
+\frac{1}{2} \left(1 \mp \sqrt{1-4|{\cal B}|^2 }\right)  n_1 ({\bf 1}-  n_2)
\\
&+&
\frac{1}{2}\left(1 \pm  \sqrt{1-4|{\cal B}|^2 } \right)({\bf 1}-  n_1)  n_2 
+\delta  n_1  n_2\\
\nonumber
&  = &  
 {\cal B}  a^\dagger_2   a_1 
+ {\cal B}^\dagger  a^\dagger_1   a_2 
+\frac{1}{2} \left(1 \mp \sqrt{1-4|{\cal B}|^2 }\right)  n_1
\\
&+&
\frac{1}{2}\left(1 \pm  \sqrt{1-4|{\cal B}|^2 } \right)  n_2 
+(\delta-1)  n_1  n_2  .
\end{eqnarray}
\end{subequations}
The associated conservative quantum logic gate can also be rewritten by inserting (\ref{idempotent-solution1}) and (\ref{idempotent-solution2}) into (\ref{idempotent-conservative-q-logic-gate-form2}):
\begin{widetext}
\begin{subequations}
\begin{eqnarray}
\nonumber
 \Upsilon(z)&=&
 {\scriptsize
\left(
\begin{matrix}
1 & 0 & 0 & 0 \cr 
0 &  \frac{1}{2}(e^z+1) \pm  \frac{1}{2}(e^z-1)\sqrt{1-4|{\cal B}|^2 }& (e^z-1) {\cal B}& 0 \cr 
0 & (e^z-1) {\cal B}^\dagger &  \frac{1}{2}(e^z+1) \mp \frac{1}{2} (e^z-1)\sqrt{1-4|{\cal B}|^2 } & 0 \cr 
0 & 0 & 0 &  (e^z-1)\delta+1 
\end{matrix}
\right) 
}
\\
\\
\nonumber
 & = &  {\bf 1} 
+ (e^z-1)\Big[
 {\cal B}  a^\dagger_2   a_1 
+ {\cal B}^\dagger  a^\dagger_1   a_2
+
\frac{1}{2} \left(1 \mp \sqrt{1-4|{\cal B}|^2 }\right)  n_1
+\frac{1}{2}\left(1 \pm  \sqrt{1-4|{\cal B}|^2 } \right)  n_2 
+(\delta-1)  n_1  n_2 
\Big].
\qquad\ \ 
\end{eqnarray}
\end{subequations}
%
%
\end{widetext}

\subsubsection{ {\sc swap} gate and entangling $\sqrt{\text{\sc swap}}$ gate}

Finally, for $z=i\pi$ we get the  quantum swap gate
\begin{subequations}
\begin{eqnarray}
\label{quantum-swap-gate}
 \Upsilon(i\pi) 
 & =&  
  {\scriptsize
\begin{pmatrix}
1 & 0 & 0 & 0 \cr 
0 & 0 & e^{-i\xi} & 0 \cr 
0 & e^{i\xi} & 0 & 0 \cr 
0 & 0 & 0 & 1-2\delta 
\end{pmatrix}
}
\\ 
\nonumber
 &=& 
  {\bf 1} 
-\left(   a^\dagger_1-e^{-i\xi}  a^\dagger_2\right)\left(  a_1-e^{i\xi}  a_2\right)
-2(\delta -1)  n_1   n_2 .
\\
\end{eqnarray}
\end{subequations}
For $\xi=0$ and $\delta=0$, (\ref{quantum-swap-gate}) is a classical  {\sc swap} gate.

To satisfy the unitary condition for our quantum logic gate, $\Upsilon \Upsilon^\dagger=1$, we must restrict the real-valued component $\delta$ by the following constraint equation:
\begin{equation}
(1-2\delta)^2 = 1,
\end{equation}
which implies that either $\delta=0$ or $\delta =1$.   Then, our quantum swap gate (\ref{quantum-swap-gate}) can be rewritten as:
\begin{equation}
\label{quantum-swap-gate-fermion-boson}
 \Upsilon(i\pi)  =  
  {\scriptsize
\begin{pmatrix}
1 & 0 & 0 & 0 \cr 
0 & 0 & e^{-i\xi} & 0 \cr 
0 & e^{i\xi} & 0 & 0 \cr 
0 & 0 & 0 & \pm 1 
\end{pmatrix}
},
\end{equation}
where the plus sign applies for the $\delta=0$ case and the minus sign for the $\delta =1$ case.  For $z=\frac{i\pi}{2}$ we get the entangling $\sqrt{\text{\sc swap}}$ gate
\begin{widetext}
\begin{equation}
 \Upsilon\left(\frac{i\pi}{2}\right)  =  
\left(
\begin{matrix}
1 & 0 & 0 & 0 \cr 
0 & \frac{1}{2}+\frac{i}{2}& \left(\frac{1}{2}-\frac{i}{2}\right)e^{-i\xi} & 0 \cr 
0 &\left(\frac{1}{2}-\frac{i}{2}\right)e^{i\xi} & \frac{1}{2}+\frac{i}{2} & 0 \cr 
0 & 0 & 0 &  (i-1)\delta+1 
\end{matrix}
\right) 
=
{\bf 1} 
+(i-1)\left[\frac{1}{2} \left(   a^\dagger_1-e^{-i\xi}  a^\dagger_2\right)\left(  a_1-e^{i\xi}  a_2\right) +(\delta -1)  n_1  n_2\right].
\end{equation}
\end{widetext}
%
%

\subsubsection{$H^3=H$ case}

There exists an alternative Hamiltonian that is not idempotent but has a similar property at third order, $H^3=  H$ but neither idempotent nor an involution (i.e. $H^2\neq  H$  and  $H^2\neq 1$), which can generate a conservative quantum logic gate of the form (\ref{idempotent-conservative-q-logic-gate-matrix}). 
In this second case,  the series expansion of the quantum gate (\ref{upsilon-series-expansion}) reduces to the form (\ref{HcubedEqualHcase}), which is
\begin{displaymath}
  \Upsilon(z) = {\bf 1} + (\cosh z-1)   H^2 + \sinh z   H.
\end{displaymath}
Our approach will be to assume the Hamiltonian still has the form (\ref{idempotent-hamiltonian-matrix-form2}) and that its square has a diagonal matrix form:
\begin{subequations}
\begin{eqnarray}
\label{non-idempotent-hamiltonian-matrix-squared}
  H^2 
&=&
 {\scriptsize
\left(
\begin{matrix}
0 & 0 & 0 & 0 \cr 
0 & {\cal A} & {\cal B} & 0 \cr 
0 & {\cal B}^\dagger & {\cal D} & 0 \cr 
0 & 0 & 0 & \delta 
\end{matrix}
\right)
\cdot
\left(
\begin{matrix}
0 & 0 & 0 & 0 \cr 
0 & {\cal A} & {\cal B} & 0 \cr 
0 & {\cal B}^\dagger & {\cal D} & 0 \cr 
0 & 0 & 0 & \delta 
\end{matrix}
\right)
}
 =
  {\scriptsize
\left(
 \begin{matrix}
0 & 0 & 0 & 0 \cr 
0 & 1& 0 & 0 \cr 
0 & 0 & 1 & 0 \cr 
0 & 0 & 0 & \delta 
\end{matrix}
\right)
} 
\qquad
\ 
\\
&  = &  
   n_1 ({\bf 1}-  n_2)
+({\bf 1}-  n_1)  n_2 
+\delta   n_1  n_2\\
&  = &  
  n_1
+  n_2 
+(\delta-2)  n_1  n_2,
\end{eqnarray}
\end{subequations}
where as in the previous case either $\delta = 0$ or $\delta=1$.
This imposes the following constraint equations on the components:
\begin{subequations}
\begin{eqnarray}
{\cal A}^2 & = & 1- |{\cal B}|^2 \\
{\cal A} + {\cal D}& = &0\\
{\cal D}^2 & = & 1- |{\cal B}|^2 ,
 \end{eqnarray}
\end{subequations}
 which admit the solutions:
\begin{subequations}
 \begin{eqnarray}
 \label{non-idempotent-solution1}
{\cal A} & =& \pm \sqrt{1-|{\cal B}|^2 } \\
 \label{non-idempotent-solution2}
{\cal D} & =& \mp \sqrt{1-|{\cal B}|^2 } .
\end{eqnarray}
\end{subequations}
%
%
Then, the Hamiltonian has the form
\begin{subequations}
\begin{eqnarray}
\label{non-idempotent-hamiltonian-matrix}
  H 
&= &
 {\scriptsize
\left(
\begin{matrix}
0 & 0 & 0 & 0 \cr 
0 & \pm \sqrt{1-|{\cal B}|^2 } & {\cal B} & 0 \cr 
0 & {\cal B}^\dagger &\mp \sqrt{1-|{\cal B}|^2 }  & 0 \cr 
0 & 0 & 0 & \delta 
\end{matrix}
\right)
}
\\
\nonumber
&  = &  
 {\cal B}\,  a^\dagger_2   a_1 
+ {\cal B}^\dagger  a^\dagger_1   a_2 
\mp \sqrt{1-|{\cal B}|^2}\,   n_1 ({\bf 1}-  n_2)
\\
&\pm&  \sqrt{1-|{\cal B}|^2}\,({\bf 1}-  n_1)  n_2 
+\delta   n_1  n_2\\
\nonumber
&  = &  
 {\cal B}\,  a^\dagger_2   a_1 
+ {\cal B}^\dagger  a^\dagger_1   a_2 
 \mp \sqrt{1-|{\cal B}|^2 }\,  n_1
\\
& \pm&
  \sqrt{1-|{\cal B}|^2 }\,   n_2 
+\delta  n_1  n_2,
\end{eqnarray}
\end{subequations}
and hence, using (\ref{HcubedEqualHcase}), the matrix representation of the conservative quantum gate becomes
%
\begin{widetext}

\begin{subequations}
\begin{eqnarray}
\label{non-idempotent-conservative-q-logic-gate-matrix}
\nonumber
  \Upsilon(z) 
& = & 
 {\scriptsize
\left(
\begin{matrix}
1 & 0 & 0 & 0 \cr 
0 & \cosh z \pm \sqrt{1-|{\cal B}|^2 } \sinh z & {\cal B}\sinh z & 0 \cr 
0 & {\cal B}^\dagger\sinh z &\cosh z \mp \sqrt{1-|{\cal B}|^2 } \sinh z & 0 \cr 
0 & 0 & 0 & (e^z -1)\delta+1 
\end{matrix}
\right)
}\\
&&\\
\nonumber
& = &  {\bf 1} 
+ (\cosh z-1) 
\left[
  n_1
+  n_2 
+(\delta-2)  n_1  n_2 
\right]
 \\
\nonumber
& & 
+
\sinh z \left[
{\cal B} \,  a^\dagger_2   a_1 
+ {\cal B}^\dagger\,   a^\dagger_1   a_2 
 \mp \sqrt{1-|{\cal B}|^2 }  n_1
 \pm  \sqrt{1-|{\cal B}|^2 }   n_2 
+\delta  n_1  n_2
\right]
\\
\\
\nonumber
& = &  {\bf 1} 
+ \sinh z {\cal B}\,   a^\dagger_2   a_1 
+ \sinh z {\cal B}^\dagger \,  a^\dagger_1   a_2 \\
\nonumber
&+&
 (\cosh z-1 \mp \sqrt{1-|{\cal B}|^2 }) \,  n_1
+(\cosh z -1\pm \sqrt{1-|{\cal B}|^2 })\,   n_2 \\
&+&\left[(e^z-1)\delta - 2(\cosh z -1)\right]  n_1  n_2 .
\end{eqnarray}
\end{subequations}
%
\end{widetext}
A useful special case occurs for ${\cal B} = ie^{-i\xi}$.  Then,
\begin{subequations}
\begin{eqnarray}
\label{non-idempotent-hamiltonian-matrix-B=i}
  H 
&=& 
{\scriptsize
\left(
\begin{matrix}
0 & 0 & 0 & 0 \cr 
0 & 0 & ie^{-i\xi} & 0 \cr 
0 &-ie^{i\xi} &0 & 0 \cr 
0 & 0 & 0 & \delta 
\end{matrix}
\right) }
\\
 &=&   
 ie^{-i\xi}  a^\dagger_2   a_1 
-ie^{i\xi}  a^\dagger_1   a_2 
+\delta   n_1  n_2
\\
\nonumber
&  =&   
 \left(  a^\dagger_1 + i e^{-i\xi}  a^\dagger_2 \middle) \middle(  a_1 - i e^{i\xi}  a_2\right) -  n_1 -   n_2
+\delta   n_1  n_2.
\\
\label{non-idempotent-hamiltonian-matrix-B=i-form2}
\end{eqnarray}
\end{subequations}
The quantum gate has the form: 
\begin{subequations}
\begin{eqnarray}
\label{non-idempotent-conservative-q-logic-gate-matrix-B=i}
  \Upsilon(z) 
& = & 
{\scriptsize
\left(
\begin{matrix}
1 & 0 & 0 & 0 \cr 
0 & \cosh z & i e^{-i\xi}\sinh z  & 0 \cr 
0 & -i e^{i\xi}\sinh z  &\cosh z & 0 \cr 
0 & 0 & 0 & (e^z -1)\delta+1 
\end{matrix}
\right)}
\qquad
\\
\nonumber
& = &  {\bf 1} 
+i \sinh z  \left(e^{-i\xi}  a^\dagger_2   a_1 
-e^{i\xi}   a^\dagger_1   a_2\right) 
\\
\nonumber
&+&
 (\cosh z-1 ) (  n_1
+  n_2 )
\\
&+&
\left[(e^z-1)\delta - 2(\cosh z -1)\right]  n_1  n_2.
\end{eqnarray}
\end{subequations}
%
\subsubsection{{\sc aswap} gate and entangling $\sqrt{\text{\sc aswap}}$ gate}

Finally, for $z=\frac{i\pi}{2}$ we get the asymmetric quantum gate
\begin{eqnarray}
\label{antisymmetric-quantum-swap-gate}
  \Upsilon\left(\frac{i\pi}{2}\right) 
&=&
\left(
\begin{matrix}
1 & 0 & 0 & 0 \cr 
0 & 0 & -e^{-i\xi} & 0 \cr 
0 & e^{i\xi} & 0 & 0 \cr 
0 & 0 & 0 & (i -1)\delta+1 
\end{matrix}
\right) 
\\
\nonumber
& = &
  {\bf 1} 
+e^{i\xi}  a^\dagger_1   a_2 
-e^{-i\xi}  a^\dagger_2   a_1 
\\
& -&  n_1
-  n_2
+
[(i-1)\delta +2]  n_1   n_2.
\end{eqnarray}
For $\xi=0$ and $\delta=0$, (\ref{antisymmetric-quantum-swap-gate}) is the classical antisymmetric swap gate. 

For $z=\frac{i\pi}{4}$ we get the entangling $\sqrt{\text{\sc aswap}}$ gate
\begin{widetext}
\begin{subequations}
\begin{eqnarray}
\label{antisymmetric-quantum-square-root-of-swap-gate}
  \Upsilon\left(\frac{i\pi}{4}\right) 
& = & 
\left(
\begin{matrix}
1 & 0 & 0 & 0 \cr 
0 & \frac{1}{\sqrt{2}} & -\frac{1}{\sqrt{2}}e^{-i\xi} & 0 \cr 
0 & \frac{1}{\sqrt{2}}e^{i\xi} & \frac{1}{\sqrt{2}} \cr 
0 & 0 & 0 & (e^{\frac{i\pi}{4}} -1)\delta+1 
\end{matrix}
\right)\\
\nonumber
& = &  {\bf 1} 
+\frac{1}{\sqrt{2}} \left(e^{i\xi}  a^\dagger_1   a_2 
- e^{-i\xi}  a^\dagger_2   a_1 \middle) +
 \middle(\frac{1}{\sqrt{2}}-1 \middle) \middle(  n_1
+  n_2 
-2  n_1  n_2 \middle)
+ \middle(e^{\frac{i\pi}{4}} -1\right)\delta  n_1  n_2.
\\
\end{eqnarray}
\end{subequations}
\end{widetext}
%

\section{Chiral symmetry breaking operator for the Dirac equation  algorithm}
\label{Collide_operator_for_the_Dirac_equation}

Here we derive a quantum gate representation of the collide operator used in the quantum lattice gas algorithm for a system of Dirac particles in 1+1 dimensions.  The matrix representation of the collision operator that acts on a Dirac 2-spinor is
\begin{equation*}
U_C 
=
\begin{pmatrix}
  \sqrt{1-\left(\frac{mc^2\tau}{\hbar}\right)^2}    &
-i   \frac{mc^2\tau}{\hbar}\, e^{- i \frac{mc\ell }{\hbar}\sqrt{\gamma^2-1} }  
      \\
-i \,     \frac{mc^2\tau}{\hbar} e^{ i \frac{mc\ell}{\hbar}\sqrt{\gamma^2-1}  }   
      & 
       \sqrt{1-\left(\frac{mc^2\tau}{\hbar}\right)^2}
\end{pmatrix}.
\end{equation*}
In 1+1 dimensions, a 2-spinor field is sufficient to describe the  Dirac particle
\begin{equation*}
\psi = 
\begin{pmatrix}
      \psi_\uparrow    \\
      \psi_\downarrow
\end{pmatrix}.
\end{equation*}

Let us start with a change of variables
\begin{equation*}
{\cal B} = e^{- i \frac{mc\ell }{\hbar}\sqrt{\gamma^2-1} }  
\qquad
\text{and}\qquad
\cosh z  =    \sqrt{1-\left(\frac{mc^2\tau}{\hbar}\right)^2},
\end{equation*}
which implies
\begin{equation*}
-i   \frac{mc^2\tau}{\hbar} = \sinh z .
\end{equation*}
Then the collide operator is
\begin{equation*}
U_C 
=
\begin{pmatrix}
  \cosh z   &
 {\cal B} \sinh z
      \\
{\cal B}^\ast \sinh z
      & 
    \cosh z
\end{pmatrix}
=
\cosh z 
+
\begin{pmatrix}
 0   &
 {\cal B}
      \\
{\cal B}^\ast
      & 
    0
\end{pmatrix}
\sinh z .
\end{equation*}
Since 
\begin{equation*}
\begin{pmatrix}
 0   &
 {\cal B}
      \\
{\cal B}^\ast
      & 
    0
\end{pmatrix}^2 = \bm{1},
\end{equation*}
we may use Euler's identity to write collide operator as

\begin{equation*}
U_C 
=
\exp\left[
z
\begin{pmatrix}
 0   &
 {\cal B}
      \\
{\cal B}^\ast
      & 
    0
\end{pmatrix}
\right].
\end{equation*}
Hence, in the perpendicular subspace, we see that the hermitian generator of an entangling gate representation  of $U_C$ should have the form
\begin{equation*}
N = 
\begin{pmatrix}
   0   & 0 & 0 & 0    \\
    0   & 0 & {\cal B} & 0\\
    0 & {\cal B}^\ast & 0 & 0\\
    0 & 0 & 0 & 1 
\end{pmatrix}
\quad
\longrightarrow
\quad
N^2 = 
\begin{pmatrix}
   0   & 0 & 0 & 0    \\
    0   & 1 & 0 & 0\\
    0 & 0 & 1 & 0\\
    0 & 0 & 0 & 1 
\end{pmatrix}.
\end{equation*}
From this matrix representation, we see that the generator is the tri-idempotent type and not the involution nor idempotent type (i.e. $N^3=N$ and $N^2\neq \bm{1}$ and $N^2 \neq N$).

Using the matrix representation from part (c) as a guide, the tri-idempotent generator can now be written in an analytical form in terms the qubit creation and annihilation operators. This form  is useful for many-body quantum simulations of a system of Dirac particles.  So from part (c) above, we see that $N$ and $N^2$ both have three terms, which we write down by inspection
\begin{equation*}
N_{\alpha\beta} = 
{\cal B} \, a^\dagger_\beta a_\alpha 
+ {\cal B}^\ast \, a^\dagger_\alpha a_\beta + n_\alpha n_\beta
\end{equation*}
and
\begin{equation*}
N_{\alpha\beta}^2 = 
n_\alpha +  
n_\beta - n_\alpha n_\beta,
\end{equation*}
where $\alpha$ is the label of a qubit at some point where the qubit encodes the occupancy of a spin-up Dirac particle at that point and $\beta$ is the label of another qubit at that same point and this other qubit encodes the occupancy of a spin-down Dirac particle at that point.

For a tri-idempotent generator,  again in terms qubit creation and annihilation operators, we know the entangling gate has the analytical form
\begin{widetext}
\begin{eqnarray*}
\Upsilon_{\alpha\beta} 
& =& 
e^{z N_{\alpha\beta}}
\\
& =& 
 1 + \left(\cosh z -1\right) N_{\alpha\beta}^2 +  N_{\alpha\beta}   \sinh z 
\\
& = &
1+ 
\left(\cosh z -1\right)\left(
n_\alpha +  
n_\beta - n_\alpha n_\beta
\right)
+
\sinh z \left( 
{\cal B} \, a^\dagger_\beta a_\alpha 
+ {\cal B}^\ast \, a^\dagger_\alpha a_\beta + n_\alpha n_\beta
\right)
\\
& = &
1
-n_\alpha -  
n_\beta + n_\alpha n_\beta 
+
\sinh z \left( 
{\cal B} \, a^\dagger_\beta a_\alpha 
+ {\cal B}^\ast \, a^\dagger_\alpha a_\beta 
\right)
+
\cosh z \left(
n_\alpha +  
n_\beta 
- 2 \,n_\alpha n_\beta
\right)
+
e^z\, n_\alpha n_\beta
\\
& = &
1
-n_\alpha -  
n_\beta + n_\alpha n_\beta 
-
i   \frac{mc^2\tau}{\hbar}\, \left( 
e^{- i \frac{mc\ell }{\hbar}\sqrt{\gamma^2-1} }  \, a^\dagger_\beta a_\alpha 
+ h.c.
\right)
+
\text{\scriptsize $\sqrt{1-\left(\frac{mc^2\tau}{\hbar}\right)^2} $}
\Big(
n_\alpha +  
n_\beta - 2\,n_\alpha n_\beta
\Big)
\\
&+&
\left(
\text{\scriptsize $\sqrt{1-\left(\frac{mc^2\tau}{\hbar}\right)^2} $}
-
i   \frac{mc^2\tau}{\hbar}
\right)
 n_\alpha n_\beta.
\end{eqnarray*}
So to scatter the $\psi_\uparrow$ and $\psi_\downarrow$ particles at a point, one should use the collide operator
\begin{eqnarray*}
\therefore
\quad
\Upsilon 
&=&
1
-n_\uparrow -  
n_\downarrow + n_\uparrow n_\downarrow 
-
\frac{i\,mc^2\tau}{\hbar}\, \left( 
e^{- i \frac{mc\ell }{\hbar}\sqrt{\gamma^2-1} }  \, a^\dagger_\downarrow a_\uparrow 
+ h.c.
\right)
+
\text{\scriptsize $\sqrt{1-\left(\frac{mc^2\tau}{\hbar}\right)^2} $}
\Big(
n_\uparrow +  
n_\downarrow - 2\,n_\uparrow n_\downarrow
\Big)
\\
&+&
\exp\left[{-i \cos^{-1} \text{\scriptsize $\sqrt{1-\left(\frac{mc^2\tau}{\hbar}\right)^2} $}  }\right]
 n_\uparrow n_\downarrow.
\end{eqnarray*}
In the $(\uparrow,\downarrow)$ subspace at a point, the entangling gate has the matrix representation
\begin{eqnarray*}
\Upsilon 
& =&
\begin{pmatrix}
1 & 0 & 0 & 0\\
0 &   \cosh z   &  {\cal B} \sinh z & 0 \\
0& {\cal B}^\ast \sinh z &  \cosh z & 0\\
0 & 0 & 0 & e^z
\end{pmatrix}
\\
&=&
\begin{pmatrix}
1 & 0 & 0 & 0\\
0 &   \sqrt{1-\left(\frac{mc^2\tau}{\hbar}\right)^2}    &
-i   \frac{mc^2\tau}{\hbar}\, e^{- i \frac{mc\ell }{\hbar}\sqrt{\gamma^2-1} }  & 0 \\
0& -i \,     \frac{mc^2\tau}{\hbar} e^{ i \frac{mc\ell}{\hbar}\sqrt{\gamma^2-1}  }   
      & 
       \sqrt{1-\left(\frac{mc^2\tau}{\hbar}\right)^2} & 0\\
0 & 0 & 0 & \exp\left[{-i \cos^{-1}  \sqrt{1-\left(\frac{mc^2\tau}{\hbar}\right)^2}  }\right]
\end{pmatrix}.
\end{eqnarray*}
\end{widetext}

\end{document}